\documentclass[12pt, reqno, titlepage]{amsart}

\usepackage{thmtools}

\usepackage[thm_section]{macros}%
\usepackage{amssymb}
\usepackage{enumitem}
\usepackage{clipboard}
\newclipboard{output-draft}
\usepackage{booktabs,subcaption}

\renewcommand{\theequation}{\thesection.\arabic{equation}}
\newcommand{\rone}{\text{I}}
\newcommand{\rtwo}{\text{II}}
\newcommand{\rthree}{\text{III}}
\newcommand{\rfour}{\text{IV}}
\newcommand{\rfive}{\text{V}}

\newgeometry{margin=1in}
\usepackage[most,breakable]{tcolorbox}
\allowdisplaybreaks

\renewcommand{\emptyset}{\varnothing}

\tcbset{
    frame code={}
    center title,
    breakable,
    left=1em,
    right=1em,
    top=0pt,
    bottom=0pt,
    colback=gray!5,
    colframe=white,
    width=\dimexpr\textwidth\relax,
    fontupper=\footnotesize,
    enlarge left by=0mm,
    boxsep=5pt,
    arc=0pt,outer arc=0pt,
}

\newcommand\DoToC{%
  \startcontents
  \printcontents{}{0}{\textbf{Contents}\vskip1em\hrule\vskip1em}
  \vskip1em\hrule\vskip5pt
}

\newenvironment{myproof}{
    \begin{proof}
    
}{
    \qedhere
    \end{proof}
}
\usetikzlibrary{arrows.meta, decorations.pathreplacing}

\title{Nonparametric Treatment Effect Identification in School
Choice}
\author{Jiafeng Chen \\ Department of Economics, Stanford University}

\date{\today.
\href{mailto:jiafeng@stanford.edu}{jiafeng@stanford.edu}. I am especially indebted to
 David C. Parkes, Scott Duke Kominers, Isaiah Andrews, and Elie Tamer for their valuable
 guidance. I thank Aureo de Paula, an anonymous associate editor and four anonymous
 referees for helpful comments and suggestions. Additionally, I thank Marinho Bertanha,
 St\'ephane Bonhomme, Xavier D'Haultfoeuille, Bryan Graham, Jeff Gortmaker, Peter Hull,
 Guido Imbens, Patrick Kline, Ashesh Rambachan, David Ritzwoller, Brad Ross, Jonathan
 Roth, Jesse Shapiro, Neil Shephard, Winnie van Dijk, Christopher Walker, as well as
 participants of the Harvard Econometrics Workshop and SOLE 2021. This paper subsumes my
 undergraduate thesis at Harvard College \citep{ChenJiafeng2019CIiM}, as well as SSRN
 submissions 3510897, 3510899, and 3510903 in January, 2020. ChatGPT-o3 provided
 excellent research assistance. All errors are my own. }

\begin{document}

\begin{abstract} 
This paper studies nonparametric identification and estimation of causal
 effects in centralized school assignment. In many centralized assignment algorithms,
 students face  both lottery-driven variation and regression discontinuity- (RD) driven
 variation. We characterize the full set of identified \emph{atomic treatment effects}
 (aTEs), defined as the conditional average treatment effect between a pair of schools
 given student characteristics. Atomic treatment effects are the building blocks of more
 aggregated treatment contrasts, and common approaches to estimating aTE aggregations can
 mask important heterogeneity. In particular, many aggregations of aTEs put zero weight
 on aTEs driven by RD variation, and estimators of such aggregations put asymptotically
 vanishing weight on the RD-driven aTEs. We provide a diagnostic and recommend new
 aggregation schemes. Lastly, we provide estimators and asymptotic results for inference
 on these aggregations.

    \vspace{3em}
    {\slshape \noindent Keywords: School choice,
    nonparametric identification, regression discontinuity, heterogeneous
    treatment effects
        
        \noindent JEL codes: C1, D47, J01}
\end{abstract}
\thispagestyle{empty}

 \maketitle

\section{Introduction}

\Copy{specificstudy}{A rapidly growing empirical literature studies causal
effects of schools using centralized school assignment.\footnote{See, among others,
\citet
 {kapor2020heterogeneous,agarwal2018demand,fack2019beyond,calsamiglia2020structural,angrist2020simple,beuermann2018good,abdulkadirouglu2017research,abdulkadiroglu2017impact,abdulkadirouglu2020parents,abdulkadroglu2017regression,marinho2022causal,kirkeboen2016field,angrist2021credible,ketel2023unimport,angrist2022race}.}
 In several centralized assignment systems, two sources of variation determine how
 schools enroll applicants and drive quasi-experimental identification. For instance, in
 New York City, studied by
\citet{abdulkadiroglu2017impact}, some schools randomize priority to students, where
students with higher lottery numbers are favored over others. Other schools use
non-lottery
tiebreakers, such as test scores, to distinguish students with otherwise similar
characteristics. 

These sources of variation provide opportunities for causal inference. For lottery
schools, comparing lucky students who receive favorable priorities to those less
fortunate estimates a causal effect. For test-score schools, comparing students just above a cutoff to those just below also identifies causal effects.
\citet{abdulkadiroglu2017impact} use both sources of variation to measure school
effectiveness in New York City. School districts elsewhere, such as Chicago and Boston, also feature schools with lottery and non-lottery
tiebreakers.\footnote{\citet{angrist2023methods} write ``Some centralized assignment
schemes, such as those used for Boston and New York City exam schools and New York City
screened schools, employ non-lottery tiebreakers such as test scores instead of, or
alongside, lottery numbers.'' Schools in the Chicago Public School High School Choice
program use either a lottery system or a points system for giving priority to students,
where the points system is a composite of test scores and other student characteristics.
See \url{https://www.cps.edu/gocps/high-school/results/choice/}. }

We refer to the first source of variation as lottery variation and the second as
regression-discontinuity (RD) variation. Researchers often estimate aggregate causal
quantities, such as contrasts between sets of ``treatment'' and ``control'' schools,
pooling over lottery and RD variation and over pairs of schools. \citet
{abdulkadiroglu2017impact} do so for New York City, estimating aggregate effects
of enrolling in a school receiving an ``A'' grade on the school district's report card
for quality, compared to enrolling in non-Grade-A schools. }

 These aggregate estimates provide convenient causal summaries but can mask rich
 heterogeneity by pooling many different comparisons. These contrasts may involve
 different pairs of schools, different student characteristics, or different sources of
 identifying variation. Aggregate causal estimands are often interpreted as simple
 average treatment effects between treatment and control schools. Because these aggregate
 effects blend together different conditional average treatment effects, their
 interpretation should be more nuanced.

In particular, economically, heterogeneity along RD variation versus lottery variation
could be meaningful. Consider a lottery school $s_1$ and a test-score school $s_0$. In many centralized algorithms, lottery variation between these schools occurs only for students preferring $s_{1}$ to $s_{0}$, driven by whether they lose the $s_{1}$ lottery. Meanwhile, RD
variation exists only for students who prefer $s_0$ to $s_1$---driven by whether these
students narrowly qualify or fail the test used by $s_0$. If submitted
preferences select positively on gains, the causal contrasts identified by lottery (resp.
RD) variation  represent subpopulations with more (resp. less) favorable potential
outcomes under $s_1$. Different aggregations  thus put different weights
on subpopulations with different attitudes toward $s_1$.

Motivated by the rich heterogeneity, this paper seeks to reduce causal effects to their
smallest unit, which we call \emph{atomic treatment effects} (aTEs). We define an atomic
treatment effect as a conditional average treatment effect between two schools,
conditional on all observed student characteristics. Our first contribution is to
characterize the set of point-identified aTEs based solely on lottery and RD variation
from the assignment mechanism. Since any identified average treatment effect is thus a
weighted average over aTEs, this characterizes the set of estimable treatment effect
parameters in these settings.

Our characterization yields a clean classification of aTEs as driven by RD or lottery
variation. Building on this classification, one concerning observation is that common
regression estimators---often derived under homogeneous treatment effect
assumptions---implicitly aggregate aTEs using weights that are not chosen by the researcher and can depend on the sample size. Moreover, these
implicit weighting schemes can be at odds with the interpretation of these
estimates. Starkly, aggregate treatment effects that pool over lottery and RD variation
asymptotically put \emph{zero weight} on the atomic treatment effects identified by RD
variation. Our previous observation then also implies that these aggregates only reflect
treatment contrasts for test-score schools \emph{for students who disprefer these
test-score schools}. Nevertheless, such estimators are often interpreted as partly
reflecting the RD variation on a broader set of students.\footnote{\label
{fn:intro}For instance, \citet{abdulkadiroglu2017impact} employ such an approach, and the
title of \citet{abdulkadiroglu2017impact} suggests that RD-variation is central. To be
sure, there is evidence that in finite samples, their estimator puts nontrivial weight on
the RD-driven causal comparisons, even though the weight put on these comparisons
converges to zero asymptotically---see \cref{sub:diagnostic}.}

The reason for this result is simple: Since RD variation can only occur at a cutoff,
atomic treatment effects identified by such variation make comparisons for a ``thin'' set
of student characteristics that have population measure zero
\citep{khan2010irregular}---namely, those students with scores at a cutoff. Translated to
estimation, this means that the number of students contributing to RD variation is a
vanishing fraction compared to the number of students subject to lottery variation.
As a result, treatment effect aggregations that weigh each student equally put vanishing
weight on the RD-identified aTEs.

Our second set of contributions speaks to these concerns: We provide practical
recommendations, a new diagnostic, and theoretical guarantees. First, we introduce a
diagnostic that assesses---in finite samples---how much RD-variation contributes to a
particular regression estimator. In some cases, in finite samples, the weight on RD
variation for these estimators may be non-zero and substantial. This diagnostic assesses
whether particular estimators are subjected to the starkest problems arising from their
implicit weighting schemes. 

\Copy{aggregationlit}{Second, researchers are encouraged to take an explicit stand on the
 aggregation of aTEs. Our identification result for aTEs helps practitioners reason about
 the behavior of identified aggregate estimands. We also provide some reasonable default
 choices: We first aggregate identified aTEs for each ordered pair of schools $s_1, s_0$,
 among those that prefer $s_1 \succ s_0$, into parameters $\tau_{s_1\succ s_0}$. This
 aggregation respects identifying variation, and thus lottery-driven aTEs do not
 dominate. We then propose further aggregation to school-level value-added by explaining
 the variation in $\tau_{s_1 \succ s_0}$ through the difference in $s_1$ and $s_0$'s
 value-added, similar to Bradley--Terry models of tournament rankings. In spirit, our
 call for taking an explicit stand on aggregation echoes \citet
 {cattaneo2016interpreting} and \citet{bertanha2020regression}, who consider aggregation
 of regression discontinuity treatment effects at multiple cutoffs. Our setting is
 additionally complex due to the presence of lotteries and of centralized assignment. }

To implement a user-chosen aggregation of aTEs, we also provide asymptotic theory for
estimators of lottery- and RD-identified aTEs---which can then be further aggregated
according to a user-chosen weighting. As a theoretical contribution, relative to the
existing literature \citep{abdulkadiroglu2017impact,abdulkadirouglu2017research}, our
asymptotic theory more accurately reflects the fact that cutoffs have nontrivial
randomness in finite samples, and only converge to fixed quantities asymptotically
\citep{azevedo2016supply}. \Copy {introdiff}{Nevertheless, we show that data-dependent
 cutoffs do not affect the first-order asymptotic behavior of estimators for these
 treatment effects, using a more refined analysis than in
\citet {abdulkadiroglu2017impact,abdulkadirouglu2017research}. This is a novel
contribution to the statistical theory of estimators in settings with school assignment
algorithms.}

This paper applies to any setting where school assignments are made using
deferred acceptance-like algorithms and researchers seek causal identification based on
the centralized assignments
\citep{beuermann2018good,abdulkadirouglu2017research,marinho2022causal,kirkeboen2016field,angrist2021credible}.
It contributes to a growing methodological literature on causal inference in centralized
school assignment settings
\citep{singh2023kernel,marinho2022causal,munrojmp,abdulkadiroglu2017impact,abdulkadirouglu2017research,narita2021theory,narita2021algorithm,che2023leveraging,arkhangelsky2025evaluating}.
This paper is most related to \citet{abdulkadiroglu2017impact}. We complement their work
by investigating the ramifications of heterogeneous treatment effects on their
identification strategy and by formally characterizing the set of atomic treatment effects
identified by centralized school assignment.\footnote{In particular, \citet{abdulkadiroglu2017impact}
``look forward to a more detailed investigation of the consequences of heterogeneous
treatment effects for identification strategies of the sort considered here,'' which is
precisely the theme of this paper.}  Our results also imply that the regression
estimator proposed by \citet{abdulkadiroglu2017impact} estimates a treatment effect that
puts vanishing weight on the RD-identified aTEs, though nevertheless it appears the weight
on RD-identified aTEs is nontrivial in finite samples in their empirical application.

This paper proceeds as follows. \Cref{sec:model} introduces notation, setup, a class of
school choice mechanisms, and a set of treatment effect estimands; in particular, \cref
{sec:nonstoch}  discusses in detail the motivation for our identification results and some
recommendations for aggregating atomic treatment effects. \Cref {sec:identi} characterizes
those that are identified. \Cref{sub:est} proves asymptotic properties for standard
estimators of lottery- or RD-driven estimands. \Cref {sec:monte_carlo} illustrates our
recommendations and results with a Monte Carlo study. \Cref{sec:conclusion} concludes.

\section{School choice mechanisms, estimands, and identification}
\label{sec:model}

Consider a finite set of students $I = \{1,\ldots, N\}$ and schools $S =
\{0, 1, \ldots, M\}.$ The schools have \emph{capacities} $q_1,
\ldots, q_M \in \N$. Assume that school $0$ represents an outside option with infinite
 capacity. Each student $i\in I$ has observed (by the analyst) characteristics $X_i$.
 $X_i$ contains characteristics that are relevant for the assignment mechanism.\footnote
 {It may also contain other observed  characteristics that the analyst may condition on.
 Since these additional covariates do not change our results materially, we suppress them
 and assume for convenience that they are not available.} For each student $i$ and each
 school $s$, there is a potential outcome $Y_i(s)$, representing the outcomes of the
 student if assigned to school $s$.\footnote{Consistent with much of the literature, this
 formulation rules out peer effects or other violations of the stable unit treatment
 value assumption(SUTVA). } The assignment mechanism takes in $X_1,\ldots, X_N$ and
 produces matchings $D_i = [D_ {i1},\ldots, D_{iM}]'$, such that $D_{is} = 1$ if and only
 if $i$ is matched to $s$, and $(D_1,\ldots, D_N)$ satisfies school capacity
 constraints.\footnote{That is, $\sum_{i} D_{is} \le q_s$ for all $s$. } The analyst
 observes $Y_i =
\sum_{s}
D_{is} Y_i(s)$, $D_i$, and $X_i$ for every individual.

To embed this setup in an asymptotic sequence, assume that $(X_i, Y_i(0),\ldots, Y_i
(M)) $ are i.i.d. draws from a superpopulation. The set of schools $S$ is fixed, but
their capacities in finite samples are generated by $q_s = \lfloor N q_s^*
\rfloor$ for some fixed value $q_s^* \in [0,\infty)$.

\subsection{Assignment mechanism}
\label{sub:assignment_mechanism} Following \citet{abdulkadiroglu2017impact}, we consider
 the assignment mechanisms that derive matchings from student-proposing deferred
 acceptance. This setup accommodates the school choice
 mechanism in New York City, and nests many mechanisms that either
 use deferred acceptance or can be reduced to deferred acceptance.\footnote{\label
 {fn:othermech}We consider the same mechanisms as \citet{abdulkadiroglu2017impact}. See
 footnote 5 in
\citet{abdulkadiroglu2017impact} for a list of school choice settings accommodated by this
setup, which includes mechanisms (e.g. immediate acceptance) that can be represented by
deferred acceptance under certain transformations of preferences.} 
Deferred acceptance requires student rankings over schools (termed \emph{preferences}) and
school rankings over students (termed \emph{priorities}). Students submit preferences, which are included in $X_i$, and school priorities are computed from $X_i$ as follows.

There are two types of schools, lottery schools and test-score schools.
School priorities are lexicographic in either $(Q_{it_s}, R_ {it_s})$ or $(Q_{i
\ell_s}, U_{i
\ell_s})$, for $R$ the set of test scores and $U$ the set of lottery numbers. The quantity
$Q_ {is} \in
\br{0,\ldots, \bar q_s}, \bar q_s \in \N \cup \br{0},$ represents certain qualifiers;
 higher values of $Q_{is}$ mean that $i$ is ranked favorably by $s$. This accommodates
 settings where, for instance, having a sibling at $s$ makes $i$ high-priority; we may
 represent such students with $Q_{is} = 1$ and others with $Q_{is} = 0$. When two
 students $ (i,j)$ have the same discrete qualifier $Q_{is} = Q_{js}$, ties are broken by
 either a test score $R_{it_s}$ or a lottery number $U_ {i\ell_s}$, where $t_s, \ell_s$
 indexes which test or lottery number school $s$ uses.

Formally, assume we partition $S$ into lottery and test-score schools. A lottery school
$s$ uses a lottery indexed by $\ell_s \in \br{1,\ldots, L}$, $L\in \N$, and a test-score
school uses a test-score indexed by $t_s \in \br{1,\ldots, T}$, $T\in \N$. Two schools may
use the same lottery or test score for tie-breaking. Assume that the assignment-relevant
information $X_i$ takes the form \[ X_i = (\succ_i, R_i, Q_i) = (\succ_i,
\underbrace{(R_{i1}, \ldots, R_{iT})}_{ \text{continuously distributed on $[0,1]^T$}},
(Q_{i0},\ldots, Q_{iM}))
\]
where $\succ_i$ represents the preferences of student $i$, $R_{it}$ represents $i$'s test
score on test $t$, and $Q_{is}$ represents $i$'s discrete qualifier at school $s$. 

Let $U_{i} = [U_{i1},\ldots, U_{iL}]' \iid F_U$, supported on $[0,1]^{L}$, be the vector of
lottery numbers for
student $i$, where components $(U_{i\ell_1}, U_{i\ell_2})$ need not be
independent.  To implement the lexicographic
priorities in $Q_{is}$ and $U_{i\ell}$ (or $R_{it}$), each
school $s$
computes a \emph{priority score} for each student $i$ \[ V_ {is} = \begin{cases}
    \frac{Q_{is} + R_{it_s}}{\bar q_s + 1}, &\text{ if $s$ is a test-score school}\\
    \frac{Q_{is} + U_{i\ell_s}}{\bar q_s + 1}, &\text{ if $s$ is a lottery school}.
\end{cases}
\] $V_{is}$ encodes lexicographic priorities in ($Q$,  $U$ or $R$), because the integer
part of $ (1+\bar q_s)V_ {is}$ is $Q_{is}$ and the fractional part is $R_{it_s}$ or $U_
{i\ell_s}$. The priority ranking for school $s$ is thus \[ i \rhd_s j \iff \bk{(Q_{is},
R_{it_s}) >_ {\text{lex}} (Q_{js}, R_{jt_s}) \text{ or } (Q_{is}, U_{i\ell_s})
>_{\text{lex}} (Q_ {js}, U_{j\ell_s})} \iff V_ {is} > V_ {js},
\numberthis \label{eq:school_priority_v}
\] 
where $i \rhd_s j$ means $i$ is more favorably ranked than $j$ by school $s$. 

The matching $D_1,\ldots, D_N$ is then computed by the student-proposing deferred
acceptance algorithm with student preferences $\succ_1,\ldots, \succ_N$ and school
priorities $\rhd_ {0},\ldots, \rhd_{M}$ \citep{GS1962,abdulkadirouglu2003school}. The
student-proposal deferred acceptance algorithm is well-known; we reproduce it in 
\cref{asec:misc}.

To make the notation concrete, consider the following setup, which is accommodated by
these mechanisms. 

\begin{exsq} 
\label{ex:notation_example}
Suppose there are four schools in addition to the outside option $s_0$. Two ($s_1, s_2$)
are
 lottery schools, and two ($s_3, s_4$) are test-score schools. Suppose the two lottery
 schools use the same lottery number ($L=1$), and the two test-score schools use the same
 test ($T=1$). Suppose school $s_1$ additionally gives priority to students who have a
 sibling ($Q_{i1} = 1$) already at the school; those students are lexicographically
 preferred to other students, and ties are broken through lottery numbers. Each student is
 characterized by \[(\succ_i, R_{i1}, (Q_{i1}, 0, 0, 0))
 \] where $\succ_i$ is her reported preferences, $R_{i1}$ is her performance on the only
  test in this setup, and $Q_{i1} \in \br{0,1}$ is whether the student has a sibling enrolled at
  $s_1$.  Each student additionally has the lottery number $U_{i1} \iid \Unif[0,1]$. 

Then, under this setup, for two students $i, j$, the school priorities are as follows:
 \begin{itemize}
     \item  School
  $s_1$ prefers $i$ to $j$ ($i \rhd_{s_1} j$) iff either $Q_{i1} > Q_{j1}$ or ($Q_{i1} =
  Q_
  {j1}$ and $U_{i1} > U_{j1}$).
  \item School $s_2$ prefers $i$ to $j$ iff $U_{i1} > U_{j1}$. 
  \item School $s_3, s_4$ prefer $i$ to $j$ iff $R_{i1} > R_{j1}$.  
 \end{itemize}  These school priorities are equivalently represented by \eqref
  {eq:school_priority_v}. Assignments are then taken by running student-proposing
  deferred acceptance on student-reported preferences $\succ_i$ and the school priorities
  $\rhd_s$.
\end{exsq}

\citet{azevedo2016supply} observe that deferred acceptance can be interpreted
 as computing a vector of \emph{cutoffs}. Let \[C_{s,N} = \begin{cases}
    \min_{i} \br{V_{is} : D_{is}=1}, &\text{ if $s$ is matched to $q_s$ students}\\
    0, & \text{ otherwise.}
\end{cases} \] be the
least-favorable priority score among those matched to $s$, if $s$ is oversubscribed. The
matching computed by deferred acceptance is rationalized by each student $i$ being matched
to their favorite school among the set of schools for which $V_{is} \ge C_{s,N}$. This
cutoff structure generates RD-driven identification.

\subsection{Atomic treatment effects and their aggregation in school choice}
\label{sec:nonstoch}

A key contribution is to clarify which conditional average treatment effects are
identified by centralized assignment without restricting treatment effect heterogeneity.
To do so, we consider the ``smallest'' building block of treatment contrasts in this
setting, which we call \emph{atomic treatment effects}. Specifically, for a pair of
schools $s_0, s_1 \in S$, define the
\emph{atomic treatment effect} (aTE) as the conditional average treatment effect between
this pair of schools for a particular value of observables $X$:
\[
\tau_{s_1, s_0} (x) = \E\bk{
    Y(s_1) - Y(s_0) \mid X=x}.
\]
For a school $s$, define the {atomic potential outcome mean} as $
\mu_s(x) = \E[Y(s) \mid X=x].
$

Our identification results characterize for which $(s_1, s_0, x)$ the aTE $\tau_
{s_1, s_0}(x)$ is point-identified under mild assumptions. This exercise reinforces some
claims in the literature,\footnote{\Cref{asub:identification_lit} shows that the values for
which $\mu_s(x)$ is identified under our formulation correspond exactly to those values
$(s, x)$ for which the local deferred acceptance propensity score
\citep{abdulkadiroglu2017impact} is positive.} separates the RD-driven and the
 lottery-driven causal effects, and clarifies that existing estimates of aggregate
 treatment effects may have poor interpretation. Our results then show how to estimate
 (certain granular aggregations of) the aTEs, thus enabling users to construct more
 interpretable estimands. We pause and discuss the motivation and limitation of this
 exercise.

\Copy{idlitmotiv}{First, knowing which aTEs are identified informs us which aggregate
treatment effects are point-identified given only the assignment algorithm.
 Existing work \citep{abdulkadiroglu2017impact,abdulkadirouglu2017research} values
 centralized assignments partly because they provide analogues of
 quasi-experimental research designs that are popular in other microeconometric
 settings.\footnote{For instance, \citet{abdulkadiroglu2017impact} write, ``Centralized
 student assignment opens new opportunities for the measurement of school quality. The
 research potential of matching markets is enhanced here by marrying the conditional
 random assignment generated by lottery tie-breaking with RD-style variation at screened
 schools.''} We establish formal identification and extends it to  heterogeneous treatment effects.\footnote{\citet{abdulkadiroglu2017impact} calculate
 what they term local deferred acceptance (DA) propensity scores and show in their
 Corollary 1 that under constant treatment effects, the treatment effect is identified.
 Our results characterize the set of identified aTEs. These identification results connect
 to \citet{abdulkadiroglu2017impact}: We show that $\mu_s (x)$ is identified if and only
 if the local DA propensity score for school $s$ at characteristics $x$ is positive. See
 \cref{prop:eligibility,asub:identification_lit}.}} Naturally,  the \emph{only} treatment
 effect parameters that are point-identified correspond to aggregate treatment effects
 that place weight only on identified aTEs: i.e., estimands of the form
 \[
    \tau \equiv \int \sum_{s_1 \neq s_0} w(s_1, s_0 \mid x) \cdot \underbrace{\E[Y(s_1)-Y
    (s_0) \mid
    X=x]}_{\tau_{s_1, s_0}(x)} dW (x), \numberthis \label{eq:aggregate_TE}
\] where $w(s_1, s_0 \mid x)$ and $W(x)$ only put nonzero weight on $\tau_{s_1, s_0}
 (x)$ that are point-identified. Thus, a
 practitioner who seeks to only rely on quasi-experimental research designs can inspect
 the class of identified aTEs and decide the aggregation that is most informative of
 their substantive economic question. 

 Second, disaggregating into aTEs helps us understand what practitioners---who may impose
 constant-treatment-effect assumptions---target under misspecification. For instance,
 regression estimators for comparing schools---commonly, estimating causal effects of one
 group of schools $S_1$ to another $S_0 = S \setminus S_1$ by regressing $Y$ on $\sum_
 {s\in S_1} D_{is}$ as well as other controls---pool over aTEs between different school
 pairs and aTEs using different sources of variation. Our analysis disaggregates such
 estimands. Our subsequent analysis cleanly separates which aTEs are identified through
 lottery variation and which are through RD variation, collected in the following
 definition:
 \begin{defn}
\label{defn:lottery_vs_rd}
Fix $x = (\succ, q, r)$. We say that an identified aTE $\tau_{s_1, s_0}(x)$ is
\emph{driven by lottery variation} if $\max_{\succ}(s_1, s_0) $ is a lottery school, and
we refer to all other identified aTEs as \emph{driven by RD variation}.
\end{defn}

\Copy{aggzero}{In disaggregating these estimands, we find that regression estimators---as
 used by, e.g., \citet{abdulkadiroglu2017impact}---pool aTEs from both sources in a way
 that asymptotically puts zero weight on RD-driven aTEs.} This arises because RD-driven
 estimands use a much smaller set of students than lottery-driven ones, so
 inverse-variance aggregation leads lottery-driven estimates to dominate. We provide some
 diagnostics for assessing the severity of this issue in finite samples. This feature
 also calls for separating out the lottery and RD driven components when aggregating
 aTEs, for which our estimation results are helpful.

 Third, granular aggregations of aTEs can yield more informative, yet still
 low-dimensional, summaries of school value-added. Each individual aTE is likely too
 imprecisely estimated to be useful, since it conditions on a single value of $X$. In
 constructing useful aggregations, one should consider what features plausibly predict
 heterogeneity. We speculate that, between schools $(s_1, s_0)$, it matters whether
 $\tau_{s_1, s_0}(X)$ is lottery- or RD-driven, and it also matters whether $X$ contains
 individuals who report preferring $s_1$ to $s_0$ or vice versa.  

 Given $s_1$ and $s_0$, for every student with $s_1 \succ s_0$, there is a
 \emph{maximal} identified aggregation \[\tau_{s_1 \succ s_0} = \int \tau_ {s_1,s_0}(x)
 dW_ {s_1\succ s_0} (x),\numberthis \label{eq:maximal}\] where the measure $W_ {s_1\succ
 s_0}$ is the conditional distribution of $X$ given that $s_1 \succ_i s_0$ and that
 $\tau_{s_1, s_0} (X)$ is identified.\footnote{Our identification results produce a set
 $\mathcal I_{s_1, s_0}$ of $X = (\succ, R, Q)$ values such that $x \in \mathcal I_{s_1,
 s_0}$ if and only if $\tau_ {s_1, s_0}(x)$ is point-identified. $W_{s_1 \succ s_0}$ is
 then the conditional distribution $(X \mid X \in \mathcal I_{s_1, s_0}, s_1 \succ s_0)$.
 For RD-driven aTEs, the set $\mathcal I_{s_1, s_0}$ has zero measure under $P$, and thus
 we take additional care to \emph{define} the conditional distribution. See
 \cref{sub:ATE_def} for defining general aggregate treatment effects, for which $\tau_{s_1
 \succ s_0}$ is a special case. } $\tau_ {s_1
 \succ s_0}$ does not aggregate across identifying variation: It is either RD-driven or
 lottery-driven depending on whether $s_1$ is a lottery school, per
 \cref{defn:lottery_vs_rd}. \cref{sub:est} provides asymptotic guarantees
 for estimating \cref{eq:maximal}.

 Practitioners can further summarize estimates $\hat \tau_ {s_1 \succ
 s_0}$ into school value-added by positing a Bradley--Terry-style \citep{firth2005bradley}
 model in which, for instance, the pairwise treatment effects are explained by differences
 in school value-added, adjusted for reported preference and source of identification \[
     \hat\tau_{s_1 \succ s_0} =  \alpha_0  + 
     (\mu_ {s_1}^L - \mu_ {s_0}^L)\one (\text{$s_1$ is
     lottery}) +
     (\mu_{s_1}^R - \mu_{s_0}^R)\one(\text{$s_1$ is test-score}) + \epsilon_ {s_1, s_0}
     \numberthis \label{eq:BT_aggregation}.
 \] Here, $\mu_s^L, \mu_s^R$ represent school value-added among lottery- and RD-based
 comparisons, respectively, and $\alpha_0$ adjusts for the fact that $s_1$ is revealed to
 be preferred to $s_0$. These parameters may be estimated via least-squares (normalizing
 some effects to zero) or further regularized under random effects or hierarchical
 Bayesian-type assumptions. Under constant effects ($Y(s) = Y(0) +
 \beta_s$ almost surely), this model is exactly well-specified, and $\mu_ {s_1}^L  - \mu_ 
 {s_0}^L =
 \mu_{s_1}^R - \mu_{s_0}^R$ recovers $\beta_{s_1}-\beta_{s_0}$. Without such an
  assumption, $\mu_ {s}^L, \mu_s^R$ can be interpreted as least-squares summaries of
  school causal effects. This cleanly separates identification of causal effects from
  their aggregation and summary.

\Copy{contlarge}{
 \Copy{cont}{This exercise has limitations. First, as we shall require in \cref
  {as:cts}, identification of RD-driven aTEs relies on continuity of $\mu_s(x)$ in test
  scores at cutoffs \citep{hahn2001identification}. \citet{marinho2022causal} note that
  this continuity assumption restricts strategic misreporting of preferences. Suppose
  students know the admission cutoffs and can strategically report preferences based on
  their test scores or lottery numbers. With strategic reporting, those who report
  $\succ$ just below a cutoff may have very different true preferences from those who
  report $\succ$ just above a cutoff. The mix of students can thus
  change sharply at the cutoff, possibly making conditional expectations discontinuous.

 For strategic misreporting to be a concern for continuity, though, students must
 be able to reliably forecast cutoffs. Prediction errors would smooth over the mix of
 students above and below a cutoff, potentially restoring continuity.\footnote{
 \label{fn:blm_fn}For
 instance, in many settings (e.g. New York City \citep{che2023leveraging} and the Chinese
 college entrance exam \citep{chen-kesten-chinese}), students submit preferences before
 learning their test scores, rendering accurate manipulation difficult. In their empirical
 application, \citet{marinho2022causal} do show evidence that students manipulate reported
 preferences: Their Figure 1 shows that students are more likely to rank school $j$ more
 than the outside option if their running variable is closer to $j$'s cutoff, but this
 probability does not appear to change discontinuously at the cutoff.}
 Moreover, such
 manipulation plausibly leads to discontinuities in the density of various covariates at
 the cutoff, and can in principle be empirically assessed.

 Second, even if  continuity holds, the aTEs may still have limited external validity.
 They are useful for the types of program evaluation exercise in
 \citet{abdulkadirouglu2017research,abdulkadiroglu2017impact}, but are perhaps less
  directly informative of counterfactuals in which school choice mechanisms change or
  school admissions policies change. The aTEs condition on reported preferences, test
  scores on existing tests, and are only identified for certain subpopulations. If any of
  them change in a counterfactual policy environment, additional extrapolative
  assumptions are needed. We refer readers to \citet{marinho2022causal} for assumptions
  and methods that partially identify conditional average treatment effects that
  condition on true preferences.}

 Despite this caveat, the maximal aggregation of aTEs (\cref{eq:maximal}) does have a
 clear policy interpretation: It measures the effects of policies on the margin. Consider
 a small increase in $s_1$'s capacity and assume such a change does not alter preference
 submission. Such an increase diverts some students from enrolling in $s_0$ as they now
 qualify for $s_1$, which they prefer. This aggregate treatment effect (\cref{eq:maximal})
 between $s_1, s_0$ exactly measures the gain of these students in terms of the outcome
 $Y$.  Thus, for \emph{marginal} changes, such as increasing the capacity of school $s_1$
 by a small amount or allocating marginal resources to $s_1$, such aggregations of aTEs
 are plausibly informative of the impact of such policies; see 
 \citet{arkhangelsky2025evaluating} for a similar argument.
 Indeed,
 \citet{abdulkadiroglu2017impact} are motivated by debates in New York City over whether students have adequate access to Grade A schools. This policy question can be viewed as evaluating marginal expansions of Grade A schools; if the impact is large, current access is suboptimally limited.}

So far, we have discussed identification heuristically. Because the school assignments
$D_i$ depend jointly on cutoffs $C_N$, themselves computed from all characteristics $
(X_1,\ldots, X_N)$, $(Y_i, D_i, X_i)$ \emph{are not i.i.d.} conditional on $C_N$. Thus,
the usual cross-sectional notion of identification does not apply.\footnote{A standard definition states that a quantity is identified if no
two observationally equivalent distributions of observable data and potential outcomes
give rise to different values of the quantity. Here, ``distributions'' refers to the
\emph{joint distribution} of the observable data, school assignments, and potential
outcomes \emph{for the finite sample of $N$ students}. However, vanishing variation of
$C_N$ around $c$ would allow for certain quantities to be ``identified'' per the standard
definition, but not consistently estimable. We give a simple example in
\cref{asub:id_example} to illustrate the conceptual difficulties.} The literature
\citep{abdulkadiroglu2017impact,marinho2022causal} often appeals to large market
 asymptotics and treats the $C_N$ as nonrandom instead in identification analysis.\footnote{Notable exceptions
 include
\citet{agarwal2018demand} and \citet{munrojmp}.} We conclude this section by making this
 heuristic precise, and we take $C_N$ as random in our estimation section.

\subsection{Identification and random cutoffs}

When students are drawn from a population and the number of students is large, the random
cutoffs $C_{s,N}$ concentrate around some population quantity, satisfying a law of large
numbers. Proposition 3 of \citet{azevedo2016supply} states that if $
\br{(\succ_{is}, V_ {is}) : s \in S}$ are i.i.d. across students, then under mild
 conditions,\footnote{Precisely speaking, \citet{azevedo2016supply} define a notion of
 deferred acceptance matching acting on the \emph{continuum economy}---which is the
 distribution of $\br{(\succ_{is}, V_ {is}) : s \in S}$. The additional regularity
 condition is that this continuum version of deferred acceptance matching admits a unique
 stable matching. In other words, it is a very mild regularity condition on the
 distribution of $\br{(\succ_{is}, V_ {is}) : s \in S}$.} the cutoffs $C_N = [C_
 {0,N},\ldots, C_{M,N}]$ concentrate to some population counterpart $c$ at the parametric
 rate: $
\norm{C_N - c}_\infty = O_P(N^{-1/2}).
$

\begin{as}
\label{as:limitcutoff}
    The population of students is such that the cutoffs $\br{C_{s,N}}$, satisfy
    \[\max_{s\in S}\, \abs{C_{s,N} - c_s} = O_P(N^ {-1/2})\] for some fixed $c = (c_s \in 
    [0,1) :
    s\in S)$.
\end{as}

We define identification relative to the population cutoff $c$. For a fixed cutoff $c$, we
can define $D_i^*(c)$ as the (fictitious) assignment $i$ would receive if the cutoffs were
set to $c$. The tuple $(X_i, D_i^*(c), Y_i (0),\ldots, Y_i(M), U_i)$ are then i.i.d., and
identification reduces to its definition in standard settings. Since as
$C_N \to c$, $D_i^*(C_N)
\to D_i^*(c)$ for $N \to \infty$, this notion of identification captures the information
content of the data for large markets.

Formally, let $P \in \mathcal P$ be the distribution of student characteristics, potential
outcomes, and lottery numbers $(X_i, Y_i(0),\ldots, Y_i(M), U_i)$, where we assume every
member of $\mathcal P$ satisfies \cref{as:limitcutoff} \citep{azevedo2016supply}. Let
$c(P)$ be the corresponding large-market cutoffs, defined as the probability limit of
$C_{N}$ when data is sampled according to $P$. Define $D_{i}^*(c)$ as the assignment made
if the cutoffs were set to $c$: i.e. $D_ {is}^*(c) = 1$ if and only if $s$ is $i$'s
favorite school among those with $V_{is} \ge c_s$. Let $Y^*_{i,\text{obs}}(c) = \sum_{s\in
S} D_{is}^*(c) Y_i(s) $ denote the observed outcome under $D_i^*$.

\begin{defn}
 Let $ P^*_{\text{obs}}(P) $ denote the distribution of $(D_i^*(c(P)),
Y^*_{i,\text{obs}}(c(P)), X_i, U_i)$. We say a parameter $\tau(P)$ is \emph{identified at
$P$} if there does not exist $\tilde P$ such that $\tau(P) \neq
\tau(\tilde P)$ and $P,
\tilde P$ are observationally equivalent: 
$P^*_{\text{obs}}(P) = P^*_{\text{obs}}(\tilde P)$ and $c(P) = c(\tilde P)$.
\end{defn}

\section{Identification of atomic treatment effects}
\label{sec:identi}
\subsection{An extended example}
\label{ex:running_ex} We begin with a simplified setting that contains most of the intuition for our formal results.

\begin{exsq}[Extended example setup]
Suppose there are four schools $s_0, s_1, s_2, s_3$ and three types of students
denoted by $(A, B, C)$. The students have the following preferences and the same
discrete priorities $Q_{is} = 0$:
\begin{center}
\vspace{1em}
\begin{tabular}{ccc}
\toprule & Preferences   \\\midrule $A$ & $s_2 \succ s_3 \succ
 s_1 \succ s_0$ \\ $B$ & $s_2 \succ s_1 \succ s_3 \succ s_0$ \\ $C$ & $s_3 \succ
 s_2 \succ s_1 \succ s_0$ \\
 \bottomrule
\end{tabular}
\vspace{1em}
\end{center}
Additionally, consider the following setup, illustrated in \cref{fig:illustration}:
\begin{enumerate}[wide]
\item The schools $s_1, s_2$ are
test-score schools using the same continuously distributed test score $R \in [0,1]$.

\item $s_3$ is a lottery school, and $s_0$ is an undersubscribed 
(lottery) school with sufficient capacity. Assume that $s_3$ is oversubscribed---the
probability of qualifying for $s_3$ for any student is not zero or one.\footnote{Note that
if certain students had discrete priority over others, then it may be the case that
they qualify for $s_3$ with probability one, even if $s_3$ is oversubscribed.}

\item Assume the number of students is large enough that cutoffs $c_1, c_2$ can be treated as
 fixed.

\item 
Since everyone prefers $s_2 \succ s_1$, school 2 has a more stringent cutoff:
$c_2 > c_1$.

\item Since the distribution of $R$ is unspecified, assume without loss that $c_1 =
\frac13, c_2 = \frac23$. 
\end{enumerate}
\begin{figure}[bth]
\centering
\begin{tikzpicture}
    \draw[thick] (0,0) -- (5,0);
    \draw[thick] (0,-0.2) -- (0,0.2);
    \draw[thick] (5,-0.2) -- (5,0.2);
    
    \filldraw (5/3,0) circle (1.5pt) node[below=2mm] {\(c_1\)};
    \filldraw (10/3,0) circle (1.5pt) node[below=2mm] {\(c_2\)};
    
    \node[below] at (0,-0.5) {$0$};
    \node[below] at (5,-0.5) {$1$};

    \node[above] at (2.5, 0.3) {$R$};
    
    \node[right=1cm] at (6,.5) {\(s_0\) - lottery, undersubscribed};
    \node[right=1cm] at (6,0) {\(s_1\) - test $R$, cutoff $c_1$};
    \node[right=1cm] at (6,-.5) {\(s_2\) - test $R$, cutoff $c_2$};
    \node[right=1cm] at (6,-1) {\(s_3\) - lottery, oversubscribed};
\end{tikzpicture}
\caption{Illustration of the example setup}
\label{fig:illustration}
\end{figure}

Our Monte Carlo results in \cref{sec:monte_carlo} revisit this example, with uniformly
distributed $R,U$ that determines the cutoffs $C_{1,N}, C_{2,N}$.
\end{exsq}

For this setup, we can write $X_i = (\succ_i, R_i)$ and omit $Q_{is}$.  Note
that $\tau_{s,s'}(\succ, R)$ is identified if and only if both $\mu_{s}(\succ, R)$
and $\mu_{s'}(\succ, R)$ are. For a given preference $\succ$ and a school $s$, consider
the
\emph{$s$-eligibility set} \[ E_{s}(\succ, c) = \br{ r:
\P(D^*_{s} (c) = 1 \mid {\succ}, R=r) > 0 }. \numberthis \label{eq:s_elig_def} \] $E_s
(\succ, c)$ collects the test scores $r$ such that someone with $X = (\succ, r)$ has
positive probability of being matched to school $s$. As a result, $\mu_s(\succ, r)$ is
identified if  $r \in E_s(\succ, c)$.

Computing these sets for each type yields:
    \begin{center}
    \vspace{1em}
    \begin{tabular}{ccccc}
    \toprule Preference type & $E_{s_0}$ & $E_{s_1}$ & $E_{s_2}$ & $E_
    {s_3}$ \\
     \midrule 
     $A$ & $[0, \frac13)$  & $[\frac13,\frac23)$  & $[\frac23, 1]$ & $[0, \frac23)$ \\ 
     $B$ & $[0, \frac13)$  & $[\frac13,\frac23)$  & $[\frac23, 1]$ & $[0, \frac13)$  \\ 
     $C$ & $[0, \frac13)$  & $[\frac13,\frac23)$  & $[\frac23, 1]$ & $[0, 1]$
     \\
     \bottomrule
    \end{tabular}
    \vspace{1em}
    \end{center} 

To illustrate the computation, consider students with preference $\succ_A$. For a given
value of $R=r$ and a school $s$, we ask whether $r \in E_s(\succ_A, c)$:
\begin{itemize}
\item When $R \in [0, \frac13)$, students of type $A$ do not qualify for either $s_1$ or
 $s_2$, and they are assigned to $s_3$ if they win the lottery at $s_3$. Otherwise, they
 are assigned to $s_0$. Thus $R \in E_{s_0}(\succ, c)$ and $R \in E_{s_3}(\succ, c)$, but
 $R \not \in E_{s_2}$ and $R\not\in E_{s_1}$.

\item When $R \in [\frac13, \frac23)$, they do not qualify for $s_2$, but
they do qualify for $s_1$. Since they prefer $s_3$ to $s_1$, they are assigned to $s_3$ if
they win the lottery at $s_3$. Otherwise, they are assigned to $s_1$. 
Thus $R \in E_{s_1}(\succ, c)$ and $R \in E_{s_3}(\succ, c)$, but $R \not\in E_{s_2}$ and
$R\not\in E_{s_0}$.

\item When $R \in [\frac23, 1]$, they qualify for $s_2$. Since $s_2$ is
their favorite school, they are assigned to $s_2$. Since they have positive probability of
being assigned only to $s_2$, $R$ is only in $E_{s_2}$, and all of $E_{s_0}, E_{s_1}, E_
{s_3}$ exclude $[\frac23, 1]$.
\end{itemize}
Our subsequent results make the above calculation systematic.

If we further assume that $r\mapsto \mu_s(\succ, r)$ is continuous in $r$ for every
preference and school $s$, then we can extend identification to the \emph{closure} of
$E_{s_j}(\succ, c)$ in $[0,1]^T$. Continuity assumptions allow us to take
sequences and extend identification to their limits: $\mu_s(\succ, r_k) \to \mu_s(\succ,
r)$ if $r_k \to r$. Thus, under continuity
of the atomic potential outcome means, we can compute regions on which each aTE
$\tau_{s,s'}(\succ, R)$ is identified: For a pair of schools $(s, s')$ and preference type
in $\br{A,B,C}$, we tabulate the set of $R$ values for which $\tau_{s,s'} (\succ, R)$ is
identified. The following table tabulates $\bar E_{s} \cap \bar E_{s'}$ for choices of
$s, s'$ and $\succ$:
     \begin{center}
    \vspace{1em}
    \begin{tabular}{cccccccc}
    \toprule Preference type & $(s_0, s_1)$ & $(s_0, s_2)$ & $(s_0, s_3)$ &
$(s_1, s_2)$ & $(s_1, s_3)$ & $(s_2, s_3)$    \\
     \midrule 
     $A$ & $\br{\frac13}$ & $\emptyset$ & $[0, \frac13]$ & $\br{\frac23}$ & $[\frac13,\frac23]$ &
     $\br{\frac23}$\\
     $B$ & $\br{\frac13}$ & $\emptyset$ & $[0, \frac13]$ & $\br{\frac23}$ & $\br{\frac13}$ &
     $\emptyset$\\
     $C$ & $\br{\frac13}$ & $\emptyset$ & $[0, \frac13]$ & $\br{\frac23}$ & $[\frac13,\frac23]$ & 
     $[\frac23,1]$ \\
     \bottomrule
    \end{tabular}
    \vspace{1em}
    \end{center} 

    \noindent This calculation shows that the interpretation of aggregate treatment
     effects is complex in two senses, a complexity often masked by simple aggregations.

    First, aggregate treatment effects may mask heterogeneity in terms of which pairs of
    schools are compared, since different pairs of schools $(s_i, s_j)$ are comparable on
    different regions of the test score $R$. For instance, we might wish to estimate the
    treatment effect of being enrolled in an inside option ($s_1,s_2,s_3$) relative to the
    outside option $s_0$. Naturally, corresponding aggregate treatment effects are some
    weighted average of the identified aTEs $\tau_{s_1, s_0}, \tau_{s_2,s_0}, \tau_
    {s_3, s_0}$.     In this case, interpreting these aggregates as general inside-versus-outside effects overstates their generality:
    \begin{itemize}
         \item For one, since $\tau_{s_2, s_0}$ is identified
    for no value of $X$, all identified aggregations of the aTEs must exclude comparisons
    between $s_2$ and $s_0$. 
    \item For another, among aggregations over $\tau_{s_1, s_0},
    \tau_ {s_2,s_0}, \tau_{s_3, s_0}$, there is a unique maximal aggregation that weighs
    each student equally, namely the $s_3$-against-$s_0$ treatment effect for those with
    low test scores: $\E\bk{Y(s_3) - Y(s_0)
    \mid R \le \frac{1}{3}}.$ 
     \end{itemize} Therefore, in this case, interpreting aggregate treatment effects as an
      inside-versus-outside option effect drastically overstates the generality of these
      estimands in the presence of heterogeneous effects.

    Second, aggregate treatment effects may mask heterogeneity in terms of which students
    are compared for a particular pair of schools, since regions of $R$ that admit
    comparisons for a given pair of schools differ substantially across student types.
    This is true in this example for comparisons between $(s_2,s_3)$: \begin{itemize}
        \item Students of type $A$ have identified aTEs between $s_2, s_3$ at a single
    point $\br{\frac23}$
    \item There are no identified aTEs for students of type $B$ between $s_2, s_3$.
    \item Students of type $C$ have identified aTEs for $(s_2, s_3)$ for the
     set $[\frac23,1]$, which has positive measure.
    \end{itemize} In this case, no identified aggregations of aTEs between $s_2$ and $s_3$
     take into account students of preference type $B$. Moreover, the maximal identified
     aggregation of aTEs between $s_2$ and $s_3$ that weighs each student equally is the
     $s_2$-against-$s_3$ effect among students of type $C$ with high test scores:
     $\E[Y(s_2) - Y(s_3) \mid {\succ_C}, R\in [2/3,1]]$, since the set of students of type
     $A$ with $R = 2/3$ is a measure-zero set. Again, interpreting these estimands as
     blanket causal comparisons between $s_3$ and $s_2$ assumes that treatment effects for
     other students are similar to those of type $C$ with test scores in $[2/3, 1]$.

    We should expect the heterogeneity in both senses to be even    more complex in
    general, since this example only includes 3 out of the 24 possible preferences and
    only a single type of test score. Given this heterogeneity, aggregations that pool
    over many schools, preference types, and test scores can obscure the implicit weights
    on aTEs. To understand these estimates, the next subsection characterizes the
    eligibility sets $E_{s}$ as well as pairwise treatment contrasts formally.

\subsection{Identification of atomic treatment effects}
\label{sub:identification_of_atomic_treatment_effects}
\Copy{idintro}{
Like the example above, characterizing
identification of atomic treatment effects amounts to computing the $s$-eligibility sets
(\cref{eq:s_elig_def}). Our main result, \cref{prop:eligibility}, computes them in the
general setting. The behavior of intersections of $s$-eligibility sets formalizes the
distinction between aTEs that are driven by lottery variation and those driven by RD
variation. \Cref{cor:agg} then shows that any aggregations of aTEs that aggregate over a
non-vanishing subpopulation must place zero weight on the RD-driven aTEs.

In the general setting, recall that $X_i = (\succ_i, R_i, Q_i)$, where we let $R_i$
collect the test scores $(R_ {i1},\ldots,R_{iT})$ and $Q_i$ collect the discrete
qualifiers $(Q_{i0},\ldots, Q_{iM}).$ Fix $(\succ_i, Q_i)$, we first define the
$s$-eligibility sets as the set for $R_i$ on which assignment to $s$ has positive
probability.

\begin{defn}
Define the $s$-eligibility set at $(\succ, Q)$ and cutoffs $c$ as \[
E_s(\succ, Q, c) = \br{r \in [0,1]^T : \P(D_{is}^*(c) = 1 \mid R=r, \succ, Q) >
0}. \]
\end{defn}
}
\begin{as}
\label{as:cts}
For all $s$ and $(Q, \succ)$ with positive probability, the map $[0,1]^T \to \R$ 
\[ r \mapsto \E[Y(s)
\mid {\succ}, R=r, Q]\] exists and is continuous on $[0,1]^T$. Moreover, $\E[|Y(s)|
\mid
{\succ}, R=r, Q] < \infty$. 
\end{as}

\Cref{as:cts} assumes that atomic mean potential outcomes are continuous in the test
scores (and that moments exist). This is a standard assumption in regression discontinuity
\citep{hahn2001identification}, but it does imply restrictions on strategic misreporting
of preferences, as we discuss in \cref{sec:nonstoch}. In so far as \cref{as:cts}
holds---intuitively, this is plausible when students fail to accurately forecast cutoffs
so that student type mixes do not change abruptly at a cutoff---our setting accommodates
mechanisms beyond standard deferred acceptance, since these other mechanisms can be
represented as deferred acceptance on transformed preferences \citep[as pointed out by] []
{abdulkadiroglu2017impact}.

The next proposition verifies that the $s$-eligibility sets $E_s(\succ, q, c)$ collect
values of $r$ such that $\mu_s (\succ, q, r)$ is identified. Continuity (\cref{as:cts})
extends identification to the closure of the $s$-eligibility sets.

\begin{restatable}{prop}{propid}
\label{prop:id} Consider some value $q$ of $Q_i$ and $\succ$ of $\succ_i$ with positive
 probability at $P$. Consider schools $s_0, s_1, s$ and some value $r$ in the support of
 $R_i \mid (Q_i=q, {\succ_i}={\succ})$. Under \cref{as:cts}, $\mu_s(\succ, r, q)$ is
 identified at $P$ if and only if \[r \in \bar E_s(\succ, q, c(P)),\] where $\bar E_s$ is
 the closure of $E_s$ in $[0,1]^T$. The aTE $\tau_{s_1, s_0}(\succ, r , q)$ is identified
 at $P$ if and only if $r \in \bar E_{s_0} (\succ, q, c(P)) \cap \bar E_{s_1}(\succ, q, c
 (P)).$
\end{restatable}

\newcommand{\lbr}[1]{\smash{\underline{r_{#1}}}}
\newcommand{\lbR}[1]{\smash{\underline{R_{#1}}}}

Our main result is the following characterization of all identified atomic treatment
effects: We compute the $s$-eligibility sets for both lottery and test-score schools, and
find the intersection of closures of $s$-eligibility sets between two schools.

Fix some school $s_0$ and some student with characteristics $x=(\succ, r, q)$. $\mu_{s_0}
(x)$ is identified if such a student has positive probability of being assigned to $s_0$.
Suppose $s_0$ is a lottery school. Naturally, in order for a student to have
positive probability to be assigned $s_0$, we require that
\begin{enumerate}[label=(\roman*),wide]
    \item The probability that $s_0$ is the student's favorite lottery school that she
    qualifies
    for is positive: \[\pi^*_{s_0}(\succ, q, c) \equiv \P\bk{
        s_0 = \argmax_{\succ} \br{s \text{ is a lottery school}: V_{is} > c_s} \mid {\succ},
        R=r,Q=q
    } > 0.\]
    \item The student does not prefer any test-score school $s$ that she qualifies for to
    $s_0$. This requires that the student's test scores $r_{t}$ are lower than  some value
    $\lbR{t}(s_0, \succ, q; c)$, representing the most lenient cutoff among schools
    that use test $t$ that the student prefers to $s_0$. 

    Formalizing this is complicated by the presence of $Q_{is}$. For a given value
    of $q_s$, consider the worst test score for students with $Q_{is} = q_s$ that clears
    the cutoff $c_s$---such a test score is 0 if all students with $Q_{is} = q_s$ clears
    the cutoff, and 1 if no students do so:
     \[
\lbr{t}(s, q_s, c_s) = \min\br{\max\br{(1 +\bar q_s) c_s - q_s, 0}, 1}.
\numberthis
\label{eq:test_score_space_cutoff_q}
\]
Correspondingly, let the most lenient cutoff
for test score $t$ among those
preferred to $s_0$ be \[
\lbR{t}(s_0, \succ, q; c) = \min_{s_1: s_1 \succ s_0, t_{s_1} = t} \lbr{t}(s_1, q_{s_1},
c_
{s_1}).
\numberthis
\label{eq:most_lenient_cutoff}
\]
If the student has $r_t > \lbR{t}(s_0, \succ, q; c)$, then they qualify for some school
that they prefer to $s_0$. 
\end{enumerate}

\medskip 

On the other hand, suppose that $s_0$ is a test-score school that uses test $t_0$. Then
the corresponding
requirements for identification of $\mu_{s_0}(x)$ are:
\begin{enumerate}[label=(\roman*),wide]
    \item The student does not \emph{always} win the lottery at some school $s$ preferred
     to $s_0$. Let $L(q,c)$ be the set of schools where students with qualifier $q$ win
     the lottery with probability 1. This requirement can be written as $s_0 \succ L
     (q,c)$.
    \item The student does not qualify for any test-score school $s$ preferred to $s_0$.
    As before, this can be written as $r_t < \lbR{t}(s_0, \succ, q; c)$ for all $t$. 
    \item The student qualifies for $s_0$: $r_{t_0} \ge \lbr{t_0}(s_0, q_{s_0}, c_{s_0})$.
\end{enumerate}

It is useful for our estimation results to additionally define $r_{s,t}(c)$ as the unique
value among $ \br{
\lbr{t} (s, q, c_s): q = 1,\ldots,
\bar q_s}$ that is in $(0,1)$; if no such value exists, then set $r_ {s,t} (c) = 0$: \[
r_{s,t} (c) = \max\pr{\br{\lbr{t}(s, q, c_s): q = 1,\ldots,
\bar q_s} \cap [0,1)}.
\numberthis
\label{eq:test_score_space_cutoff}
\]
Intuitively, \cref{eq:test_score_space_cutoff_q,eq:test_score_space_cutoff}
translate cutoffs in $V_s$-space to cutoffs in $R_{s,t_s}$-space for students of different
discrete priority types ($Q_{is}$), as illustrated in
\cref{fig:illustration_testscore_cutoff}.

\begin{figure}[tb]
    \centering
    \begin{tikzpicture}
    \definecolor{colorQ0}{RGB}{200,100,100}
    \definecolor{colorQ1}{RGB}{100,200,100}
    \definecolor{colorQ2}{RGB}{100,100,200}

    \draw[thick] (0,-1.5) -- (6,-1.5);
    \foreach \x in {2, 4} {
        \draw[thick] (\x,-1.6) -- (\x,-1.4);
    }
        \node[above, color=colorQ0] at (1,-1) {$q_s = 0$};
    \node[above, color=colorQ1] at (3,-1) {$q_s = 1$};
    \node[above, color=colorQ2] at (5,-1) {$q_s = 2$};

    \node[below] at (2,-1.5) {$\frac{1}{3}$};
    \node[below] at (4,-1.5) {$\frac{2}{3}$};
    \node[below] at (6,-1.5) {1};
    \node[below] at (0,-1.5) {0};

    \node[above] at (0,-1.2) {$V_s$};

    \fill[color=colorQ0] (0,-1.5) rectangle (2,-1.4);
    \fill[color=colorQ1] (2,-1.5) rectangle (4,-1.4);
    \fill[color=colorQ2] (4,-1.5) rectangle (6,-1.4);

    \fill (3,-1.5) circle (2pt);
    \node[below] at (3,-1.8) {$c_s$};

    \foreach \i/\y/\colorr/\pos/\q in 
    {1/0/colorQ0/1/0,2/-1.5/colorQ1/0.5/1,3/-3/colorQ2/0/2} {
        \draw[thick, color=\colorr, shift={(8, \y)}] (0,0) -- (6,0);
        \fill[color=\colorr, shift={(8, \y)}] (\pos*6,0) circle (2.5pt);
        \node[below, color=\colorr, shift={(8, \y)}] at (\pos*6,0) {$\underline{r_t}
        (s,\q)$};
        
        \foreach \x/\lab in {0/0, 6/1} {
            \draw[thick, shift={(8, \y)}] (\x,-0.1) -- (\x,0.1);
            \node[above, shift={(8, \y)}] at (\x,0.1) {\lab};
        }
        \node[above, shift={(8.5, \y-0.1)}] at (6,0.3) {$R_{t_s}$};
    }
\end{tikzpicture}
    \caption{Illustration of (\cref{eq:test_score_space_cutoff_q}) and
    (\cref{eq:test_score_space_cutoff}). The left axis displays the priority scores $V_s$
    for a school with three levels of $Q_s$ and a cutoff at $1/2$. The right axes displays
the test scores for students of different levels of $Q_s$.    Every student with $q=0$
would not qualify for $s$, and so the corresponding cutoff in test-score space is
$\lbr{t} =  1$. Every student with $q=2$ would qualify for the school, and so the
corresponding test-score cutoff is $\lbr{t} = 0$. Students with $q=1$ have interior
test-score cutoffs at $\lbr{t} = 1/2 = r_{s,t_s}$. }
    \label{fig:illustration_testscore_cutoff}
\end{figure}

Finally, the region on which the aTE $\tau_{s_1, s_0}(x)$ is identified is an intersection
between regions on which $\mu_{s_j}(x)$ is identified. The latter region turns out to
involve hyperrectangles in $R_{i}$. To compactly describe the intersection, it is useful
to define intersecting on coordinate $t$: Given $S_1 \subset [0,1]^T$ and $S_2
\subset [0,1]$, let $
\mathrm{Slice}_t(S_1,
    S_2) = \br{s \in S_1 : s_t \in S_2}$ be the set that takes the intersection of the
    $t$\th{} coordinate of $S_1$ with $S_2$.  The following figure illustrates slicing an
    ellipse $S_1 \subset [0,1]^2$ on the first dimension onto an interval $S_2$:

\begin{center}
\begin{tikzpicture}
\draw[thick, ->] (0,-0.1) -- (0,5);
\draw[thick, ->] (-0.1,0) -- (5,0);
\foreach \i in {1,...,4}
{
    \draw (-0.1,\i) -- (0.1,\i);
    \draw (\i,-0.1) -- (\i,0.1);
}
\node[left] at (0,5) {$[0,1]$};
\node[below] at (5,0) {$[0,1]$};

\fill[blue, opacity=0.3] (2.5,2.5) ellipse (2cm and 1.5cm);
\node[blue] at (4,4) {$S_1$};

\draw[thick, red] (1.5,0) -- (3.5,0);
\foreach \i in {1.5, 3.5}
    \fill[red] (\i,0) circle (2pt);
\node[below, red] at (2.5,-0.5) {$S_2$};

\begin{scope}
  \clip (1.5,0) rectangle (3.5,5);
  \fill[green, opacity=0.5] (2.5,2.5) ellipse (2cm and 1.5cm);
\end{scope}

\draw[dashed, red] (1.5,0) -- (1.5,5);
\draw[dashed, red] (3.5,0) -- (3.5,5);
\node[right, color=green] at (5,2.5) {Slice$_1(S_1, S_2)$};
\end{tikzpicture}
\end{center}

The following theorem formalizes this intuition and computes the regions on which $\tau_
{s_1, s_0}(x)$ is identified.

\begin{restatable}{theorem}{propeligibility}
\label{prop:eligibility} 
Fix $\succ, q, c$ and suppress their appearances in $\lbr{t}, \lbR{t}, L,
\pi^*_s$. Then we have:
\begin{enumerate}[wide]
    \item For a lottery school $s_0$, \[E_{s_0}(\succ, q, c) = \begin{cases}
    \bigtimes_{t=1}^T [0, \lbR{t}(s_0))
        , &\text{ if $\pi_{s_0}^* > 0$ and $\lbR{t}(s_0) > 0$
        for all $t$} \\
        \emptyset & \text{ otherwise}.
    \end{cases}
    \]

    \item  
    For a test-score school $s_0$ using test $t_0$, \[ E_{s_0}(\succ, q, c) =
\begin{cases}
    \mathrm{Slice}_{t_0}\pr{\bigtimes_{t=1}^T [0, \lbR{t}(s_0)),  [
    \lbr{t_0}(s_0), \lbR{t_0}(s_0))
    } \\ \quad \quad \quad
    \text{if $s_0 \succ L(q,c)$, $\lbR{t}(s_0) > 0$ for all $t$, and $\lbR{t_0}(s_0) > 
    \lbr{t_0}(s_0)$}\\
\emptyset \quad\quad \text{ otherwise}
\end{cases}
    \]

    \item Consider two schools $s_0, s_1$ where $s_1 \succ s_0$. Assume that neither $E_
    {s_0}$ nor $E_{s_1}$ is empty.
    If $s_1$ is a lottery school, regardless of whether $s_0$ is test-score or lottery,
    then \[
\bar E_{s_0} \cap \bar E_{s_1} = \bar E_{s_0}.
    \] 
    Otherwise, if $s_1$ is a test-score school with test score $t_1$, regardless of
    whether $s_0$ is test-score or lottery, \[
\bar E_{s_0} \cap \bar E_{s_1} = \begin{cases}
    \mathrm{Slice}_{t_1} \pr{
    \bar E_{s_0}, \br{\lbr{t_1}(s_1)}
} , &\text{ if $\lbr{t_1}(s_1) = \lbR{t_1}(s_0) > 0$}\\
    \emptyset & \text{ otherwise}.
\end{cases}   \numberthis \label{eq:cutoff_variation_intersection}\]
\end{enumerate}
\end{restatable}

The first claim of \cref{prop:eligibility} characterizes the eligibility set for a lottery
school $s_0$. The second claim analogously characterizes the eligibility set for a
test-score school $s_0$. Both formalize the verbal intuition we have described. The third
claim computes the intersection of the closures of the eligibility sets between two
schools $s_0, s_1$. Depending on whether the more-preferred school $s_1$ is a lottery
school, the intersection takes different shapes. If $s_1$ is a lottery school, then the
intersection is either empty or $\bar E_{s_0}$. Otherwise, the intersection is either
empty or the slice of $\bar E_ {s_0}$ that equals $s_1$'s cutoff on test $t_1$. 

\Copy{idlit}{These identification results closely relate to the literature. They follow
\citet{abdulkadiroglu2017impact} closely and extend their Corollary 1 to settings with
heterogeneous treatment effects. In particular, the set of students for whom $R \in \bar
E_s(\succ, Q; c)$ is precisely the set of students for whom the local DA propensity score
for school $s$ is positive. Separately, when there are only test-score schools and no
discrete qualifies ($Q_{is} = 0$), \cref{prop:eligibility} characterizes the same set---up
to a conditionally measure zero set of students---of identified pairwise RD effects as
\citet{marinho2022causal}'s Proposition 1. See \cref{asub:identification_lit} for details.

Despite \cref{prop:eligibility} identifying the same set of effects as shown elsewhere,
disaggregating these effects clarifies their interpretation and highlights potential
drivers of heterogeneity. In particular, \cref{prop:eligibility}(3) rationalizes what we
preview in \cref{defn:lottery_vs_rd} as lottery- or RD-variation. When the more-preferred
school is a lottery school, variation between $s_0$ and $s_1$ is driven by the student
potentially \emph{losing} the lottery at $s_1$. On the other hand, when the more-preferred
school is a test-score school, variation between $s_0$ and $s_1$ is driven by the RD
variation in whether the student just qualifies for $s_1$ or just fails to qualify for
$s_1$. Therefore, the two types of aTEs are different in economically meaningful ways:
There is no lottery variation between a less-preferred lottery school and a more-preferred
test-score school, and no test-score variation between a less-preferred test-score school
and a more-preferred lottery school.}

Examining the disaggregated aTEs in turn illustrates certain perils of simple aggregation
schemes. \Cref{prop:eligibility}(3) shows that aTEs driven by lottery variation are
identifiable for a set of students with positive measure, but aTEs driven by test-score
variation (\cref{eq:cutoff_variation_intersection}) are only identifiable for a set of
students with zero measure. As a result, when we aggregate treatment effects such that
each student receive equal weights (see \cref{def:ATE_def} for details), we would
effectively only aggregate the lottery-driven aTEs.  We state this result informally as
\cref{cor:agg}, which is formally stated and proved as \cref{cor:aggApp}.%

\begin{cor}
\label{cor:agg}
Under suitable assumptions, an identified aggregated treatment effect that weighs each
student equally either puts weight solely on aTEs driven by lottery variation or
aggregates over a set of students of measure zero.
\end{cor}

\Cref{cor:agg} is a concerning observation for interpreting estimates of aggregate
treatment effects, as many of these estimands necessarily put zero weight on the aTEs
driven by the RD variation. In contrast, our recommended aggregation \cref{eq:maximal}
avoids this feature.

Translated to estimation, \cref{cor:agg} implies that estimators for aggregate treatment
effects put \emph{vanishing} weight on the aTEs driven by RD variation, since these
estimators can only take comparisons local to a cutoff. While in finite samples these
weights can still be positive and nontrivial, these implicit weights do depend on the
sample size (and on bandwidth tuning parameters); moreover, the larger these weights are
(equivalently, the larger the bandwidth), the more susceptible to bias the estimator is.
The share of influence from RD-driven aTEs is larger in smaller samples than in larger samples, which makes such aggregate treatment effects potentially difficult to interpret. Indeed, the vanishing weight issue affects popular estimation
approaches proposed in the literature.\footnote{\label{fn:yata}\Copy{fnyata}{These
issues are not
unique to
aggregation in the school-choice context. In settings---for instance studied in 
\citet{narita2021algorithm}---where the distribution of treatment assignment is a function
 of $X$, under continuity assumptions, RD-type treatment effects are identified on the
 boundary of sets of the form $\br{X : \P(D=1\mid X) > 0}$. Similarly, aggregating these
 effects with effects in the interior may place vanishing weight on these RD-type
 effects, since the boundary has zero measure.}}

The next subsection returns to our extended example and illustrate that the regression
estimator in \citet{abdulkadiroglu2017impact} puts vanishing weights on the RD-driven
aTEs. To be sure, these estimators can be simple to implement and more efficient when
treatment effects are homogeneous \citep{goldsmith2021estimating,angrist1998estimating};
motivated by these advantages of regression, we also develop a simple diagnostic for the
weight put on RD-driven aTEs for linear estimators.

\subsection{Implications for regression estimators and a diagnostic}
\label{sub:prop_scores}

The estimation approach proposed by \citet{abdulkadiroglu2017impact} converges to
aggregations of treatment effects that ignore the RD variation in the limit. We illustrate
this with our extended example in \cref{ex:running_ex}. First, we make a few additional
assumptions on the example.

\begin{exsq}[Additional setup for the extended example]
Here, we additionally assume that the probability of qualifying for $s_3$ for any student
is approximately equal to $1/2$, in order to compute the local DA propensity scores
\citep{abdulkadiroglu2017impact}. Suppose we are interested in the treatment effect of
$s_2, s_3$ relative to $s_1, s_0$. Consider the treatment indicator $\tilde D_i =
\one(\text{$i$ is assigned to either $s_2$ or $s_3$ at cutoff $c$})$. For simplicity, we
remove
heterogeneity of potential outcomes at the school level and assume $Y_T = Y(s_2) = Y(s_3)$
and $Y_C = Y(s_1) = Y(s_0)$.
\end{exsq}

Roughly speaking, for a region of test score $A \subset [0,1]$,
\citet{abdulkadiroglu2017impact} compute $\P(\tilde D_i = 1 \mid {\succ}, R \in A)$ by
counting those $R$'s near a cutoff as having $1/2$ probability of falling on either side.
``Near a cutoff'' is determined by a bandwidth parameter $h > 0$. 
To be more concrete,    we partition the space of test scores into five regions, with a
bandwidth parameter $h > 0$. Regions \rtwo{} and \rfour{} are small bands around the
cutoffs $c_1, c_2$, and regions $\rone, \rthree,\rfive$ are large regions in between the
cutoffs:

        \begin{center}

\begin{tikzpicture}
    \def\delt{0.04} %
    
    \pgfmathsetmacro{\totalwidth}{5*3} %
    \pgfmathsetmacro{\third}{\totalwidth/3}
    \pgfmathsetmacro{\posA}{0}
    \pgfmathsetmacro{\posB}{\third - \delt*\totalwidth}
    \pgfmathsetmacro{\posC}{\third}
    \pgfmathsetmacro{\posD}{\third + \delt*\totalwidth}
    \pgfmathsetmacro{\posE}{2*\third - \delt*\totalwidth}
    \pgfmathsetmacro{\posF}{2*\third}
    \pgfmathsetmacro{\posG}{2*\third + \delt*\totalwidth}
    \pgfmathsetmacro{\posH}{\totalwidth}
    
    \draw[thick,->] (-0.1,0) -- (\totalwidth+0.1,0);
    
    \fill[blue!20] (\posA, -0.2) rectangle (\posB, 0.2);
    \fill[red!20] (\posB, -0.2) rectangle (\posD, 0.2);
    \fill[green!20] (\posD, -0.2) rectangle (\posE, 0.2);
    \fill[yellow!20] (\posE, -0.2) rectangle (\posG, 0.2);
    \fill[orange!20] (\posG, -0.2) rectangle (\posH, 0.2);
    
    \foreach \x/\label in {\posA/0, \posB/{\frac{1}{3}-h}, \posD/{\frac{1}{3}+h}, \posE/{\frac{2}{3}-h}, \posG/{\frac{2}{3}+h}, \posH/1} {
        \draw[thick] (\x,0.2) -- (\x,-0.2) node[below] {$\label$};
    }
    
    \node at ({(\posA+\posB)/2},0.5) {I};
    \node at ({(\posB+\posD)/2},0.5) {II};
    \node at ({(\posD+\posE)/2},0.5) {III};
    \node at ({(\posE+\posG)/2},0.5) {IV};
    \node at ({(\posG+\posH)/2},0.5) {V};
\end{tikzpicture}

        \vspace{1em}
    \end{center} 
    The local deferred acceptance propensity scores, as a function of the region that the test score $R$ falls into,
    are as follows:\footnote{To explain the propensity score calculation, consider a student of type $B$ with test
scores in \rtwo:
\begin{itemize}[wide]
    \item They fail to qualify for $s_2$ when her test score is in \rtwo{}
    with probability one. 
    \item Heuristically, they qualify for $s_1$ with probability 0.5 since her test
    score is in \rtwo{} and near the cutoff for $s_1$. 
    \item $s_1$ is the only control school preferred to $s_3$
    \item They qualify for $s_3$ (via lottery) with probability 0.5.
    \item Thus their probability of being treated is $\psi = 0 + (1-0.5)
    \cdot 0.5 =
    0.25$.
\end{itemize}
This computation is heuristic with $h > 0$, but as we take the limit $h \to 0$,
the probability $\P(\tilde D_i =1 \mid \rtwo, {\succ_B}) \to 0.25$. }
    \begin{center}
        \vspace{1em}
           \begin{tabular}{cccccc}
           \toprule 
          & \rone & \rtwo & \rthree & \rfour & \rfive \\ \midrule 
        $A$ &0.5&0.5&0.5&0.75&1 \\ 
        $B$ &0.5&0.25&0&0.5&1 \\
        $C$ &0.5&0.5&0.5&0.75& 1  \\\bottomrule
        \end{tabular} 
    \vspace{1em}
\end{center}

Let $v \in \br{A, B, C} \times \br{\rone, \rtwo, \rthree,\rfour, \rfive}$ denote a student
type, according to preferences and coarsened test scores. Let $\psi(v)$ denote the
corresponding local propensity score collected in the above table. Consider the population
regression\footnote{If $\psi(V)$ is exactly equal to $\E[D \mid V]$, then this regression
is equivalent to the more familiar control-for-propensity-score regression
\citep{angrist1998estimating} \[ Y = \alpha + \tau \tilde D + \gamma \psi(V) + \epsilon
\]
by Frisch--Waugh. This latter specification conforms with the  specification used for
intent-to-treat effects in \citet{abdulkadiroglu2017impact} (i.e., the reduced form in
their IV specification).} \[Y = \tau (\tilde D - \psi(V)) + \epsilon \numberthis
\label{eq:ols_spec}
\text{ such that }
\tau = \frac{\E[(\tilde D - \psi(V)) Y]}{\E[(\tilde D - \psi(V))^2]}.
\]
We compute in \cref{asub:prop_score_app} that
\begin{align*}
\tau = \sum_{v \in \br{A, B, C} \times \br{\rone, \rtwo, \rthree,\rfour, \rfive}} 
\frac{\P(v)(1-\psi(v))\psi(v) }{\sum_u \P(u) (1-\psi(u))\psi(u)} \E[Y_T - Y_C \mid 
(\succ, R)
\in v] +
\text{Bias}(h).
\numberthis \label{eq:bias_prop_score}
\end{align*}
where $\text{Bias}(h) \to 0$ as $h \to 0$.\footnote{We note that because of the regression
specification (\cref{eq:ols_spec}), the estimand weighs according to $(1-\psi (v))\psi(v)$.
As a result, this aggregation does not weigh each student equally, but nevertheless the
conclusion of \cref{cor:agg} continues to hold for such weighting schemes.}

\Cref{eq:bias_prop_score} shows that the implicit aggregation recovers weighted averages
of treatment effects (over preference-test score region cells) up to a bias that vanishes
as $h \to 0$. However, the weights on regions $\rtwo$ and $\rfour$ also vanish as $h \to
0$, since the corresponding $\P(v) \to 0$. Translated to estimation, this
implies that the asymptotic unbiasedness of propensity-score estimators requires $h = h_N
\to 0$ at appropriate rates, yet the part of the estimator driven by variation from
 regression discontinuity then becomes asymptotically negligible, as long as there is
 lottery-driven variation.

To further relate to finite samples, note that under typical assumptions---as confirmed by
our estimation results in \cref{sub:est}---the popular locally linear regression estimator
for regression-discontinuity-type variation converges at the rates no faster than
$N^{-2/5}
\gg N^{-1/2}$, reflecting that identification for the conditional average treatment effect
at the cutoff is \emph{irregular} \citep{khan2010irregular}. As a result, asymptotically,
estimators for the RD aTEs are much noisier than those for the lottery-driven aTEs, and
pooling them with inverse-variance-type weights in (\cref{eq:bias_prop_score}) results in
diminishing weight on the RD aTEs.

\subsubsection{Diagnostic}
\label{sub:diagnostic}

Motivated by this decomposition, we introduce a simple diagnostic for regression-based
procedures that gives upper and lower bounds on the weight put on students who are
subject to RD variation. For simplicity, we limit to considering binarized comparisons.
That is, there is some treatment $\tilde D_i = \sum_{s \in S_1 \subset S} D_
{is}$ corresponding to a subset $S_1$ of the schools $S$, where a student is considered
treated if they are matched to a school in $S_1$ and untreated otherwise.

We consider linear estimators of the form \[
\hat\tau = \sum_{i=1}^n \hat w_i \cdot Y_i, 
\]
where $\hat w_i$ is a function of $X_{1},\ldots, X_N, \tilde D_1,\ldots, \tilde D_N$. Any
linear regression estimator with $Y_i$ on the left-hand side and functions of $X_i, \tilde
D_i$ on the right-hand side can be written this way. We will also assume the estimator is
associated with some chosen bandwidth parameter $h_N$. 

Using the characterization in \cref{prop:eligibility}, we label student observations by
whether they are subject to lottery variation:
\begin{defn}
\label{defn:diagnostic}
For some user-chosen bandwidth parameter $h_N$ used implicitly to construct
$\hat\tau$, we will consider an observation $i$ to be \emph{possibly subjected} (denoted
by $\overline{\mathrm{RD}}_i = 1$) to RD variation if there exists $s_1 \succ_i s_0$ such
that:
\begin{itemize}
    \item $s_1$ is a test-score school using test score $t_1$
    \item Exactly one of $s_1$ and $s_0$ belongs to the treated group $S_1$
    \item $\bar E_{s_0}(\succ_i, Q_i, C_N) \cap \bar E_{s_1}(\succ_i, Q_i, C_N) \neq
    \emptyset$.
    \item $R_i \in \mathrm{Slice}_{t_1}\pr{
        \bar E_{s_0}(\succ_i, Q_i, C_N), [\lbr{t_1}(s_1, Q_{is_1},
        C_{s_1, N}) - h_N, \lbr{t_1}(s_1, Q_{is_1},
        C_{s_1, N}) + h_N]
    }.$
\end{itemize}
We consider observation $i$ to be \emph{definitely subjected} (denoted by $\underline{
\mathrm{RD}}_i = 1$) to RD variation if it is
possibly subjected and there \emph{does not exist} $s_1 \succ_i s_0$ where: 
\begin{itemize}
    \item $s_1$ is a lottery school
    \item Exactly one of $s_1$ and $s_0$ belongs to the treated group $S_1$
    \item $R_i \in \bar E_{s_0}(\succ_i, Q_i, C_N) \cap \bar E_{s_1}(\succ_i, Q_i, C_N).$
\end{itemize}
\end{defn}
Intuitively, those with $\overline{\mathrm{RD}}_i = 1$ are individuals who are close
enough to the cutoff used by $s_1$ such that, \emph{on some lottery realizations}, they
would be assigned to a test-score school $s_1$ if they clear the cutoff, and to some
school $s_0$ of the opposite treatment type otherwise. Those with
$\underline{\mathrm{RD}}_i = 1$ are those who do so on \emph{all} lottery realizations.

We can then define \[
\bar p_{\text{RD}} \equiv \frac12 \sum_{i=1}^n \hat w_i (2 \tilde D_i - 1) 
\bar{\mathrm{RD}}_i \text{ and } \underline p_{\text{RD}} \equiv \frac12 \sum_{i=1}^n \hat
w_i (2 \tilde D_i - 1)
\underline{\mathrm{RD}}_i
\]
as the upper and lower bounds for the weight assigned to the RD variation, and these can
be implemented by using $(2 \tilde D_i - 1) {\mathrm{RD}}_i$ as left-hand side variables
in the regression. Formally, these estimates are interpreted as the change in $\hat\tau$
if every treated student with $\bar{\mathrm{RD}}_i = 1$ (resp. $\underline{\mathrm{RD}}_i
= 1$) has their outcome increase by 1/2 unit, and every untreated student with $\bar{
\mathrm{RD}}_i = 1$ has their outcome decrease by 1/2 unit.\footnote{Our assumptions on
linear estimators do not rule out estimators that overly extrapolate (i.e. some $\hat w_i$
has the wrong sign: It is negative for $\tilde D_i = 1$ and positive for $\tilde D_i =
0$). As a result, it is possible that one or both of  $\bar p_{\text{RD}}$ and $\underline
p_{ \text{RD}} $ are negative, or that $\bar p_{\text{RD}} < \underline p_{\text{RD}} $.
These unpleasant realizations serve as an additional diagnostic. They cast doubt on the
interpretation of the linear estimator $\hat\tau$, as they reveal that a subpopulation
chosen solely on the basis of $X_i$ has weights that are wrong-signed. See \citet{chen2025potentialweightsimplicitcausal} for general diagnostics in regression estimators. }

\citet{abdulkadiroglu2017impact} study the impact of attending a ``Grade A school'' (one
that receives grade A on the school district's report card for school quality) in New York
City versus attending a non-Grade A school. While we do not have access to their data,
\citet{abdulkadiroglu2017impact} (Appendix Figure D.1) report that about 9,000 students
out of 32,866 students have local propensity scores exactly equal to $1/2$, under their
bandwidth choice $h_N$. This means that these students are solely subject to RD
variation between the treatment schools $S_1$ (Grade A schools in New York City) and the
control schools. Correspondingly, these individuals would be \emph{definitely subjected}
according to \cref{defn:diagnostic}. Thus, this puts a lower bound of about
$9,000/32,866\approx 0.3$ on $\underline p_{\text{RD}}$ for their empirical
application.\footnote{$9,000/32,866\approx 0.3$ would be the weight on these observations
if every observation were weighted equally. Since the specification
(\cref{eq:ols_spec}) weighs students proportionally to $(1-\psi)\psi$, those with
propensity score at exactly $1/2$ receive the highest weights. Thus, in this case, 0.3
serves as a lower bound.} Assuming that the bias term is sufficiently small, we may
conclude that \citet{abdulkadiroglu2017impact} estimate aggregations of aTEs that put
nontrivial weight on the RD-driven aTEs.

In general---though especially when $\bar p_{\text{RD}}, \underline p_{\text{RD}}$ imply
unreasonably small weight on the RD-driven aTEs---researchers may wish to unpack
heterogeneity further and isolate aTEs that are driven by RD variation, along the lines
in \cref{sec:nonstoch}. If researchers have in mind weights for an aggregate treatment
effect that they prefer, they can also aggregate these finer treatment effects manually.
The next section discusses estimation and inference for maximal aggregations atomic treatment effects between two schools (\cref{eq:maximal}).

We close this section with two miscellaneous discussions.

\begin{rmksq}[Design-based inference]
In some school choice markets (e.g. Denver Public Schools and Boston Public Schools), all
schools are lottery schools, and so there is no RD-driven variation. In such settings, we
can directly use the lottery variation to conduct design-based estimation and inference.
Design-based approaches have the benefit that we do not need to assume assignment
mechanisms have the cutoff structure in \cref{as:limitcutoff}, nor do we need to consider
any asymptotic notions of identification. \Copy{rubin}{A previous draft of this paper
(\url{https://arxiv.org/abs/2112.03872v2}) contains results on design-based estimation and
connect propensity-score based regression estimators to Horvitz--Thompson estimators.
These results apply to arbitrary school assignment algorithms, not limited to the deferred
acceptance algorithms that we consider. On the other hand, a design-based approach conditions on the set of students observed and considers randomness solely through the lottery; thus it would not capture the uncertainty in generalizing to matchings of future students \citep{rubin1974estimating}. }
\end{rmksq}

\begin{rmksq}[Noncompliance] Our results define the treatment as the school that a student
 is matched to by the assignment mechanism. This may not be the school that a student
 eventually attends---some students may attend a school that is not in the centralized
 assignment system. Atomic treatment effects are thus interpreted only as intent-to-treat
 effects. Studying treatment effects of schools that students eventually attend amounts
 to analyzing an instrumental variable setting with heterogeneous treatment effects,
 multiple treatments, and a multivalued discrete instrument. Such settings---even without
 the additional complication of school choice mechanisms---remain an area of active
 research
\citep{lee2018identifying,mogstad2020policy,behaghel2013robustness,kirkeboen2016field,kline2016evaluating}.
We leave such analyses to future work.\footnote{\Copy{fnhomo}{In contrast,
\citet{abdulkadiroglu2017impact} are able to consider noncompliance because they assume
no unobserved heterogeneity in treatment effects \citep{mogstad2024instrumental}.}}
\end{rmksq}

\section{Estimation and inference} 
\label{sub:est}

\newcommand{\di}{D^\dagger}

Having characterized the identified atomic treatment effects, we advocate that
practitioners first estimate more granular aggregations of aTEs and then summarize these
aggregations further. Towards this goal, this section provides asymptotic theory for the
maximal aggregation between two schools (\cref{eq:maximal}).

To that end, consider two schools $s_1, s_0$ and all students who declare $s_1 \succ_i
s_0$. We are interested in the estimand \[
    \tau_{s_1 \succ s_0}(c) = \E[Y(s_1) - Y(s_0) \mid R \in \bar E_{s_0}(\succ, Q; c)
    \cap \bar E_{s_1}(\succ, Q; c), s_1 \succ s_0].
\]
Constructing estimators for $\tau_{s_1 \succ s_0}$ depends on whether $s_1$ is a lottery
school, but they share some common structure. 

For a given set of cutoffs $c$, let $J_i(c) = 1$ indicate the set of $X$ values such that 
\begin{enumerate}
    \item $s_1 \succ s_0$
    \item $R_i \in \bar E_{s_0}(\succ_i, Q_i; c)
    \cap \bar E_{s_1}(\succ_i, Q_i; c)$, if $s_1$ is a lottery school
    \item $R_{i,t} \in \br{r_t : r \in \bar E_{s_0}(\succ_i, Q_i; c)
    \cap \bar E_{s_1}(\succ_i, Q_i; c)}$ for all $t \neq t_1$, if $s_1$ is a test-score
    school that uses test $t_1$.
\end{enumerate}
We can thus write \[
    \tau_{s_1 \succ s_0}(c) = \begin{cases}
        \E[Y(s_1) - Y(s_0) \mid J_i(c) = 1] , &\text{ if }s_1 \text{ is a lottery
        school}\\
        \E[Y(s_1) - Y(s_0) \mid J_i(c) = 1, R_{it_1}=r_{s_1,t_1}(c)] &\text{ if }s_1 
        \text{ is a test-score school.}
    \end{cases}
    \numberthis
    \label{eq:estimands_estimation}
\]
recalling \cref{eq:test_score_space_cutoff}. $\tau_{s_1\succ s_0}(c)$ corresponds to the
maximal aggregation of aTEs for $s_1\succ s_0$ when the population cutoffs equal $c$. 

To construct analogue estimators for $\tau_{s_1\succ s_0}$, we account for the fact
that---depending on the lottery results---students with $J_i(c) = 1$ may be matched to
neither $s_1$ nor $s_0$, e.g., if they qualify for some lottery school they prefer. Let
$D_ {i1} (U_i; c)$ indicate the event---as a function of the lottery numbers $U_i$---that
student $i$ fails to qualify for any lottery school $s
\succ_i s_1$ and---if $s_1$ is lottery---additionally qualifies for $s_1$. Let
$D_ {i0} (U_i; c)$ indicate
the event that student $i$ fails to qualify for any lottery school $s\succ_i s_0$ and---if
$s_0$ is lottery---additionally qualifies for $s_0$. Likewise, let $\pi_{ij}(c) = \int D_
{ij} (u;
c)\,dF_U(u)$ be the fixed-cutoff expectation of these events.\footnote{These objects are
defined formally in \eqref{eq:def_d1}.} By construction, for all
fixed cutoff $c$,  we have that $\E\bk{ J_i(c)
\frac{D_{i1}(U_i; c) Y_i}{\pi_{i1}(c)} } = \E\bk{J_i(c) Y_i(s_1)}
$ and that $\E\bk{ J_i(c)
\frac{D_{i0}(U_i; c) Y_i}{\pi_{i0}(c)} } = \E\bk{J_i(c) Y_i(s_0)}$.
Denote $Y_i^{(j)}(c) \equiv \frac{D_{ij}(U_i; c) Y_i}{\pi_{ij}(c)} $.

We thus have two natural estimators for $\tau_{s_1\succ s_0}(c)$, depending on whether
$s_1$ is a lottery school. If $s_1$ is a lottery school, then an inverse propensity
estimator takes the form \[
    \hat\tau_{s_1 \succ s_0} = \pr{\frac{1}{N}\sum_{i=1}^N J_i(C_N)}^{-1}\pr{\frac{1}
    {N}\sum_ {i=1}^N J_i (C_N) \br{Y_i^
    {(1)}(C_N) - Y_i^{(0)}(C_N)}}. \numberthis
    \label{eq:ipw_lottery}
\]
\Cref{eq:ipw_lottery} is simply the inverse-weighting estimator among those with $J_i(C_N)
= 1$.

\Copy{llr}{On the other hand, if $s_1$ is a test-score school and we shorthand $\rho(c) =
r_ {s_1,
t_1}(c)$, then \eqref{eq:estimands_estimation} is akin to an RD estimand among those with
$J_i(c) = 1$. We consider the analogue of local linear regression for this estimand. For
technical reasons, we limit our consideration to a uniform kernel. Specifically, fix some
bandwidth $h_N \to 0$, we let \[
    \hat\tau_{s_1 \succ s_0} = \hat\beta_+(h_N) - \hat \beta_-(h_N)
    \numberthis
    \label{eq:llr_estimand}
\]
where, for $J(c,h)$ defined in \cref{eq:jdef}, \begin{align*}
    \hat \beta_+(h_N) =
    \argmin_{b_0} \min_{b_1} \sum_{i: R_{i, t_1} \in [\rho(C_N),
    \rho(C_N) + h_N]} J_i(C_N, h_N) \pr{Y_i^{(1)}(C_N) - b_0 - (R_{it_1} - \rho(C_N))
    b_1}^2.
\end{align*} The estimator for the left-limit, $\hat \beta_-(h_N)$, is defined
 analogously. \Cref{eq:llr_estimand} implements local linear regression among those with
 $J_i(C_N, h_N) = 1$, which essentially indicates those with $J_i(C_N) = 1$ and have test
 $R_{it_1}$ within bandwidth $h_N$ of the cutoff $\rho(C_N)$. We focus on local linear
 regression since it is a standard choice in the regression-discontinuity
 literature \citep{cattaneo2022regression}. We speculate that the same analysis likely
 extends to local polynomial regression with a uniform kernel as well.}

\Copy{estimation}{ Relative to standard setups, estimation is complicated by the fact that
 the estimators feature the finite-sample cutoffs $C_N$, computed from all the data.
 Since $C_N$ enters all terms in sample averages, these terms are no longer independent,
 precluding a standard asymptotic argument. Our asymptotic results account for the effect
 of a stochastic $C_N$; they can be understood as applications of two-step GMM analyses
 where we verify certain stochastic equicontinuity conditions in $C_N \to c$.
 Interestingly, the asymptotic distributions of the estimators do not depend on the
 asymptotic distribution $\sqrt{N}(C_N - c)$, meaning that, for instance, we do not need
 to adjust for the fact that $C_N$ is random in calculating standard errors. Our results
 thus justify procedures in the literature that treat $C_N$ as fixed.\footnote{The sense
 in which they are valid is nuanced. See the discussion after \cref
 {thm:lottery_main_estimation}. }}

Having introduced the estimator, we turn to assumptions. The key assumption is a
substantive restriction on school capacities and the population distribution of student
characteristics, such that the large-market cutoffs are not in certain knife-edge
configurations. This does limit the uniform validity of our asymptotic results.

\begin{restatable}[Population cutoffs are
interior]{as}{interior}
    \label{as:interior}
        The distribution of student observables satisfies 
        \cref{as:limitcutoff}, where the population cutoffs $c$ satisfy:  
        \begin{enumerate}[wide]
        \item School $s_1$ is neither undersubscribed nor impossible to qualify for: $\rho
         (c)
        \in
        (0,1)$. If $s_1$ uses the same test as $s_0$, then its cutoff is more stringent
        $\rho (c) > r_ {s_0, t_1}(c)$.
            \item For each school $s$, the cutoff \[
c_s \not\in \br{\frac{1}{\bar q_s + 1}, \ldots, \frac{\bar q_s}{\bar q_s + 1}, 1}.
            \]
            
            \item If $c_s =0$, then $s$ is eventually undersubscribed: $\P
            \pr{C_
            {s,N} = 0} \to 1.$
            
            \item If two schools $s_3, s_4$ use the same test $t$, then their test score
            cutoffs are different, unless both are undersubscribed: If $r_{s_3,t}(c) = r_{
            s_4,t}(c)$ then $c_{s_3} = c_{s_4} = 0,$ where $r_{s,t}$ is defined in
            \cref{eq:test_score_space_cutoff}.
        \end{enumerate}
    \end{restatable}
    
    \cref{as:interior} rules out populations where the large-market cutoffs from
    \citet{azevedo2016supply} are on the boundary of certain sets. The first assumption
    simply says that the school $s_1$ is not undersubscribed and not strictly easier to
    qualify for than $s_0$. The second assumption rules out a scenario where everyone with
    $Q=q$ does not qualify for $s$ regardless of their tiebreakers and everyone with
    $Q=q+1$ does qualify for $s$ regardless of their tiebreakers. The third assumption
    says that undersubscribed schools are eventually undersubscribed, for which it
    suffices to impose that the population capacity of a school is not \emph{exactly} at
    the threshold making the school undersubscribed.\footnote{This assumption is stronger
    than the $O_p(N^{-1/2})$ convergence of the cutoffs that we assume. However, the
    assumption holds generically for sufficiently large population school capacities.
    Precisely speaking, suppose some school $s$ with population capacity $q_s$ is
    undersubscribed in the population, but the probability that it is undersubscribed in
    samples of size $N$ does not tend to one (and so violates the third assumption). Then
    we may add a little slack to the school capacity---for any $\epsilon > 0$, making the
    capacity $q_s +
    \epsilon$ instead---to guarantee that the third assumption holds.
    Intuitively, adding $\epsilon$ to the capacity adds $O(\epsilon N)$ seats to the
    school in finite samples, but the random fluctuation of the market generates variation
    in student assignments of size $O (\sqrt{N}).$} Lastly, the fourth assumption assumes
    that the population cutoffs in test-score space are not exactly the same for two
    schools that uses the same test.\footnote{This is Assumption 2 in
    \citet{abdulkadiroglu2017impact}.} Collectively, these assumptions rule out
    adversarial scenarios where, for instance, the population intersection is empty, $\bar
    E_ {s_0}
    ({\succ_i}, Q_i,
    c) \cap \bar E_{s_1} ({\succ_i}, Q_i, c) = \emptyset$, but the sample intersection is
       nonempty with probability non-vanishing in $N$, $\P\bk{\bar E_ {s_0} (\succ_i, Q_i,
       C_N)
       \cap \bar
       E_{s_1} (\succ_i, Q_i, C_N) \neq \emptyset} \not\to 0$.

 Additionally, we maintain a few technical assumptions,
    \cref{as:bounded_density,as:ctsdensity,as:second_moment,as:cts_diff,as:moment_strong},
    stated in \cref{asec:estimation}. These assumptions assert that the distribution of
    test scores and lotteries are suitably smooth, have suitably smooth conditional means,
    and have bounded moments.

We have the following result for lottery comparisons. 
\begin{restatable}{theorem}{thmlotterymain}
\label{thm:lottery_main_estimation} Suppose $s_1$ is a lottery school. Suppose $s_1, s_0$
    are comparable under the limiting cutoffs, i.e., $\E[J_i(c)] > 0$. Then, under
    \cref{as:limitcutoff,as:bounded_density,as:moment_strong,as:interior}, for $g(X_i,
    Y_i; c) \equiv J_i (C_N) \br{Y_i^{(1)}(C_N) - Y_i^{(0)}(C_N)}$,
    \begin{align*} &\sqrt{N}
    \pr{\hat\tau_{s_1 \succ s_0} - \tau_{s_1\succ s_0}(C_N)} \\& = \frac{1}{
    \sqrt{N}}
    \sum_{i=1}^N \br{\frac{g(X_i, Y_i; c) - \E[g(X_i, Y_i; c)]}{\E[J_i(R_i; c)]} - 
    \frac{\E [g
        (X_i, Y_i; c)]}{(\E J_i(R_i; c))^2} (J_i(R_i; c) - \E J_i(R_i; c))}  + o_P(1)
        \numberthis \label{eq:influence}\\
        &\dto  \Norm\pr{
            0, \frac{\var\pr{g(X_i, Y_i; c) \mid J_i(R_i; c) = 1}}{\E[J_i(R_i; c)]}
        }.
        \numberthis \label{eq:clt}
    \end{align*}
\end{restatable}

\Cref{thm:lottery_main_estimation}, proved in \cref{asub:lottery_ate}, shows that the
scaled estimation error
$ \sqrt{N}
\pr{\hat\tau_{s_1 \succ s_0} - \tau_{s_1\succ s_0}(C_N)}$ is equivalent to a scaled
 sample mean of i.i.d. random variables (\cref{eq:influence}), which attains a central
limit theorem (\cref{eq:clt}). The influence function representation (\cref{eq:influence})
is conducive to deriving joint convergence statements (\cref{rmk:joint}).

One subtlety here is that the distribution of $\hat\tau_{s_1 \succ s_0}$ is centered at
the random estimand $\tau_{s_1 \succ s_0} (C_N)$, corresponding to the causal effect
holding fixed the cutoff at the random value $C_N$, rather than at the population cutoff
$c$, $\tau_ {s_1 \succ s_0}(c)$. The difference between the two estimands is of order $O
(1/\sqrt{N})$,
$
\sqrt{N} (\tau_ {s_1
\succ s_0}(C_N) - \tau_{s_1 \succ s_0} (c)) = O_P(1)$, and cannot be ignored. Controlling
 this difference is feasible since the asymptotic distribution $\sqrt{N}(C_N - c)$ can be
 analyzed along the lines in \citet{agarwal2018demand}, and we leave it to future work.
 It is reasonable to consider $\tau_{s_1\succ s_0}(C_N)$ as the target of inference,
 since whether a student chooses between $s_1$ and $s_0$ is ultimately a function of
 $C_N$ rather than of $c$. Doing so has the additional convenience that the asymptotic
 distribution does not depend on the limit distribution of $\sqrt{N}(C_N -c)$---and thus
 standard errors do not need to be adjusted.

Analogously, we have the following result for RD-type comparisons. To facilitate its
statement, let
$
    \check\tau_
{s_1\succ s_0} = \check\beta_+(h_N) - \check \beta_-(h_N)
$
where
\begin{align*}
    \check \beta_+(h_N) \equiv \argmin_{b_0} \min_{b_1} \sum_{i:\one(R_{i, t_1} \in [\rho
    (c),
    \rho(c) + h_N])} J_i(c)  \pr{Y_i^{(1)}(c) - b_0 - (R_{it_1} - \rho(c))
    b_1}^2
\end{align*}
and similarly for $\check\beta_-$. The local linear regression estimator $\check\tau_
{s_1\succ s_0}$ uses the population cutoff $c$ to construct $J_i$ and $\rho(c)$, and thus
its asymptotic properties are well-understood \citep{hahn2001identification}.

\begin{theorem}
\label{thm:estimation_main}
    Under 
\cref{as:limitcutoff,as:interior,as:bounded_density,as:ctsdensity,as:second_moment,as:cts_diff}, assuming $N^{-1/2} = o
(h_N)$ and $h_N = O(N^{-d})$ with $1/5 < d < 1/4$, \begin{align*}
\sqrt{Nh_N} (\hat \tau_{s_1\succ s_0} - \tau_{s_1\succ s_0}(c)) &= \sqrt{Nh_N} 
(\check\tau_
{s_1\succ s_0} - \tau_{s_1\succ s_0}(c)) + o_p (1) \\
&\dto \Norm\pr{0, 
\frac{4}{\P(J_i(c) = 1) f(\rho(c))} (\sigma_+^2 + \sigma_-^2) }
\end{align*}
for
$\sigma_+^2 = \var(Y_i(s_1) \mid J_{i}(c) = 1, R_{it_1} = \rho(c))$, $
\sigma_-^2 = \var(Y_i(s_0) \mid J_{i}(c) = 1, R_{it_1} = \rho(c))$, and $f(r)$ the
density of $R_{it_1} \mid J_i(c) = 1$.

\end{theorem}

The proof of \cref{thm:estimation_main} is relegated to \cref{asec:estimation}
(\cref{athm:equiv}).

Like \cref{thm:lottery_main_estimation}, \cref{thm:estimation_main} shows that $
\sqrt{Nh_N} (\hat \tau_{s_1\succ s_0} - \tau_{s_1\succ s_0}(c))$ is asymptotically
equivalent to a quantity that is easier to analyze, $\sqrt{Nh_N} (\check\tau_ {s_1\succ
s_0} - \tau_{s_1\succ s_0}(c))$. A central limit theorem---standard in the RD literature
\citep{hahn2001identification}---on this quantity is then used to derive the asymptotic
distribution. Compared to \cref{thm:lottery_main_estimation}, we state this theorem by
centering at $\tau_{s_1\succ s_0} (c)$. However, since we scale by $\sqrt{Nh_N} \ll
\sqrt{N}$, the difference between the estimands is immaterial, as $\sqrt{Nh_N} (\tau_{s_1
\succ s_0}(c) - \tau_{s_1\succ s_0}(C_N)) = o_P (1)$.

\Copy{rdliterature}{\Cref{thm:estimation_main} is related to the literature on regression
 discontinuity with unknown or estimated cutoffs \citep
 {hansen2000sample,porter2015regression}. In some settings \citep{card2008tipping}, the
 cutoff that determines treatment assignment is unknown and must be estimated. In those
 settings the cutoff turns out to be estimable at faster-than-$\sqrt{N}$ rates, and its
 estimation has no effect on subsequent asymptotics. This setting differs from ours
 since, here, treatment assignment is determined by $C_N$ and not $c$, and $C_N$ is not
 superconsistent for $c$. Nevertheless, because nonparametric estimators for RD effects
 converge at slower-than-$\sqrt{N}$ rates, the randomness in $C_N$ also does not affect
 the asymptotics of the RD estimators. }

We conclude this section by outlining how these asymptotic results are useful to construct
inferential statements for further aggregations of aTEs.

\begin{rmksq}[Joint convergence and further aggregation]
\label{rmk:joint}
    The representation results in \cref{thm:lottery_main_estimation,thm:estimation_main}
    enable us to consider further aggregation, along the lines described in
    \cref{sec:nonstoch}. Both results represent the scaled estimation error as the
    following \emph{asymptotically linear} representation: For some $r_N \to \infty$ and
    $\E\psi_N (X_i, Y_i;
    c) = 0,$\footnote{See \cref{eq:influence_fn_rd} for the influence function
    representation of the local linear regression estimator.}
    \[
        r_N (\hat\tau_{s_1 \succ s_0} - \tau_{s_1 \succ s_0}) = \frac{1}{r_N} \sum_{i=1}^N
        \psi_N(X_i, Y_i; c) + o_P(1).
    \]
    The subsequent asymptotic normality is derived by observing that $\frac{1}{r_N} \sum_
    {i=1}^N
        \psi_N(X_i, Y_i; c) \dto \Norm(0, \sigma^2_\psi)$. Thus, for a vector of
        estimators $\hat\tau_1,\ldots,\hat\tau_K$---where each $\tau_k$ is an aggregation
        of pairwise aTEs---this representation implies that for some vector-valued mean
        zero function $\bm{\psi}(X_i, Y_i; c)$, we have the joint convergence: For $\odot$
        the entrywise product, \[
            \colvecb{3}{r_{N,1} (\hat\tau_1 - \tau_1)}{\vdots}{r_{N,K} (\hat\tau_K -
            \tau_K)} = \colvecb{3}{r_{N,1}^{-1}}{\vdots}{r_{N,K}^{-1}} \odot \sum_{i=1}^N
            \bm{\psi}(X_i, Y_i; c) + o_P(1) \dto \Norm(0, \Sigma_{\bm{\psi}}).
        \]
        The $(j,k)$\th{} entry of $\Sigma_{\bm{\psi}}$ is equal to $
            \lim_{N\to\infty} \E\bk{
                \frac{N}{r_{N,j}r_{N,k}}\psi_j(X_i, Y_i; c) \psi_k(X_i, Y_i; c)
            }, $ which is consistently estimable by an analogue estimator $\frac{1}{r_
            {N,j} r_ {N,k}}
        \sum_{i=1}^N
        \psi_j (X_i, Y_i; C_N) \psi_k(X_i, Y_i; C_N)$. For instance, we show in \cref
         {lemma:variance_est,thm:variance_est} that the asymptotic variance of the RD estimators is
         consistently estimable. In fact, since both \cref{eq:ipw_lottery} and \cref
         {eq:llr_estimand} can be written as regression estimators on a subsample of
         units with $J_i(C_N) = 1$, we implement the standard error estimators in \cref
         {sec:monte_carlo}  by simply reading off the Eicker--Huber--White standard
         errors of the corresponding regression.

        Motivated by the joint convergence, for $\hat\Sigma_{\bm\psi}$ the estimated
        variance, we can approximate the sampling distribution of the estimators as a
        joint Gaussian, centered at the corresponding estimands: \[
            \colvecb{3}{\hat\tau_1}{\vdots}
            {\hat\tau_K} \overset{a}{\sim} \Norm\pr{
                \colvecb{3}{\tau_1}{\vdots}{\tau_K}, \Lambda^{-1}_N\hat \Sigma_{\bm\psi} \Lambda^
                {-1}_N
            }
            \quad \Lambda = \diag(r_{N,1},\ldots, r_{N,K}).
        \]
        This joint convergence facilitates inference for some user-chosen aggregation
        $\Gamma
        \tau$. For
        instance, the Bradley--Terry-style aggregation scheme (\cref{eq:BT_aggregation})
        defines some matrix $\Gamma$ that maps pairwise school effects to the vector of
        school value added estimates. Given $\Gamma$, we may base inference on the
        following approximation to the sampling distribution of $\Gamma\hat\tau$: \[
            \Gamma \hat\tau \overset{a}{\sim} \Norm(\Gamma\tau, \Gamma \Lambda^
            {-1}_N\hat \Sigma_{\bm\psi} \Lambda^
                {-1}_N \Gamma').
        \]
        Our Monte Carlo results in the next section illustrate these results. 
\end{rmksq}

\section{Monte Carlo study}
\label{sec:monte_carlo}

\begin{figure}[tb]
    \centering
    \includegraphics[width=\textwidth]{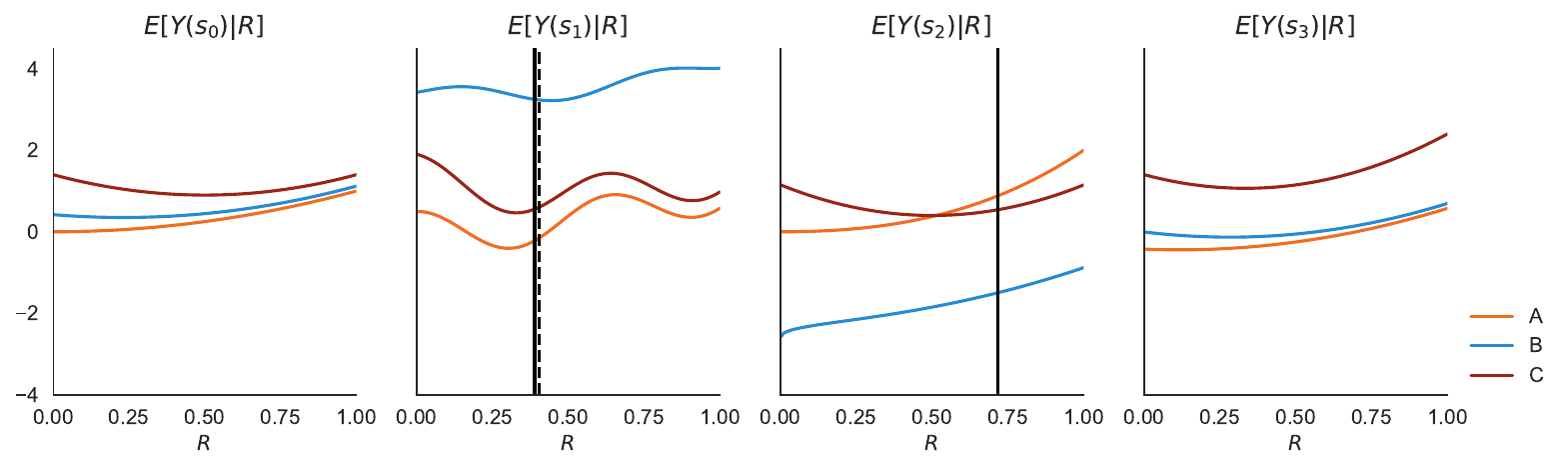}
    \caption{Mean potential outcomes given $X = (\succ, R)$. The black solid line is the
    analytical large market cutoffs, and the dashed line is $C_N$ for a sample of 1000
    students.}
    \label{fig:mean_pos}
\end{figure}

We return to the example in \cref{ex:running_ex} and set up a Monte Carlo study. Suppose
the preference types $\br{A, B, C}$ are equally probable. School capacities are $\br{N,
0.25N, 0.25N, 0.25N}$, respectively for $s_0$ through $s_3$. Suppose the test score $R_i
\iid \Unif[0,1]$, independently of preferences. Suppose the lottery $U_i \iid \Unif
[0,1]$ independently as well. The potential outcomes are heterogeneous. Their conditional
means given preference type and test scores are described by \cref{fig:mean_pos}. Detailed
construction of the simulated data is documented in the code repository
(\url{github.com/jiafengkevinchen/school-choice-monte-carlo}).

\begin{figure}[tb]
    \centering
    \includegraphics[width=\textwidth]{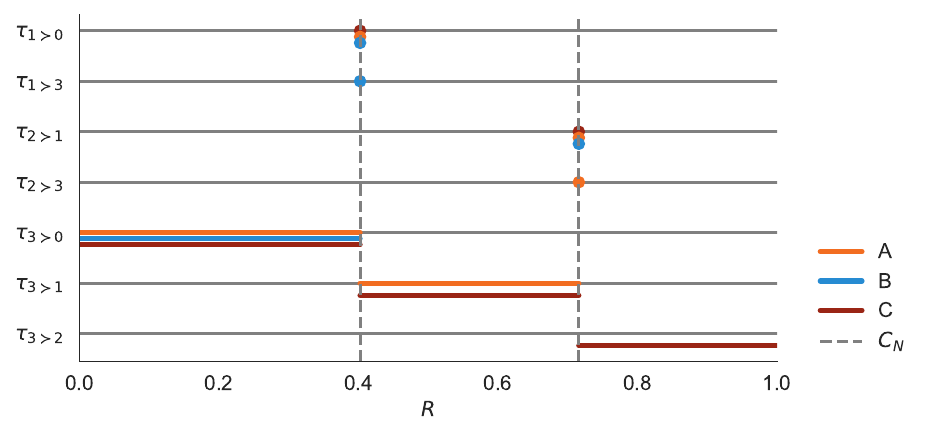}
    \caption{Aggregation region of the maximal treatment effect $\tau_{s_1 \succ s_0}$}
    \label{fig:maximal_agg}
\end{figure}

In this setup, there are seven pairs of schools with nontrivial identified maximal
treatment effects. For each pair $s_1 \succ s_0$ and each preference type,
\cref{fig:maximal_agg} plots the region where $R_i \in \bar E_{s_1}(\succ, C_N) \cap \bar
E_{s_0}(\succ, C_N)$. The maximal treatment effect ($\tau_{s_1\succ s_0}$ in
\cref{eq:maximal}) is then an aggregation of aTEs with $R_i \in \bar E_{s_1}(\succ, C_N)
\cap \bar E_{s_0}(\succ, C_N)$. For instance, the effect $\tau_{3\succ 1}$ is an average
of effects between type $A$ and type $C$ students with test scores between the two
cutoffs.

\begin{table}[tb]
    \caption{Estimates of $\tau_{s_1 \succ s_0}$}
    \label{tab:estimates}
    \centering

    \begin{tabular}{rrrrrrrr}\toprule
$s_1$ & $s_0$ & Estimate & Analytic SE & Bootstrap SE & True value & Null t-statistic &
Coverage
\\\midrule
1 & 0 & $0.71$ & $0.34$ & $0.40$ & $0.74$ & $-0.08$ & 0.912\\
1 & 3 & $3.30$ & $0.23$ & $0.19$ & $3.34$ & $-0.18$ & 0.958\\
2 & 1 & $-1.79$ & $0.41$ & $0.38$ & $-2.01$ & $0.54$& 0.946 \\
2 & 3 & $0.32$ & $0.40$ & $0.38$ & $0.85$ & $-1.32$ & 0.952\\
3 & 0 & $-0.15$ & $0.10$ & $0.10$ & $-0.29$ & $1.37$& 0.944 \\
3 & 1 & $-0.30$ & $0.16$ & $0.16$ & $-0.33$ & $0.21$& 0.951 \\
3 & 2 & $0.96$ & $0.40$ & $0.41$ & $1.11$ & $-0.36$ & 0.939\\
\bottomrule
\end{tabular}

\begin{proof}[Notes]
    All columns except for the last one are based on one draw with $N = 1000$ with 1000
    bootstrap samples. We use bandwidth $h_N = 0.3 N^{-0.24} \approx 0.057$. The last
    column computes the proportion that the null $t$-statistic is between $[-1.96, 1.96]$
    over 1000 draws of the data.
\end{proof}

\end{table}

For one draw of the data with $N=1000$, \cref{tab:estimates} shows estimates and standard
errors of $\tau_{s_1 \succ s_0}$, using the estimators in \cref{sub:est}.   Wald
inference based on the analytic standard error appears accurate, and the bootstrap
variance is close to the analytic estimated SEs. The estimates are approximately
independent: The sampling correlations between different estimates---themselves estimated
by the nonparametric bootstrap---are negligible, with all pairwise correlations less than
0.07.

\begin{figure}[tb]
    \centering
    (a)  Bounds on RD variation

    \includegraphics[width=\textwidth]{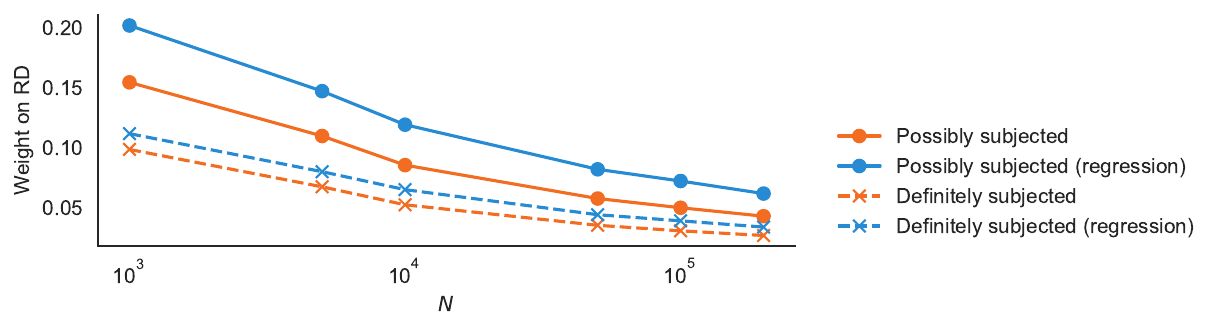}

(b) Regression estimates
    \includegraphics[width=\textwidth]{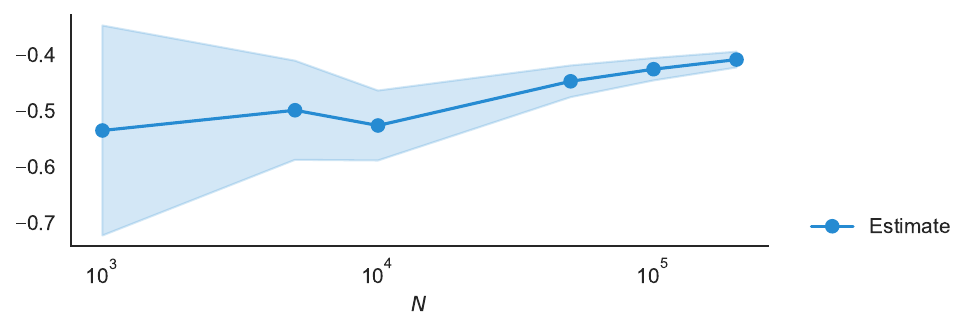}

    \caption{Pooled regression estimate and bounds on RD variation}
    \label{fig:rd_bounds}

    \begin{proof}[Notes]
    We compute large market propensity scores analogously to \cref{sub:prop_scores}, where
    we use the same bandwidth choice $h_N = 0.3 N^{-0.24}$ throughout. The treatment
    effect is estimated by the coefficient of $\tilde D_i$ in a regression of $Y_i$ on
    $\tilde D_i$ and $\hat\psi_i$, where $\tilde D_i$ indicates whether $i$ is assigned to
    schools 2 and 3, and $\hat\psi_i$ is the sum of the large market propensity scores for
    schools 2 and 3 for student $i$. This plot shows diagnostic values defined in
    \cref{sub:diagnostic} for this regression. We also show the proportion of students
    possibly and definitely subject to RD variation at this bandwidth sequence.
    \end{proof}
\end{figure} At this sample size, the RD-driven estimates ($\tau_{1\succ *}$ and $\tau_
 {2\succ *}$) are not significantly noisier than the lottery estimates. Thus in
 aggregations they receive nontrivial weight at $N=1000$. To illustrate vanishing
 weights, for a variety of sample sizes, \cref{fig:rd_bounds} in turn shows our
 diagnostic in \cref{sub:diagnostic} for the regression estimator with large market
 propensity scores, treating schools 2 and 3 as treatment schools. As expected, the
 weight put on RD variation depends on and vanishes with the sample size. 

 To accentuate the concerns regarding aggregation, we choose the potential outcome
 distributions so that $s_1$ is particularly good and $s_2$ is particularly bad for
 students of preference type $B$. This makes students of type $B$ have large negative
 RD-driven treatment effects for the treated schools $2$ and $3$. Correspondingly, in
 \cref{fig:rd_bounds}(b), we observe that the estimated treatment effect becomes less
 negative as sample size increases, since less of these RD-driven effects is aggregated.

\begin{table}[htb]
  \caption{Aggregations of pairwise school effects}
  \label{tab:agg}
  \centering
  
  \begin{subtable}[t]{0.48\linewidth}
    \centering
    \caption*{(a) Test score–driven value‑added}
    \begin{tabular}{lrrrrrr}
    \toprule
    & $\tau_{1\succ0}$ & $\tau_{1\succ3}$ & $\tau_{2\succ1}$ & $\tau_{2\succ3}$ & Estimate & SE \\ \midrule
    $\mu_1^R$ & 1 & 0   & 0 & 0  &  0.71 & 0.42 \\
    $\mu_2^R$ & 1 & $-0.33$ & 0.67 & 0.33 & $-1.48$ & 0.51 \\
    $\mu_3^R$ & 1 & $-0.67$ & 0.33 & $-0.33$ & $-2.19$ & 0.47 \\
    \bottomrule
  \end{tabular}
  \end{subtable}
  \hfill
  \begin{subtable}[t]{0.48\linewidth}
    \centering
    \caption*{(b) Lottery‑driven value‑added}
    \begin{tabular}{lrrrrr}
      \toprule
      & $\tau_{3\succ0}$ & $\tau_{3\succ1}$ & $\tau_{3\succ2}$ & Estimate & SE \\ \midrule
      $\mu_1^L$ & 1 & $-1$ & 0 & 0.15 & 0.18 \\
      $\mu_2^L$ & 1 & 0 & $-1$ & $-1.12$ & 0.41 \\
      $\mu_3^L$ & 1 & 0 & 0 & $-0.15$ & 0.10 \\
      \bottomrule
    \end{tabular}
  \end{subtable}

  \vspace{1em}

  \caption*{(c) Treatment‑ vs. control‑school aggregation}
  \begin{tabular}{lrrrrrrrrr}
    \toprule
    & $\tau_{1\succ0}$ & $\tau_{1\succ3}$ & $\tau_{2\succ1}$ & $\tau_{2\succ3}$ & $\tau_{3\succ0}$
    & $\tau_{3\succ1}$ & $\tau_{3\succ2}$ & Estimate & SE \\ \midrule
    $\tau^R$ & 0.00 & $-0.50$ & 0.50 & 0.00 & 0.00 & 0.00 & 0.00 & $-2.55$ & 0.22 \\
     $\tau^L$ & 0.00 & 0.00 & 0.00 & 0.00 & 0.50 & 0.50 & 0.00 & $-0.23$ & 0.09 \\
    \bottomrule
  \end{tabular}
\begin{proof}[Notes]
    Panels (a) and (b) show the estimates in the least-squares specification \[\tau_{s_1
    \succ s_0} = \one (
    \text{$s_1$ is lottery}) (\mu_{s_1}^L - \mu_{s_0}^L) + \one(
    \text{$s_1$ is test-score}) (\mu_{s_1}^R - \mu_{s_0}^R) + \epsilon_{s_1, s_0},\] which
    is
    \cref{eq:BT_aggregation} without the intercept. Because we did not include the
    intercept, we normalize $\mu_0^L = \mu_0^R = 0$, making $\mu_s^L, \mu_s^R$ not
    comparable to each other. 

    Panel (c) estimates instead \[
        \tau_{s_1 \succ s_0} =J(s_1, s_0) S_
        {s_1}  \bk{
        \one (
    \text{$s_1$ is lottery}) \tau^L + \one(
    \text{$s_1$ is test-score}) \tau^R} + \epsilon_{s_1, s_0}
    \]
    where $J(s_1, s_0) = 1$ if $s_1, s_0$ belongs to different treatment groups and zero
    otherwise, and $S_{s_1} = 1$ if $s_1$ is a treatment school and $-1$ otherwise. Again,
    the treatment group is defined as schools $2$ and $3$.  In each panel, the left pane
    shows how each parameter aggregates the pairwise effects determined by the
    least-squares regression, which is represented by the matrix $\Gamma$ in
    \cref{rmk:joint}: For instance, $\mu_2^R = \tau_ {1\succ 0} - \frac{1} {3}
    \tau_{1\succ 3} + \frac{2}{3} \tau_{2\succ 1} + \frac{1}{3} \tau_{2\succ 3}$. The
    standard errors in the aggregation are computed from $\Gamma \hat\Sigma \Gamma'$ where
    $\hat\Sigma$ is the bootstrap estimate of the variance-covariance matrix of the
    pairwise estimates $\hat\tau_{s_1\succ s_0}$.
\end{proof}
\end{table}

We illustrate alternative methods for aggregation in \cref{tab:agg}, which shows estimates
of $\mu_s^L, \mu_s^R$ in a specification like \cref{eq:BT_aggregation} in panels (a)--(b).
Each value-added should be interpreted as a comparison of school $s$ to school 0, where
\cref{tab:agg} shows how each is computed from the pairwise effects $\tau_{s_1 \succ
s_0}$. Generally speaking, the RD-driven effects $\mu_s^R$ are noisier than the
lottery-driven effects. Lottery-driven value-added estimates can be different from the
test-score driven ones. For instance, for school $3$, the pairwise effect $\tau_ {1 \succ
3}$ is large and positive, and
$\tau_{2\succ1}$ is negative, since they include variation in students of type $B$.
These effects are aggregated in the value-added $\mu_3^R$, contributing to its large
negative value relative to $\mu_3^L$. 

In \cref{tab:agg}(c), we fit a similar regression to \cref{eq:BT_aggregation}, but we
further aggregate effects by treatment schools (2 and 3) and control schools (0 and 1).
This regression specification implicitly defines the RD-driven effect $\tau^R$ as $
\frac{1}{2} (\tau_ {2\succ 1} - \tau_ {1\succ 3})$ and   $\tau^L$ as $\frac{1}{2}(\tau_
{3\succ 0}
+ \tau_ {3\succ1})$. We find that $\tau^R$ is much more negative than $\tau^L$, partly
driven by students of preference type $B$. The corresponding estimate from the local DA
propensity score regression is about $-0.53$, lying between these two estimates.

\section{Conclusion}
 \label{sec:conclusion}
 
 Detailed administrative data from school choice settings provide an exciting frontier
 for causal inference in observational data. Remarkably, school choice markets are
 engineered \citep{roth2002economist} to have desirable properties for market
 participants, and yet they may yield natural-experiment variation that inform program
 evaluation and policy objectives. Credibility of empirical studies using such variation
 demands an understanding of the limits of the data---an understanding of what
 counterfactual queries the data can and cannot answer (absent further assumptions). Our
 analyses here provide a step towards that understanding.
 
 As a review, we provide a detailed analysis of treatment effect identification in school
 choice settings. We characterize the identification of aTEs, the building blocks of
 aggregate treatment effects. We find that pooling over lottery- and RD-driven aTEs leads
 the former to dominate the latter asymptotically. We provide a simple regression
 diagnostic for the weight put on RD-driven aTEs, as well as some suggestions for
 aggregating aTEs. Lastly, we contribute asymptotic theory for estimating aggregations of
 RD-driven aTEs.

\vspace{3em}
\noindent \textbf{Declaration of generative AI and AI-assisted technologies in the writing
process}

During the preparation of this work the author(s) used ChatGPT and Google Gemini in order
to structure ideas, copy-edit, and produce code to implement procedures. After using this
tool/service, the author(s) reviewed and edited the content as needed and take(s) full
responsibility for the content of the publication.

\tiny
\singlespacing
\bibliographystyle{aer}
\bibliography{main.bib}

\onehalfspacing

\newpage 

\footnotesize
    \appendix

    \begin{center}
        \textbf{Appendix to ``Nonparametric Treatment Effect
        Identification in School Choice''}
        
        \vspace{1em}
        
        Jiafeng Chen 
        
        Stanford University
        
        \href{mailto:jiafeng@stanford.edu}{jiafeng@stanford.edu}
        
    \end{center}

\DoToC

\newpage

    \small

    \section{Miscellaneous discussions and proofs}

    \subsection{Deferred acceptance}

    \label{asec:misc}
    The following describes the deferred acceptance algorithm given student preferences
    $\succ_i$ and student priorities $\rhd_s$. 

\begin{enumerate}
\item Initially, all students are unmatched, and they have not been
rejected from any school.
\item At the beginning of stage $t$, every unmatched student proposes to
her favorite school, according to $\succ_i$, from which she has not been
rejected.
\item Each school $s$ considers the set of students tentatively matched to
$s$ after stage $t-1$, as well as those who propose to $s$ at stage $t$, and
{tentatively} accepts the most preferable students, up to capacity $q_s$, ranked
according to $\rhd_s$, while rejecting the rest.
\item Stage $t$ concludes. If there is an unmatched student who has not
been rejected from every school on her list, then stage $t+1$ begins and we
return to step (2); otherwise, the algorithm terminates, and outputs the
tentative matches at the conclusion of stage $t$.
\end{enumerate}
    
    \subsection{An example illustrating that identification notions are
    not straightforward}
    \label{asub:id_example}

    \Copy{median}{
    Consider $X_1,\ldots, X_N \iid \Norm(0,1)$ and a
    treatment where everyone above median is treated: 
    \[D_i = \one(X_i \ge 
    \mathrm{Med}(X_1,\ldots, X_N)).\]
    Note that for any finite $N$, the conditional ATE \[\tau(x) = \E[Y (1) - Y(0) \mid X =
    x]\] is identified, in the sense that $\tau(x)$ is a function of the joint
    distribution of $ (Y_ {1:n}, X_ {1:n}, D_ {1:n}) \sim P_n$.

    To see this, let us fix unit $1$ and write $X_{-1} = (X_2, \ldots, X_N)$. Note that,
    for all $x \in \R$, both $\E[Y_1 \mid D_1=0, X_1=x]$ and $\E[Y_1 \mid D_1 = 1, X_1 =
    x]$ are identified from $P_n$, since $P_n(D_i = d \mid X_i = x) > 0$.
    Then, for any $d \in \br{0,1}$ and $x$, 
    \begin{align*}
    &\E[Y_1 \mid D_1 = d, X_1=x]\\&=\E[Y_1(d) \mid D_1 = d, X_1 = x] \tag{Potential outcomes}\\
    &= \E[\E[Y_1(d) \mid D_1 = d, X_{-1},
    X_1 = x] \mid D_1 = d, X_1 = x]\\
    &= \int_{\br{x_{-1}: d = \one(x \ge \mathrm{Med}(x,x_{-1}))}} \E[Y_1(d) \mid X_
    {-1}=x_{-1},
    X_1 = x , D_1=d] p(x_{-1} \mid D_1 = d, X_1 = x) dx_{-1} \tag{Law of iterated
    expectations} \\
    &=  \E[\E[Y_1(d) \mid X_{-1},
    X_1 = x] \mid D_1 = d, X_1 = x] \tag{$D_1=d$ is measurable with respect to $X_1, X_
    {-1}$ and thus can be dropped} \\
    &= \E[\E[Y_1(d) \mid X_1 = x] \mid D_1 = d, X_1 = x]  \tag{$Y_1(d) \indep X_
    {-1} \mid X_1=x$} \\
    &= \E[Y_1(d) \mid X_1 = x],
    \end{align*}
Therefore we can identify $\tau(x)$ as \[
    \tau(x) = \E[Y_1 \mid D_1 = 1, X_1=x] - \E[Y_1 \mid D_1 = 0, X_1=x].
\]

 However, there is a sense in which the only \emph{morally} identified
    parameter is $\tau(0)$ since $
    \mathrm{Med}
    (X_1,\ldots, X_N) \asto 0$. For any $c < 0$, it becomes vanishingly unlikely that a
    unit with $X_i = c$ is treated, and similarly for $c > 0$. However, we would need $N
    \to \infty$ for the conditional expectations to be estimated, and hence $\tau(x)$
    cannot be consistently estimated if $x \neq 0$.
    }

\subsection{Precise definition of aggregate treatment effects}

\label{sub:ATE_def}
Let $x = (\succ, q, r)$ be a value of the covariate and fix cutoffs at $c$. An empirical
researcher would like
to target an aggregate treatment effect over aTEs
\[ \int \sum_{s_1\neq s_0} w(s_1, s_0 \mid x) \tau_{s_1 , s_0}(x) \,d W_{x \mid \mathcal
I}(x)\]
integrated over a certain probability measure $W_{x\mid \mathcal I}(x)$.
We impose the following assumptions 
\begin{enumerate}
    \item Either the researcher ignores $x$ or the weights sum to 1:
    \[
    \sum_{s_1 \neq s_0} w(s_1, s_0 \mid x) \in \br{0,1}
\]
for all values of $x$. The individual weights are nonnegative: $w(s_1, s_0 \mid x) \ge 0$.
We can accommodate negative weights by considering differences in such aggregations.
\item The researcher only aggregates over identified aTEs: 
$w(s_1, s_0 \mid x) > 0$ only
if $\tau_ {s_1 \succ s_0} (x)$ is identified.
Equivalently, $w(s_1, s_0 \mid x) > 0$ only if $r \in \bar E_{s_1}(\succ, q; c) \cap \bar
E_{s_0}(\succ, q; c)$. 

\item Let $\mathcal I = \br{x: \sum_{s_1 \neq s_0} w(s_1, s_0 \mid x) = 1}$.

\end{enumerate}
Intuitively, we would like $W_{x \mid \mathcal I}(x)$ to be the ``conditional
distribution'' of $W(x)$ conditioned on $x \in \mathcal I$. For $\mathcal I$ of positive
measure under $W$, this is well-defined. However, we would like to extend the definition
to potentially measure-zero sets $\mathcal I$. We would like the conditional
measure  to capture the following: Intuitively, suppose $\mathcal I = \br{
(\succ, q, r_1), (\succ, q, r_2)}$ contains two points, then we would like the
conditional measure to be such that \[
    W(\succ, q, r_1 \mid \mathcal I) = \frac{f(r_1 \mid {\succ}, q)}{f(r_1 \mid {\succ}, q)
    + f(r_2 \mid {\succ}, q)}
\]
where $f(r \mid {\succ}, q)$ is the density of $r$ given $\succ ,q$ with respect to the
Lebesgue measure over $[0,1]^T$, assumed to exist. The right notion is to define the
conditional measure $W(r  \mid {\succ} ,q, \mathcal I)$ to have density proportional
$f (r \mid
{\succ}, q)$ with respect to the \emph{Hausdorff measure} over the set $\br{r : (\succ
,q ,r )\in \mathcal I}$.  In this example, the Hausdorff measure of $\br{r_1, r_2}$
 (of Hausdorff dimension zero) is exactly the counting measure. We formalize below, and
 we shall restrict to sets $\mathcal I$ with integer Hausdorff dimension and positive but
 finite corresponding Hausdorff measure. This rules out $\mathcal I$ that are
 pathologically complex.

\begin{enumerate}[resume]
\item Let $W(x)$ be some measure over $x$, such that $W(r \mid {\succ}, q)$ has a positive
and continuous
density $f_{r \mid \succ, q}$ over $[0,1]^T$ for all $\succ, q$ values. 

\item Fix a value of $\succ, q$, and let \[
    \mathcal I_{\succ, q} = \br{r : \sum_{s_1\neq s_0} w(s_1, s_0 \mid {\succ}, q, r) = 1}
    \subset [0,1]^T
\]
collect the values of $r$ such that $(\succ, q, r) \in \mathcal I$. 

Assume that $\mathcal
I_{\succ, q}$ is closed (and hence Borel measurable), has integer Hausdorff dimension $ k
= k (\succ, q)$, and
$0 < \mathcal
H^ {k} (\mathcal
I_{\succ, q}) < \infty$, for $\mathcal H^k$ the Hausdorff measure of dimension $k$. 
\end{enumerate}

By (5), we may define a measure $W(r \mid {\succ}, q, \mathcal I)$ over the set $\mathcal I_
{\succ
,q}$ and its induced Borel $\sigma$-algebra: 
\[
    dW(r\mid {\succ}, q, \mathcal I) = \one(r \in \mathcal I_{\succ, q})\frac{f_{r\mid
    \succ, q} (r) d \mathcal H^ {k (\succ,
    q)}}{\int_{\mathcal I_{\succ, q}}  f_{r\mid \succ, q}(r) d \mathcal H^{k(\succ,
    q)}}.
\]
By (4), $\int_{\mathcal I_{\succ, q}}  f_{r\mid \succ, q}(r) d \mathcal H^{k(\succ,
    q)} > 0$ and thus the measure is well defined. Now, let \[
        \bar k = \max_{\succ, q} k(\succ, q). 
    \]
    We define the measure where \[
        W(\succ, q \mid \mathcal I) \propto \mathcal H^{\bar k}(\mathcal I_{\succ, q}).
    \] Thus, we can define the measure $W_{x\mid \mathcal I}(x)$ through the conditional
     probability formula\[ W((\succ, q, r) \in A \mid \mathcal I) = \sum_{\succ, q} W
     (\succ, q \mid \mathcal I) W\pr{
            \br{(\succ ,q, r) : r \in \mathcal I_{\succ, q}} \cap A \mid {\succ}, q,
            \mathcal I
        }.
    \]

Now that we have defined $W_{x \mid \mathcal I}$, we can now proceed to define
aggregations of aTEs.

\begin{defn}
\label{def:ATE_def}

We call \[
    \E_{W_{x \mid \mathcal I}} \bk{
        \sum_{s_1 \neq s_0} w(s_1, s_0 \mid X) \tau_{s_1 \succ s_0}(X)
    }
\]
an identified aggregate treatment effect. 

We say that such a treatment effect \emph{weighs each student equally} if the measure $W
(x)$ equals the distribution of $X$ under $P$. 

Among a class of identified aggregations that weigh students equally, indexed by $w(s_1,
s_0 \mid x)$ and $\mathcal
I$, we say that an aggregation is \emph{maximal} if no other aggregation in the class has
a strictly larger $\mathcal I$ in the set inclusion sense.

\end{defn}

\Cref{def:ATE_def} formalizes an aggregation of atomic treatment effects by defining it
relative to a weighting scheme $(w(s_1, s_0 \mid x), W(x))$.  An aggregation weighs
students equally if the weight $W(\cdot)$ is equal to the distribution of the observed
data $X$, thereby simply encoding relative frequencies of values of $X$ arising in the
data. Fixing a class of identified aggregate treatment effects of interest, which all
weigh students equally, effects with larger $\mathcal I$ aggregate over more
individuals---and so effects with the largest $\mathcal I$'s can be called ``maximal'' in
this sense.

As an application of this definition, we might consider the aggregation where $w(s, s'
\mid {\succ}, q, r) = 1$ if and only if $s =s_1, s'=s_0$, $s_1 \succ s_0$,  and $r \in 
\bar E_{s_1} \cap \bar E_{s_0}$ and $W(\cdot)$ is the distribution of $X$ under $P$. Then
 the corresponding aggregate effect is the maximal identified treatment effect $\tau_
 {s_1
\succ s_0}.$

\subsection{Proofs of \cref{prop:eligibility}, \cref{prop:id}, and \cref{cor:agg}}
    \label{asub:eligibilityproof}

\propid*

\begin{myproof}
    We prove the only if parts first. 

    Suppose $r \not \in \bar E_s(\succ, q, c(P))$, we would like to show that $\mu_s$ is not identified. To do so, we would like to find $\tilde P$ that yield different value of $\mu_s$, but is observationally equivalent to $P$. There exists an open set $B \subset [0,1]^T$ (relative to $[0,1]^T$) such that $r \in B \subset \pr{\bar E_s(\succ, q, c(P))}^C$. 

    Let $\tilde P$ differ from $P$ only in terms of the conditional distribution \[
Y(s) \mid R \in B, Q_i=q, {\succ_i}={\succ}
    \]
    such that \[
\E_{\tilde P}[Y(s) \mid {\succ}, R=r, Q=q]\neq \mu_s(\succ, r, q).
    \]
    and $\E_{\tilde P}[Y(s) \mid {\succ}, R=r, Q=q]$ is continuous in $r$.
    Since \[
\P(D_{is}^* = 1 \mid R \in B, Q_i=q, {\succ_i}={\succ} ) = 0,
    \]
    $P$ and $\tilde P$ are observationally equivalent. 

    For the second part, the proof is similar, except that we alter both $Y(s_1) \mid R, Q_i=q, {\succ_i}={\succ}$ and $Y(s_0) \mid R, Q_i=q, {\succ_i}={\succ}$ on $R \in B' \subset \pr{\bar E_{s_0}(\succ, q, c(P))\cap \bar E_{s_1}(\succ, q, c(P))}^C$.

    Now, for the if parts, it suffices to observe that the IPW outcome \[
\frac{D_{is}^* Y^*_{\text{obs}}}{\P(D_{is}^* = 1 \mid {\succ}, R, Q)}
    \] 
    is known whenever $\P(D_{is}^* = 1 \mid {\succ}, R, Q) > 0$ and conditionally unbiased for $\E[Y(s) \mid {\succ}, R, Q]$. Hence if $r \in E_s(\succ, r, q)$, $\mu_s(\succ, r, q)$ is identified by the conditional expectation of the IPW outcome. We extend identification to $\bar E_s$ by invoking continuity in \cref{as:cts}. 
\end{myproof}

    \propeligibility*
    
    \begin{myproof}
       The claims (1) and (2) simply reflects the definition of $E_{s}(\succ, q, c)$ and
       are explained in the main text. Here, we focus on verifying (3) given the
       expressions for $\bar E_s$. 

       Consider the case where $s_1$ is a lottery school. Then since $s_1 \succ s_0$, by
       definition, for every test $t$, \[
\lbR{t}(s_0) \le \lbR{t} (s_1). 
       \]
       As a result, since $s_1$ is a lottery school, \[
E_{s_0}(\succ, q, c) \subset E_{s_1}(\succ, q, c).
       \]
       This proves that the intersection of the closures is equal to $\bar E_{s_0}$. 

       If $s_1$ is a test-score school, we continue to have that for every test $t$, \[
\lbR{t}(s_0) \le \lbR{t} (s_1). 
       \]
       We additionally have that \[
\lbR{t_1}(s_0) \le \lbr{t_1}(s_1).
       \]

       If $\lbR{t_1}(s_0) < \lbr{t_1}(s_1)$, then by inspecting the intersection on
       dimension $t_1$, we find that the intersection $\bar E_{s_1} \cap \bar E_{s_0} =
       \emptyset.$  Otherwise, if $\lbR{t_1}(s_0) = \lbr{t_1}(s_1)$, then the intersection
       on dimension $t_1$ is $\br{\lbr{t_1}(s_1)}$. The intersection on all other
       dimensions is the same as the case where $s_1$ is a lottery school. This completes
       the proof. 
    \end{myproof}

\begin{restatable}{cor}{coragg}
\label{cor:aggApp}
Assume the distribution of student characteristics is such that $R_i \mid {\succ_i}, Q_i$
is measurable with respect to the Lebesgue measure on $[0,1]^T$ with a continuous and
positive density almost surely. Assume \cref{as:cts}. For an identified aggregated
treatment effect that weighs each student equally \[\tau =
\E_{W_{x
\mid \mathcal I}} \bk{\sum_{s_0 \neq s_1} \tau_{s_1, s_0}(X) w(s_1, s_0 \mid X)},\] if the
set of characteristics $\mathcal I$ with identified atomic treatment effects has positive
measure, i.e. $W(\mathcal I) > 0$, then $\tau$ only puts weight on aTEs driven by lottery
variation:
\[
\tau = \E_{W_{x
\mid \mathcal I}} \bk{\sum_{s_0 \neq s_1} \one\pr{\text{$s_1$ is a lottery school}} \tau_
{s_1 \succ s_0} (X) w(s_1, s_0 \mid X)}.
\]
\end{restatable} 

\begin{myproof}
It suffices to show that \[
\E_{W_{x
\mid \mathcal I}} \bk{\sum_{s_0 \neq s_1, s_1 \text{ test}} |\tau_{s_1 \succ s_0}(X)|
\cdot w (s_1, s_0 \mid X)} = 0 \]

By \cref{def:ATE_def}, $w(s_1, s_0 \mid X) > 0$ implies $R \in \bar E_{s_1}(\succ, Q, c)
\cap \bar E_{s_0}(\succ, Q, c)$ for $X = (\succ ,Q ,R)$. Thus \begin{align*}
    &\E_{W_{x
\mid \mathcal I}} \bk{\sum_{s_0 \neq s_1, s_1 \text{ test}} |\tau_{s_1 \succ s_0}(X)|
\cdot w (s_1, s_0 \mid X)} \\
&\le \E_{W_{x
\mid \mathcal I}} \bk{\sum_{s_0 \neq s_1, s_1 \text{ test}} |\tau_{s_1 \succ s_0}(X)|
\cdot \one \pr{R \in \bar E_{s_1}(\succ, Q, c)
\cap \bar E_{s_0}(\succ, Q, c)}}. 
\end{align*}

By \cref{as:cts}, since $r\mapsto \mu_s(X)$ is continuous over a compact set, it is
uniformly continuous and hence uniformly bounded. As a result, it suffices to show that
for a test-score school $s_1$, \[
W_{x \mid \mathcal I}\bk{
    R \in \bar E_{s_1}(\succ, Q, c)
\cap \bar E_{s_0}(\succ, Q, c)
} = \sum_{\succ, q} W(\succ, q \mid \mathcal I) W(R \in \bar E_{s_1}(\succ, Q, c)
\cap \bar E_{s_0}(\succ, Q, c) \mid \succ, q, \mathcal I) = 0.
\]
Since $\mathcal I$ has positive measure under $W$, there is some $\succ ,q$ with positive
$W$-measure for which $\mathcal I_{\succ, q}$ has positive Lebesgue measure over $
[0,1]^T$. This means that it has Hausdorff dimension $\bar k = T$ and positive measure in
terms of 
$\mathcal H^ {\bar k}$. Thus, if $\mathcal I_{\succ, q}$ does not have Hausdorff dimension
$T$, $W (\succ, q \mid \mathcal I) = 0$, and we can ignore those terms in the above
display. On the other hand, for $(\succ, q)$ such that the Hausdorff dimension of
$\mathcal I_ {\succ, q}$ is $k=T$, we have that \[
    \mathcal H^T\pr{
        \bar E_{s_1}(\succ, Q, c)
\cap \bar E_{s_0}(\succ, Q, c)
    } = 0
\]
since the $t_{s_1}$\th{} coordinate of $\bar E_{s_1}(\succ, Q, c)
\cap \bar E_{s_0}(\succ, Q, c)$ is at most a singleton, by \cref{prop:eligibility}(3).
Because the Hausdorff measure is the dominating measure for $W(\cdot \mid \succ, q,
\mathcal I),$ we have that \[W(R \in \bar E_{s_1}(\succ, Q, c)
\cap \bar E_{s_0}(\succ, Q, c) = 0.\] Thus 
\[\sum_{\succ, q} W(\succ, q \mid \mathcal I) W(R \in \bar E_{s_1}(\succ, Q, c)
\cap \bar E_{s_0}(\succ, Q, c) \mid \succ, q, \mathcal I) = 0.\]
This completes the proof.

\end{myproof}

\subsection{Computation of propensity score estimand}
\label{asub:prop_score_app}
In this subsection, we continue the computation in \cref{sub:prop_scores}. Recall that we
are interested in the following estimand as a function of the bandwidth parameter $h$:
\[
\tau = \frac{\E[(\tilde D - \psi(V)) Y]}{\E[(\tilde D - \psi(V))^2]}.
\]
Now, the numerator is equal to \begin{align*}
\E[(\tilde D - \psi(V)) Y] 
&= \E[(\tilde D - \psi(V)) (Y_C + \tilde D (Y_T - Y_C))] \\ 
&= \E\bk{
    \tilde D Y_T - \psi(V) Y_C - \tilde D \psi(V)(Y_T-Y_C)
} \\
&= \E\bk{\psi(V) (Y_T - Y_C) + (\tilde D - \psi(V)) Y_T - \psi(V)^2(Y_T - Y_C) - 
(\tilde D-\psi
(V))(Y_T - Y_C)} \\ 
&= \E[\psi(V) (1-\psi(V)) (Y_T - Y_C)] + \E[(\tilde D - \psi(V)) Y_C]
\end{align*}
The denominator is equal to \begin{align*}
\E[(\tilde D - \psi(V))^2] &= \E[\tilde D + \psi^2(V) - 2 \tilde D \psi(V)] \\
&= \E\bk{
    \psi(V) - \psi^2(V) + (\tilde D - \psi(V)) - 2(\tilde D - \psi(V)) \psi(V)
} \\ 
&= \E\bk{
   \psi(V) (1-\psi(V))} + \E\bk{(\tilde D - \psi(V)) (1-2\psi(V))}
\end{align*}
Since $(a, b )\mapsto a/b$ is continuous when $b > 0$, it remains to show that \[
\lim_{h\to 0} \E\bk{(\tilde D - \psi(V)) (1-2\psi(V))} = 0 = \lim_{h\to 0}\E[(\tilde D -
\psi(V)) Y_C]. \numberthis \label{eq:no-bias}
\]
Note that on regions $\rone, \rthree, \rfive$, \[
\psi(V) = \P(\tilde D = 1 \mid {\succ}, Q, R, Y_T, Y_C)
\]
Hence \[
\E[(\tilde D - \psi(V)) f(Y_C, V) \mid \rone \cup \rthree \cup \rfive] = 0.
\]
Since $\P(\rone \cup \rthree \cup \rfive) \to 1$ as $h \to 0$, (\cref{eq:no-bias}) follows.

In fact, this calculation only uses the fact that $\P(\rtwo \cup \rfour) \to 0$---in other
words, the bias would vanish as a function of $h$ even if we were to use an unreasonable
estimator on regions $\rtwo$ and $\rfour$.

For completeness, we can also show that the conditional bias is vanishing in $h$ under
smoothness assumptions. We'll do so for those with preference $\succ_B$ and in cell
\rfour: \begin{align*} &\E[(\tilde D - \psi(V)) f(Y_C, V) \mid {\succ_B}, \rfour] \\&=
\E\bk{\pr{\P\pr{R \ge 2/3
\mid
Y_C,
\succ_B, R \in [2/3 \pm h]} - 
\frac{1}
{2}} f(Y_C, V) \mid {\succ_B}, \rfour} \\
&= \E\bk{\pr{\frac{\int_{2/3}^{2/3+h} p(r \mid Y_C, \succ_B) \,dr}{\int_{2/3-h}^{2/3+h} p
(r \mid Y_C, \succ_B) \,dr} -
\frac{1}
{2}} f(Y_C, V) \one(R \in [2/3 \pm h]) \mid {\succ_B}} \frac{1}{\P(R \in [2/3 \pm h]
\mid {\succ_B})}.
\end{align*}
If $p(r \mid Y_C, \succ_B)$ is bounded below by $\eta$ at $r=2/3$ and Lipschitz continuous
with constant $L$, then \[
\abs{p(r \mid Y_C, \succ_B) - p(2/3 \mid Y_C, \succ_B) } \le L |r-2/3|.
\]
As a result, \[
\abs[\bigg]{\int_{2/3}^{2/3+h} p(r \mid Y_C, \succ_B) \,dr - h p(2/3 \mid Y_C, \succ_B)}
\le L h^2
\]
and similarly 
\[
\abs[\bigg]{\int_{2/3-h}^{2/3+h} p(r \mid Y_C, \succ_B) \,dr - 2h p(2/3 \mid Y_C,
\succ_B)}
\le 4L h^2.
\]
Hence the discrepancy in the first term is of order $h^2$: For some $C$ a function of $L,
\eta$,  \[
\abs[\bigg]{\frac{\int_{2/3}^{2/3+h} p(r \mid Y_C, \succ_B) \,dr}{\int_{2/3-h}^{2/3+h} p
(r \mid Y_C, \succ_B) \,dr} -
\frac{1}
{2} } \le C h^2.
\]
On the other hand, there exists $c$ a function of $\eta, L$ where $\P(R \in [2/3 \pm h]
\mid {\succ_B}) \ge c h$ for all sufficiently small $h$. Hence \[
\E\bk{\pr{\frac{\int_{2/3}^{2/3+h} p(r \mid Y_C, \succ_B) \,dr}{\int_{2/3-h}^{2/3+h} p
(r \mid Y_C, \succ_B) \,dr} -
\frac{1}
{2}} f(Y_C, V) \one(R \in [2/3 \pm h]) \mid {\succ_B}} \frac{1}{\P(R \in [2/3 \pm h]
\mid {\succ_B})} = O(h).
\]

\subsection{Identification results in related literature}

\label{asub:identification_lit}

\subsubsection{Results in \citet{marinho2022causal}} \citet{marinho2022causal} study a
case where $Q_{is} = 0$ for all $(i,s)$ and all schools use test scores. 

Fix a student with preference $\succ_i$ and scores $R_i$. Fix a cutoff $c_s$ on test $t$.
For simplicity, let us assume all $c_s \in (0,1)$. 

Let $R_i (r)$ be the vector where we replace $R_{it}$ with $r$ and leave all $R_{it'}$ the
same, for $t' \neq t$. Under \citet{marinho2022causal}'s Definition 4 of \emph{local
preferences}, a student with preference $\succ_i$ and scores $R_i$ has a local preference
of $(s_1, s_0)$ at cutoff $c_s$  if, for some $\epsilon > 0$, among the schools for which
she qualifies, her favorite school is $s_1$ when her score is $R_i (r)$ for $r \in[c_s,
c_s+ \epsilon)$ and her favorite school is $s_0$ when her score is $R_i(r)$ for $r \in
(c_s-\epsilon, c_s]$.  Under their Definition 5, $(s_1, s_0)$ is a \emph{comparable pair}
(i) if $c_{s_1} \in (0,1)$ and (ii) the conditional distribution of students with test
scores $R_{it_{s_1}} = c_{s_1}$ contains positive mass of those with local preferences
$(s_1, s_0)$. Their Proposition 1 shows that under continuity assumptions, for $(s_1,
s_0)$ a comparable pair, the causal effect between $(s_1, s_0)$ of those with local
preference $(s_1, s_0)$ at $c_{s_1}$ and with $R_{it_{s_1}} = c_ {s_1}$ is identified.

For a student with local preferences $(s_1, s_0)$ at $c_{s_1}$ and with $R_{it_{s_1}} = c_
{s_1}$, her characteristics $(\succ_i, R_i)$ must satisfy the following: For $t_0 = t_
{s_0}$ and $t_1 = t_{s_1}$, if $t_0 = t_1 = t$
 \begin{enumerate}
     \item ($s_1$ has a stricter cutoff than $s_0$) $R_{it} = c_{s_1} > c_{s_0}$
     \item (Does not qualify for any school $s \succ_i s_1$ on test $t$) $c_{s_0} < \lbR
     {t} (s_0) = c_
     {s_1} < \lbR {t} (s_1)$
     \item (Prefer $s_1$) $s_1 \succ_i s_0$
     \item (Does not qualify for any school $s \succ_i s_0$ on other tests) For all $t'
     \neq t$, \[
         R_{it'} < \lbR{t'}(s_0) \le \lbR{t'}(s_1).
     \]
 \end{enumerate}

 For $t_0 \neq t_1$: 
 \begin{enumerate}
     \item  $R_{it_1} = c_{s_1}$  
     
     \item $R_{it_0} \ge c_{s_0}$
     
     \item $s_1 \succ_i s_0$

     \item $\lbR{t_1}(s_0) = c_{s_1} < \lbR{t_1}(s_1)$

     \item  For all $t' \neq t_1$, \[
         R_{it'} < \lbR{t'}(s_0) \le \lbR{t'}(s_1).
     \]
 \end{enumerate} We can also check that these conditions are sufficient. In either
  case, we can write these conditions in the following way:
 \begin{align*}
 s_1 &\succ_i s_0 \\
  R_{it} &\in [0, \lbR{t}(s_0)) \text{ for all $t \not\in \br{t_0,t_1}$}\\
  R_{it_0} &\in [c_{s_0}, \lbR{t}(s_0)) \\ 
  R_{it_1} &= c_{s_1} = \lbR{t_1}(s_0) < \lbR{t_1}(s_1).
 \end{align*}

 Note that we can write those with $s_1 \succ_i s_0$ and $R_i \in \bar E_{s_0} \cap \bar
 E_{s_1}$ in \cref{prop:eligibility}(3) 
as  \begin{align*}
 s_1 &\succ_i s_0 \\
  R_{it} &\in [0, \lbR{t}(s_0)], \lbR{t}(s_0) > 0  \text{ for all $t \not\in 
  \br{t_0,t_1}$}\\
  R_{it_0} &\in [c_{s_0}, \lbR{t}(s_0)], \lbR{t}(s_0) > c_{s_0} \\ 
  R_{it_1} &= c_{s_1} = \lbR{t_1}(s_0) < \lbR{t_1}(s_1)
 \end{align*}
 Thus, the distinction is only on whether or not to take the closure of certain sets for
 $t\neq t_1$. Hence, up to measure-zero sets under the conditional distribution of
 $R_{i,-t_1} \mid R_{i,t_1}$, the identification regions are exactly the same. The causal
 effect identified is also exactly $\tau_{s_1 \succ s_0} = \E[Y(s_1) - Y(s_0) \mid s_1
 \succ_i s_0, R_i \in \bar E_{s_1} \cap \bar E_{s_0}]$.

\subsubsection{Results in \citet{abdulkadirouglu2017research}}

\newcommand{\MID}{\mathrm{MID}}

We translate some notation in \citet{abdulkadirouglu2017research} into our setup and show
that the identification results are equivalent. For simplicity, we assume a population in
which the cutoffs satisfy \cref{as:interior}(2) and \cref{as:interior}(4). We also assume
that $U_{i\ell}$ is iid uniform over $\ell$. We show that the local DA propensity score is
positive at cutoff $c$ (as a user-bandwidth tends to zero) if and only if $s \in \bar E_s$
in \cref{prop:eligibility}.

First, \citet{abdulkadirouglu2017research} define a notion of whether a student is never,
always, or conditionally seated at a given school (p.128). Their definition depends on a
user-chosen bandwidth; here, we discuss the equivalent statements when the bandwidth
vanishes. In our notation, for a lottery school $s$, a student $(\succ_i, Q_i, R_i)$ is
never seated if $V_{is}(Q_i, U_i) > c_s$ with probability zero, always seated if
$V_{is}(Q_i, U_i) > c_s$ with probability one, and conditionally seated if $V_{is}(Q_i,
U_i) > c_s$ with probability in $(0,1)$. For a test-score school $s$ using a test $t$, a
student is never seated if $R_{it} < \lbr{t}(s, Q_{is}, c_s)$ or $\lbr{t}(s, Q_{is},
c_s)=1$, always seated if $R_ {it} >
\lbr{t}(s, Q_ {is}, c_s)$ or $\lbr{t}(s, Q_{is},
c_s)=0$, and conditionally seated if $R_{it} = \lbr{t}(s, Q_ {is}, c_s)$ and $\lbr{t}(s, Q_{is},
c_s)\in (0,1)$
(recall \cref{eq:test_score_space_cutoff_q}). By assumption, $\lbr{t}(s, Q_ {is}, c_s)$ is
either in $(\epsilon , 1- \epsilon)$ or in $\br{0,1}$.

Second, \citet{abdulkadirouglu2017research} define a notion of ``most informative
disqualification'' (p.128). For a given school $s$ and a lottery $\ell$, $\MID_{is}^\ell
< 1$ if and only if no school $s' \succ_i s$ is such that $(\succ_i, Q_i)$ always
qualifies for $s'$. On the other hand, for $s$ and a test $t$, $\MID_{is}^t < 1$ if and
only if no school $s' \succ_i s$ has $\lbr{t}(s', Q_{is'}, c_{s'}) = 0$.

Finally, inspecting the implication of Theorem 1 in \citet{abdulkadirouglu2017research},
the local DA propensity score of $(\succ_i, Q_i, R_i)$ for school $s$ is positive if and
only if the following hold \begin{enumerate}
    \item $i$ is always or conditionally seated at $s$ 
    \item $i$ is conditionally or never seated at all schools $s' \succ_i s$
    \item $\MID_{is}^\ell < 1, \MID_{is}^t < 1$ for all $\ell, t$
    \item If $s$ is a lottery school, $c_s < c_{s'}$ for all $s' \succ_i s$ that uses the
    same lottery. 
\end{enumerate}
Note that we have the equivalent statements if $s$ is a lottery school:
\begin{enumerate}
    \item The probability of $(\succ_i, Q_i)$ qualifies for $s$ is positive 
    \item For all tests $t$, $R_{it} \le \lbR{t}(s, \succ_i, Q_i, c)$. For all lottery
    schools $s' \succ_i s$, the student does not qualify for $s'$ with probability 1.
    \item $\lbR{t}(s, \succ_i, Q_{i}, c) > 0$ for all $t$
    \item $i$ does not always qualify for some lottery school  $s' \succ_i s$  when $i$
    qualifies for $s$.
\end{enumerate}
This is further equivalent to $R_i \in \bar E_{s}$ in \cref{prop:eligibility}(1).

If $s$ is a test-score school using test $t$,
\begin{enumerate}
    \item $R_{it} \ge \lbr{t}(s, Q_{is}, c_s)$ and $\lbr{t}(s, Q_{is}, c_s) < 1$. 
    \item For all tests $t$, $R_{it} \le \lbR{t}(s, \succ_i, Q_i, c)$. Since the cutoffs
    are distinct, $\lbR{t}(s, \succ_i, Q_i, c) \neq \lbr{t}(s, Q_{is}, c_s)$ unless they
    are both zero or both one. 

     For all lottery
    schools $s' \succ_i s$, the student does not qualify for $s'$ with probability 1. 
    \item $\lbR{t}(s, \succ_i, Q_{i}, c) > 0$ for all $t$
\end{enumerate}
This is further equivalent to $R_i \in \bar E_{s}$ in \cref{prop:eligibility}(2).

    \section{Estimation and Inference: Proofs}
    \label{asec:estimation}

    This section contains details for estimation and inference for \cref{sub:est}.
    
\subsection{Lottery-based aTEs}
\label{asub:lottery_ate}

Let us first consider estimation of $\tau_{s_1 \succ s_0}$ where $s_1$ is a lottery
school. The asymptotics of estimators for this object likewise faces difficulty arising
from the fact that $C_N$ is random. To that end, define the aggregate treatment effect at
a given value of the cutoffs \begin{align*} c &\mapsto \tau_{s_1 \succ s_0}(c) \\
&=\E[Y(s_1) - Y(s_0) \mid R_i \in E_{s_1}(\succ_i, Q_i, c) \cap E_{s_0}(\succ_i, Q_i, c),
s_1 \succ_i s_0] \\&= \int (y (s_1) - y (s_0)) dP (y (s_1), y (s_0)
    \mid R_i \in E_{s_1}(\succ_i,
    Q_i, c) \cap E_{s_0}(\succ_i, Q_i, c), s_1 \succ_i s_0).
\end{align*}
Note that $\tau_{s_1 \succ s_0} (C_N)$ is a random estimand, whose randomness comes from
$C_N$, but it integrates over the population treating the cutoff $C_N$ as fixed rather
than a function of data. 

Correspondingly, consider an estimator defined by \begin{align*}
 \hat \tau_{s_1 \succ s_0} = &\pr{\frac{1}{N}\sum_{i=1}^N J(\succ_i, Q_i, R_i; C_N)}^
    {-1}  \\&\quad \quad\times\frac{1} {N} \sum_ {i=1}^N \underbrace{J (\succ_i, Q_i, R_i;
    C_N) \br{
        \frac{D (s_1;
        U_i, Q_i, 
        \succ_i, C_N)} {\pi(s_1; Q_i, 
        \succ_i, C_N)} - \frac{D(s_0; U_i, Q_i, 
        \succ_i, C_N)} {\pi(s_0; Q_i, 
        \succ_i, C_N)}} Y_i}_{g(X_i, Y_i; C_N)} \\ 
    &\equiv (\Pn[N] J(\cdot; C_N))^{-1} \Pn[N] g(\cdot ; C_N).
\end{align*}
Here, we use the empirical process notation $\Pn[N]$ to denote a sample mean, and $\Pn[]$
to denote integral with respect to the distribution of $(Y_i, X_i)$. Moreover, 
\begin{align*}
J_i(R_i; c) &\equiv J_i(c) = J(\succ_i, Q_i, R_i; c) \equiv \one(s_1 \succ_i s_0, R_i \in
E_ {s_0}
(\succ_i, Q_i, c)
\cap E_{s_1}(\succ_i, Q_i, c) ) \\ 
D_{i1}(U_i; c) &\equiv D(s_1; U_i, Q_i,
    \succ_i, c) = \one(
    \underbrace{V_{is_1}(Q_i, U_i) \ge c_{s_1}}_{\text{qualifies for $s_1$}}) \prod_
    {s: \text{lottery}, s\succ_i s_1} \underbrace{\one\pr{V_ {is}
        (Q_i, U_i) < c_s}}_{\substack{\text{fails to qualify for} \\ \text{any better
        lottery school}}}. \numberthis \label{eq:def_d1}
\end{align*}
$D_{i0}(U_i; c)$ is defined similarly for $s_0$ a lottery school. If $s_0$ is a
test-score school, then $D_{i0}(U_i; c)$ is simply \[
    \prod_
    {s: \text{lottery}, s\succ_i s_0} \underbrace{\one\pr{V_ {is}
        (Q_i, U_i) < c_s}}_{\substack{\text{fails to qualify for} \\ \text{any better
        lottery school}}},
\]
since $R_i \in E_{s_0}$ already implies that individual $i$ does not qualify for any
preferred test-score school and qualifies for $s_0$. 
Finally, \[
    \pi(s; Q_i, 
    \succ_i, c) = \int D(s; u, Q_i,  \succ_i, c) dP_U(u)
\]
is the conditional probability that $D(s; U_i, Q_i, \succ_i, c) = 1$, given $(Q_i,
R_i, \succ_i)$ and treating the cutoff $c$ as fixed.

In short, $D_{i1}, D_{i0}$ denotes whether a student's lottery numbers qualify her at a
given $s_j$ and fail to qualify her at any schools she prefers to $s_j$, at a fixed cutoff
$c$. $\pi(s_j ;\cdot)$ then denotes the probability that $D_{ij}  = 1$, treating the
cutoff as fixed.

\begin{as}
\label{as:moment_strong}
 There exists some constant $0 < C < \infty$ such that $\E[|Y_i(s)|^4] < C$.
\end{as}

\thmlotterymain* 

\begin{tcbproof}
    Note that it suffices to show that the numerator and the denominator jointly obeys the
    asymptotic representation \begin{align*}
   \sqrt{N}(\Pn[N] g(\cdot; C_N) - \pr{\Pn[] J(\cdot; C_N)} \tau_{s_1 \succ s_0}(C_N) ) = 
   \sqrt{N}(\Pn[N] - \Pn[]) g
   (\cdot; C_N) = \underbrace{\sqrt{N} (\Pn[N] - \Pn[]) g(\cdot; c)}_{\text{asymptotically
   Gaussian}} + o_P (1)
\\
    \sqrt{N}(\Pn[N] - \Pn[]) J(\cdot; C_N) = \sqrt{N} (\Pn[N] - \Pn[]) J(\cdot; c) + o_P(1).
    \end{align*}

    The quantities on the right-hand side are scaled sample means of i.i.d. terms, and
    thus obey central limit theorems under \cref{as:moment_strong,as:interior}. The
    resulting asymptotic linear representation \cref{eq:influence} and central limit
    theorem \cref{eq:clt} for the ratio follow by an application of the delta method,
    since $\E [J_i (R_i; c)] > 0$.

    The remainder of the proof shows the representation for the numerator\[
        \sqrt{N}(\Pn[N] - \Pn[]) [g
   (\cdot; C_N) - g(\cdot; c)] = o_P (1).
    \]
    The representation for the denominator can be shown analogously. To this end, consider
    an event $A_N$, which occurs if the following occurs:
    \begin{enumerate}
        \item $\norm{C_N - c}_\infty \le M_N/\sqrt{N}$ for some $0 < M_N \to \infty$ and
        $M_N = o(\sqrt{N})$

         \item For some $\eta > 0$, $\pi(\cdot ;  \tilde c) > \eta$ when $ D_
         {i1}(U_i; \tilde c) = 1$ for $\tilde c \in \br{c, C_N}$
         \item There exists some region $B_T \subset [0,1]^T$ and $B_L \subset [0,1]^L$,
         both of Lebesgue measure at least $1-CM_N/\sqrt{N}$, such that whenever $U_i \in
         B_L, R_i \in B_T$, we have $J_i(C_N) = J_i(c)$,  $D_{i1}(U_i; C_N) = D_{i1}(U_i;
         c)$, $D_{i0}(U_i; C_N) = D_{i0}(U_i; c)$
 \end{enumerate}
\Cref{lemma:an_lottery} verifies that $\P(A_N) \to 1$ for some choice of $\eta$ and $M_N$ under
\cref{as:bounded_density,as:interior}. 

Since \[
    \P\bk{\abs[\big]{\sqrt{N} (\Pn[N] - \Pn[]) (g(\cdot; C_N) - g(\cdot; c))}> \epsilon }
    \le \P
    (A_N^C) + \P(A_N, \abs[\big]{\sqrt{N} (\Pn[N] - \Pn[]) (g(\cdot; C_N) - g(\cdot; c))}> \epsilon ),
\]
it suffices to bound the empirical process under the event $A_N$. 
Thus, under $A_N$, \[
    \abs[\big]{\sqrt{N} (\Pn[N] - \Pn[]) (g(\cdot; C_N) - g(\cdot; c))} \le \sup_{\tilde
    c : \norm{\tilde c - c}_{\infty} < M_N /\sqrt{N} } \abs[\big]{\sqrt{N} (\Pn[N] - \Pn
    []) (g(\cdot; \tilde c) - g(\cdot; c))}
\]
Let \[
        h_1(X_i, Y_i; c) = J_i(R_i; c) \frac{D_{i1}(U_i; c)}{\pi(s_1; Q_i, \succ_i, c)
        \vee
        \eta} Y_i
    \]
    and $h_0$ be analogously defined.
    We can further bound \[
        \abs[\big]{\sqrt{N} (\Pn[N] - \Pn[]) (g(\cdot; C_N) - g(\cdot; c))} \le \sum_
        {j\in\br{0,1}} \sup_ {\tilde
    c : \norm{\tilde c - c}_{\infty} < M_N /\sqrt{N} } \abs[\big]{\sqrt{N} (\Pn[N] - \Pn
    []) (h_j(\cdot; \tilde c) - h_j(\cdot; c))}.
    \]

   Thus it suffices to verify\[
   \sup_ {\tilde
    c : \norm{\tilde c - c}_{\infty} < M_N /\sqrt{N} } \abs{\sqrt{N}(\Pn[N] - \Pn[]) [h_1
       (\cdot; \tilde c) - h_1(\cdot; c)]} = o_P (1). \numberthis
   \label{eq:empirical_process_term}
    \] 
    The argument for $h_0$ follows analogously. 

    Let $\mathcal H = \br{
        h_1(\cdot; \tilde c) : \tilde c \in [0,1)^{S}, \norm{\tilde c - c}_{\infty} < M_N
        / \sqrt{N}
    }$. Let \[
        H_N(\cdot) = \sup_{h \in \mathcal H} \abs{h(\cdot) - h_1(\cdot; c)}
    \]
    be the envelope function of the class $\mathcal H - h_1(\cdot; c)$. By Section
    2.14.4 in \citet{vaart2023empirical}, \cref{eq:empirical_process_term} follows pending
    verification that (i) $\mathcal H$ is a class of functions with a uniformly bounded VC
    subgraph
    index and (ii) $\E[H_N^2] = o(1)$.

For (i), note first that it suffices to verify that \[
    \mathcal H(\succ, q) = \br{
        h(\succ, q, \cdot) : h \in \mathcal H
    }
\]
is a VC class for every fixed $\succ, q$, since there are only finitely many distinct $
(\succ, q)$ values. 

Fix $(\succ, q)$ and suppress it from notation, we note that 
\[
    h_1(R_i, U_i, c) = \min\pr{
       J(R_i;c), D_{1}(U_i; c)} \frac{Y_i}{\pi(s_1; c) \vee
        \eta}.
\]
By the permanence properties of VC classes (Lemma 2.6.18(i) and (vi) in 
\citet{van1996weak}),  it suffices to verify that the following sets and
functions are VC
classes, as $\tilde c $ ranges over $\norm{\tilde c - c}_{\infty} < M_N /\sqrt{N}$: 
\begin{align*}
J_{\tilde c} &= \br{r : J(r; \tilde c) = 1} \\
D_{\tilde c} &= \br{
    u: D(s_1; u, \tilde c) = 1
} 
\end{align*}

To prove $\mathcal J = \br {J_{\tilde c}: \tilde c}$ is a VC class, we can represent the
     set of test-score schools one qualifies for as a binary string $b = (b_1, \ldots,
     b_{T_s}) \in \br{0,1}^ {T_s}$, assuming there are $T_s \le M$ test-score schools.
     Now, $J(R; \tilde c) = 1$ if and only if $b$ takes certain given values, collected in
     some set $\mathcal B$, which depends on $(\succ, q)$. Thus,
     \[
         J_{\tilde c} = \bigcup_{b \in \mathcal B} \bigcap_{s \text{
         test-score school}} \br{R \in [0,1]^T : (-1)^{b_s} V_s(q, R) < (-1)^
         {b_s} \tilde c_s }.
     \] 
     This representation writes $J_{\tilde c}$ as a union over test-score school
     qualification statuses $b$. For each $b$, the region on $R$ then delineates whether
     $R_{t_s}$ clears or fails to clear the threshold for school $s$. Thus, each member of
     $\mathcal J$ is a finite union and intersection of sets of the form \[
         \br{R \in [0,1]^T : R_t < r} \text{ or } \br{R \in [0,1]^T : R_t > r}.
     \]
     Since the class of sets $\br{
         \br{R \in [0,1]^T : R_t < r} : r \in [0,1]
     }$ has finite VC dimension, $\mathcal J$ also has finite VC dimension. The argument
     that $D_{\tilde c}$ form a VC class is analogous. This verifies (i). 

     For (ii), we may decompose
\begin{align*}
\E[H_N^2] &= \E[H_N^2 \one(A_N)] + \E[H_N^2 \one(A_N^C)] \\
&\le \E[H_N^2 \one(A_N, R_i\in B_T, U_i \in B_L)] + 
\E[H_N^2 \one(R_i\not\in B_T \text{ or } U_i \not\in B_L)] + \E[H_N^2 \one(A_N^C)]
\end{align*}
On the event $A_N, R_i\in B_T, U_i \in B_L$, \[
    H_N(X_i, Y_i) = \sup_{h \in \mathcal H} |h - h_1(\cdot; c)| \le |Y_i| \sup_{
    \norm{\tilde c - c}_
    {\infty}}
    \abs[\bigg]{
        \frac{1}{\pi(s_1; Q_i, \succ_i, \tilde c) \vee\eta} - \frac{1}{\pi(s_1; Q_i,
        \succ_i,
        c) \vee \eta}
    } \lesssim_{\eta} |Y_i| M_N N^{-1/2}.
\]
Separately, \[
    \E[H_N^2 \one(K)] \le \sqrt{\E[H_N^4] \P(K)}
\]
where \[
    H_N^4 (X_i, Y_i) \lesssim_{\eta} |Y|^4.
\]
Now, since $B_T, B_L$ are sets with large Lebesgue measures, \[
    \P\pr{R_i\not\in B_T \text{ or } U_i \not\in B_L} = o(1)
\]
by \cref{as:bounded_density}. Finally, since $\E[|Y|^4]$ is assumed to be finite
(\cref{as:moment_strong}), we have that \[
    \E[H_N^2] = o(1). 
\]
This verifies (ii) and proves \eqref{eq:empirical_process_term}. 
\end{tcbproof}

\begin{lemma}
\label{lemma:an_lottery}
    In the proof of \cref{thm:lottery_main_estimation},  the event $A_N$ occurs with
    probability tending to one.

\end{lemma}

\begin{tcbproof}
    It suffices to check each individually, since the finite intersection of eventually
    almost sure events are eventually almost sure. 

    The first claim is immediate by \cref{as:limitcutoff} for any $M_N$ that diverges. 

    Note that $
\delta \equiv        \min_{q, \succ} \br{\pi(s_1; q, \succ, c) :  \pi(s; q,\succ, c) > 0}
    $ is a minimum over finitely many positive elements and thus $\delta > 0$. Moreover,
    each \[
        \pi(s_1; q, \succ, c) = \int D_{i1}(u; c) dF_U(u)
    \]
If $D_{i1}(U_i; c) = 1$, then its integral must be positive, and hence $\pi(s_1; q,
\succ, c) \ge \delta$. 

For all sufficiently large $N$, consider each multiplicand in \cref{eq:def_d1}. By 
\cref{as:limitcutoff} and since $\norm{C_N - c}_\infty \le o(1)$, we have the the
multiplicands in \cref{eq:def_d1} are either both zero with probability one or both
one with positive probability for cutoffs $c$ or
cutoffs $C_N$. Consider some $\pi(s_1; q, \succ, C_N)$ for which $\pi(s_1; q, \succ, c) >
0$. Pending verification of (3), we have that $D_{i1}(u; C_N) \neq D_{i1}(u; c)$ only if
$u \not\in B_L$, thus \[
    |\pi(s_1; q, \succ, C_N) - \pi(s_1; q, \succ, c)| \le \int |D_{i1}(u; C_N) - D_{i1}(u;
    c)| dF_U(u) \le \int_{B_L^c} dF_U(u) \lesssim M_N N^{-1/2}.
\]
Hence for all sufficiently large $N$, \[
    \pi(s_1; q, \succ, C_N) \ge \pi(s_1; q, \succ, c) - o(1) > \delta/2
\]
with probability tending to one. Thus (2) is shown by taking $\eta = \delta/2$.

Take $B_T^C = \br{r \in [0,1]^T : V_{s}(r_t, q) \text{ is between $c_s$ and $C_{N,s}$ for
some $s, q$}}$ and define $B_L^C$ similarly. By \cref{as:interior} and for all
sufficiently
large $N$, the set of schools anyone qualifies for sure is the same at $c$ and at $C_N$.
If $R_i \in B_T, U_i \in B_L$, then the qualification status of each school is also the
same under $c$ or $C_N$. Thus $D_{i1}, J_i$ would also be the same. We conclude the proof
by noting that \[
    \max(\P(B_T^C), \P(B_L^C)) \lesssim M_N /\sqrt{N}
\]
by \cref{as:bounded_density}. 
\end{tcbproof}

\subsection{RD-based aTEs: setup}

 Recall that we consider $s_1, s_0$ where the
    test-score school $s_1$ uses test $t_1$, and we consider only students who prefer
    $s_1$ to $s_0$.

    We have the the following treatment effect \[
    \tau = \tau_{s_0,s_1} = \E\bk{Y(s_1) - Y(s_0) \mid s_1 \succ s_0, R \in
    \bar E_{s_0}
    (\succ, Q;c)
    \cap \bar E_{s_1}(\succ, Q; c)}.
    \]

    For a school $s$ using test $t$, recall that $r_{s,t}(c)$
    (\cref{eq:test_score_space_cutoff}) is the unique test-score-space cutoff such that for
    some $q = 0,\ldots, \bar q_s$, \[
\frac{q + r_{s,t}(c)}{\bar q_s + 1} = c_s
    \]
     For convenience on the test-score cutoff of school $s_1$, let $\rho(c) = r_{s_1,t_1}
    (c)$.

\subsubsection{Unpacking $J_i(c)$ and defining $J_i(C_N, h_N)$.}
    Let us first decompose the conditioning event into restrictions
    on $t_1$ and restrictions not on $t_1$. Recall that the intersection \[
    \bar E_{s_0}
    (\succ, Q; c)
    \cap \bar E_{s_1}(\succ, Q; c) \quad \text{where $s_1 \succ s_0$}
    \]
    takes the form of (\cref{eq:cutoff_variation_intersection}), which is a Cartesian
    product of intervals corresponding to the following conditions on the vector of test
    scores $R = [R_1,\ldots, R_T]'$: \begin{align*} R_{t_1} &= \lbr{t_1}(s_1; Q, \succ, c)
    \\ R_{t_1} &\le \lbR{t_1}(s_0; Q, \succ, c) 
    \\
    R_t &\le  \lbR{t}(s_0; Q, \succ, c) \\ %
    s_0 &\succ L(Q, c)
    \end{align*}
Additionally, if $s_0$ is a test-score school that uses $t_0$ and $t_1 \neq t_0$, then \[
R_{t_0} \in [\lbr{t_0}(s_0; Q,
    \succ,
    c), \lbR {t_0} (s_0; Q, \succ, c)]
\]

    Define the
    following indicator random variables (functions of $R, \succ, Q$) that
    correspond to the above 
    restrictions, with the restrictions on $t_1$ relaxed with the
    bandwidth parameter $h$:
        \begin{itemize}[wide]
        \item (In-bandwidth) $I_1^+(c, h) =
    \one\pr{R_{t_1} \in [\rho(c), \rho(c) + h]}$ 
    and $I_1^{-}(c, h) = \one\pr{R_{t_1} \in [\rho(c) - h, \rho(c)]} $
    \item (Everyone in bandwidth does not qualify for something better than $s_1$ and
    everyone qualifies for $s_0$) If $s_0$ is a test-score school that uses $t_1$, then
    \[I_1(c,h) = \one(\rho(c) + h < \lbR{t_1}(s_1; Q, \succ, c), \rho(c) - h > \lbr{t_1}
    (s_0; Q, \succ, c)).\] Otherwise \[
I_1(c,h) = \one(\rho(c) + h < \lbR{t_1}(s_1; Q, \succ, c)).
    \]

    \item (No one left of cutoff qualifies for test schools better than $s_0$) $I_{10}(c)
    = \one\pr{\rho(c) \le \lbR
    {t_1}
    (s_0; Q, \succ, c) }$
        \item (Qualifies for $s_0$) If $s_0$ is a test-score school and $t_1 \neq t_0,$
        then
        \[I_0(c) = \one\pr{R_{t_0} \in [\lbr{t_0}(s_0;
    Q, \succ, c), \lbR
    {t_0}
    (s_0; Q, \succ, c)]
    }.\]
    Otherwise $I_0(c) = 1$
    \item (Does not qualify for test-score schools preferred to $s_0$ except for $s_1$)
    For $t \neq t_0, t_1$, define $I_t(c) = \one\pr{0 < R_t \le \lbR
    {t}(s_0; Q, \succ, c)}$
    \item (Does not qualify for preferred lottery schools with probability 1) $I(c) = \one
    \pr{s_0 \succ L(Q, c), s_1 \succ s_0}$
    \item Let the sample selection indicator be defined as 
    \begin{equation}
        J(c, h) = I(c)
    I_{10}(c) \cdot I_1(c, h) I_0(c) \cdot  \prod_{t\neq t_0, t_1} I_t(c),
    \label{eq:jdef}
    \end{equation}
    such that, for fixed $(R, Q,
    \succ)$,  \[
    s_1 \succ s_0 \text{ and } R \in \bar E_{s_0}
    (\succ, Q; c)
    \cap \bar E_{s_1}(\succ, Q; c) \iff \lim_{h \to 0 } J(c, h) (I_1^+(c,
        h) \vee I_1^-(c,h)) = 1.
    \]
    \end{itemize}
    Define $J_i(c, h) = J(c,h)$ where $R_i, \succ_i, Q_i$ is plugged in.

\subsubsection{Defining the proxy outcome $Y^{(j)}(C_N)$.}

Define $D_i^{(1)}(c)$ to be the indicator for failing to qualify for lottery schools
that $\succ_i$ prefers to $s_1$: \[
D_i^{(1)}(c) = \prod_{s : \ell_s \neq \emptyset, s\succ s_1} \one(V_{is}(Q_i, U_i) < c_s).
\]
Let \[
\pi_i^{(1)}(c) = \P_U(D_i^{(1)}(c) = 1 \mid Q_i, \succ_i).
\]
be the corresponding probability ($\P_U$ emphasizes that this is only a function of the
distribution of $U$). Similarly, define \[
D_i^{(0)}(c) = \begin{cases}
    \prod_{s : \ell_s \neq \emptyset, s\succ s_0} \one(V_{is}(Q_i, U_i) <
c_s), &\text{ if $s_0$ is a test-score school}\\
    \prod_{s : \ell_s \neq \emptyset, s\succ s_0} \one(V_{is}(Q_i, U_i) <
c_s) \cdot \one(V_{is_0}(Q_i, U_i) > c_{s_0}) &\text{ if $s_0$ is a lottery school}
\end{cases}
\]
and $\pi^{(0)}(c) = \P_U(D_i^{(0)}(c) = 1 \mid Q_i, \succ_i)$.

Define the proxy outcome as the IPW ratio \[
Y^{(j)}(c) = \frac{D_i^{(j)}(c) Y_i}{\pi_i^{(j)}(c)} \numberthis \label{eq:proxy_outcome}
\]
where we impose the convention $0/0 = 0$. 

Note that for all $h$, almost
surely
\[
J_i(C_N, h) I_1^+(C_N, h) Y^{(1)}(C_N) = J_i(C_N,h) I_1^+(C_N, h) \frac{D^{(1)}_i
(C_N) Y_i(s_1)}
{\pi_i^{(1)}(C_N)}.
\numberthis \label{eq:can_replace_outcome}
\]
and similarly 
\[
J_i(C_N, h) I_1^-(C_N, h) Y^{(0)}(C_N) = J_i(C_N,h) I_1^-(C_N, h) \frac{D_i^{(0)}
(C_N) Y_i(s_0)}
{\pi_i^{(0)}(C_N)}.
\]
This is because when $J_i(C_N, h) I_1^+(C_N, h) = 1$, either $i$ is assigned to $s_1$
or they are assigned to a lottery school they prefer to $s_1$. Similarly, when $J_i(C_N,
h) I_1^-(C_N, h)  = 1$, if $s_0$ is a test-score school (and $i$ wins the lottery at
   $s_0$), then either $i$ is assigned to $s_0$ or to a lottery school $i$ prefers to
   $s_0$.

\subsubsection{Estimators of the right-limit.} The rest of this section now restricts
to considering the right-limit. To simplify notation, define \[Y_i(C_N) =
\frac{\di_i
(C_N) Y_i(s_1)} {\pi_i(C_N)}\] where $\di_i (C_N) = D^{(1)}_i (C_N)$ and $\pi_i(C_N) =
\pi_i^{(1)}(C_N)$. We can replace $Y_i^{(1)}(C_N)$ by $Y_i(C_N)$ because of 
(\cref{eq:can_replace_outcome}). A
natural estimator of the right-limit, \[
        \lim_{h \to 0} \E\bk{
        Y \mid J(c,h)=1,  I_1^+(c, h) = 1
        } = \E[Y(s_1) \mid s_1 \succ s_0, R \in \bar E_{s_0}
        (\succ, Q;c)
        \cap \bar E_{s_1}(\succ, Q; c)],
    \]
    is a locally linear estimator with uniform kernel and bandwidth $h_N$:
    \[
    \hat \beta(h_N) = [\hat \beta_0, \hat \beta_1]' = \argmin_{\beta_0,
    \beta_1}
    \sum_{i=1}^N W_i(C_N, h_N)
    \bk{Y_i(C_N) - \beta_0 - \beta_1 (R_{it_1} - \rho(C_N))}^2,
    \]
    where $W_i(C_N, h_N) = J_i(C_N, h_N) I_{1i}^+(C_N, h_N)$.

    Let $x_i(C_N) = [1, R_
    {it_1} - \rho(C_N)]'$ collect the right-hand side variable in the
    weighted least-squares regression. Then the locally linear regression
    estimator is \[
    \hat \beta_0(h_N) = e_1'\pr{\sum_{i=1}^N W_i(C_N, h_N) x_i(C_N) x_i
    (C_N)'}^{-1}
    \pr{\sum_{i=1}^N W_i(C_N, h_N) x_i(C_N) Y_i(C_N)} \quad e_1 \equiv [1,0]'
    \]
    There is a natural oracle estimator \[
    \check \beta_0(h_N) = e_1'\pr{\sum_{i=1}^N W_i(c, h_N) x_i(c) x_i(c)'}^
    {-1}
    \pr{\sum_{i=1}^N W_i(c, h_N) x_i(c) Y_i(c)}
    \]
    whose asymptotic properties are well-understood.  Our goal is to show that
    the difference between the two estimators is small:\[
    \sqrt{nh_N}\pr{\hat \beta_0(h_N) - \check \beta_0(h_N)} = o_p(1).
    \]
    
   \subsubsection{Assumptions} 
   First, let us recall the following assumption to avoid certain knife-edge populations.
    \interior*
    
    Next, let us also state the following technical conditions
        \begin{restatable}[Bounded densities]{as}{boundeddensity}
        \label{as:bounded_density}
        For some constant $0 < B < \infty$,
        \begin{enumerate}[wide]
            \item         The density of $(R_i \mid {\succ_i}, Q_i, Y_i(0), \ldots, Y_i
            (M))$ with respect to
        the Lebesgue measure is positive and  bounded by $B$,
        uniformly over the conditioning variables.

\item         The density of $U_i = [U_{i\ell_s} : \ell_s \neq \emptyset]$
with respect to the Lebesgue measure is positive and bounded by $B$.

        \end{enumerate}
    \end{restatable}

    Define \[\mu_+(r) = \E[Y_i(s_1) \mid J_i(c) = 1, R_{t_1}=r]\] and $\mu_-(r) = \E[Y_i
    (s_1)
    \mid J_i(c) = 1, R_{t_1}=r]$ where $J_i(c) = J_i(c, 0)$.  
    
   \begin{restatable}[Moment bounds]{as}{secondmoment}
    \label{as:second_moment}
   
    \begin{enumerate}[wide]
        \item  Let $\epsilon_i^{(1)} = Y_i(s_1) - \mu_+(r)$. For some
        $\varepsilon > 0$, the $(2+\varepsilon)$\th{} moment exists and is
        bounded uniformly: \[\E
        [(\epsilon_i^{(1)})^{2 + \varepsilon} \mid J_i(c) = 1, R_{it_1} = r] <
        B_V(\varepsilon) <
        \infty.\]  Similarly, the same moment bounds hold for $Y_i(s_0)$. Note that this
        implies that the second moment is bounded uniformly by some $B_V = B_V(0)$.
        \item The conditional variance $\var(\epsilon_i \mid J_i(c) = 1,
        R_{it_1} = r)$ is right-continuous at $\rho(c)$ with right-limit
        $\sigma_+^2 > 0$. Similarly, the conditional variance for $Y_i(s_0)$ is
        also continuous with left-limit $\sigma_-^2 > 0$.
        \item The conditional first moment is bounded uniformly: $\E[
        \abs{Y_i(s_k)} \mid R_i,
    \succ_i, Q_i] < B_M < \infty$ for $k=0,1$.
    \end{enumerate}   
    \end{restatable}

    \begin{restatable}[Smoothness of mean]{as}{ctsdiff}
    \label{as:cts_diff}
        The maps $\mu_+(r), \mu_-(r)$ are thrice continuously
        differentiable with
        bounded third derivative $\norm{\mu_+'''(r)}_\infty, \norm{\mu_-'''
        (r)}_\infty < B_D <
        \infty$.
    \end{restatable}

    \begin{restatable}[Continuously differentiable density]{as}{ctsdensity}
    \label{as:ctsdensity}
        The density $f(r) = p(R_{it_1}=r \mid J_i(c) = 1)$ is continuously
        differentiable at $\rho(c)$ and strictly positive.
    \end{restatable}

   \subsubsection{Statement of results for \cref{thm:estimation_main}}
        
    \begin{theorem}
    \label{athm:equiv}
    Under 
\cref{as:limitcutoff,as:interior,as:bounded_density,as:ctsdensity,as:second_moment,as:cts_diff}, assuming $N^{-1/2} = o
(h_N)$ and $h_N = o
(1)$, then 
the feasible estimator and the
oracle estimator are equivalent in the first order \[
    \sqrt{N h_N} (\hat \beta_0 - \check \beta_0) = O_p\pr{h_N^{1/2} + N^
    {-1/4}h_N^{-1/2} + N^{-1/2}h_N^{-1}} = o_p(1).
\]
    \end{theorem}
    \begin{cor}
        Under \cref{athm:equiv}, we immediately have that the discrepancy $ \sqrt{N h_N}
         (\hat \beta - \check \beta) = o_p(\sqrt{N h_N}h_N^2)$ if $h_N = O(N^{-d})$ with
         $d \in (0.2, 0.25)$.
    \end{cor}
    
    Let $\check \beta = \check A_{1N}^{-1} \check A_{2N}$ and let $\hat
    \beta = \hat A_{1N}^{-1} \hat A_{2N}$ for matrices $\check A_{kN},
    \hat A_{kN}$. The theorem follows from the following proposition.
    \begin{prop}
    \label{aprop:steps}
     Under 
\cref{as:limitcutoff,as:interior,as:bounded_density,as:ctsdensity,as:second_moment,as:cts_diff}, assuming $N^{-1/2} = o
(h_N)$ and $h_N = o
(1)$, then
    \begin{enumerate}
        \item The matrix \[\check A_{1N} = \begin{bmatrix}
            O_p(1) & O_p(h_N) \\ 
            O_p(h_N) & O_p(h_N^2)
        \end{bmatrix}\] and, as a result, \[
        (\check A_{1N} + b_N)^{-1} =  \check A_{1N}^{-1} + \begin{bmatrix}
            O_p(b_N) & O_p(b_N / h_N^2) \\ 
            O_p(b_N/h_N^2) & O_p(b_N / h_N^4)
        \end{bmatrix}.
        \]
        Similarly, \[
        \check A_{2N} = \colvecb{2}{O_p(1)}{O_p(h_N)}.
        \]
        \item Let\footnote{We write $J_i(c)$ instead of $J_i(c, h_N)$ since it does not depend
        on $h_N$ for sufficiently small $h_N$---see 
        \cref{cor:J_fact}.} \begin{align*}
        \tilde \beta &\equiv 
        \pr{\frac{1}{Nh_N}\sum_{i=1}^N J_i(c) I_{1i}^+
        (C_N, h_N) x_i(C_N) x_i(C_N)'}^
    {-1}
    \pr{\frac{1}{Nh_N}\sum_{i=1}^N J_i(c) I_{1i}^+
        (C_N, h_N) x_i(C_N) Y_i(c)} \\&\equiv \tilde A_{1N}^{-1} \tilde A_
        {2N}.
        \end{align*}
        Then\[
        \tilde A_{1N} = \check A_{1N} + \begin{bmatrix}
            O_p(N^{-1/2}/h_N) & O_p(N^{-1/2}) \\ 
            O_p(N^{-1/2}) & O_p(N^{-1/2} h_N) 
        \end{bmatrix}
        \]
        and \[
        \tilde A_{2N} = \check A_{2N} + \colvecb{2}{N^{-1/2}/h_N}{N^
        {-1/2}} = \colvecb{2}{O_p(1)}{O_p(h_N)}.
        \]
        
        \item Moreover,  $\sqrt{Nh_N}(\hat \beta_0 - \tilde \beta_0) = O_p
        \pr{
                \sqrt{h_N}}.$
                \item We may write the discrepancy as \[
                \tilde A_{1N}^{-1} \sqrt{Nh_N} \tilde A_{2N} - \check A_
                {1N}^{-1} \sqrt{Nh_N} \check A_{2N} = \tilde A_{1N}^{-1}  \tilde B_{2N} - \check A_
                {1N}^{-1}  \check B_{2N}
                \]
                for some $\tilde B_{2N}, \check B_{2N}$ where (a) $\check
                B_{2N} = [O_p(1), O_p(h_N)]'$ and (b) \[
                \tilde B_{2N} = \check B_{2N} + \colvecb{2}{O_p(h_N^{3/2} +
                N^
                {-1/4 }h_N^{-1/2})}{O_p(h_N^{5/2} + N^
                {-1/4} h_N^{1/2})}.
                \]
        
    \end{enumerate}
    \end{prop}
    
    \begin{myproof}[Proof of \cref{athm:equiv} assuming \cref{aprop:steps}]
        
        We multiply out, by parts (2) and (4): \[
        \tilde A_{1N} \tilde B_{2N} = \pr{\check A_{1N} + \begin{bmatrix}
            O_p(N^{-1/2}/h_N) & O_p(N^{-1/2}) \\ 
            O_p(N^{-1/2}) & O_p(N^{-1/2} h_N) 
        \end{bmatrix}}^{-1} \pr{\check B_{2N} + \colvecb{2}{O_p(h_N^{3/2} +
                N^
                {-1/4 }h_N^{-1/2})}{O_p(h_N^{5/2} + N^
                {-1/4} h_N^{1/2})}}
                \]
                The first term is \[
                \check A_{1N}^{-1} + \begin{bmatrix}
            O_p(N^{-1/2}/h_N) & O_p(N^{-1/2}/h_N^2) \\ 
            O_p(N^{-1/2}/h_N^2) & O_p(N^{-1/2}/h_N^3) 
        \end{bmatrix}.
        \]
        Multiplying out, we have that the RHS is \[
        \check A_{1N}^{-1}\check B_{2N} + \colvecb{2}{
        O_p\pr{N^{-1/2} / h_N + h_N^{3/2} + N^{-1/4}h_N^{-1/2}}
        }{
        O_p\pr{h_N^{-1}\cdot \pr{N^{-1/2} / h_N + h_N^{3/2} + N^{-1/4}h_N^
        {-1/2}}}
        }
        \]
        The total discrepancy between $\hat \beta$ and $\check \beta$, in
        the first entry, by (3), is then \[
        O_p\pr{N^{-1/2} / h_N + h_N^{3/2} + N^{-1/4}h_N^{-1/2} + 
        \sqrt{h_N}} = O_p\pr{N^{-1/2} / h_N + h_N^{1/2} + N^{-1/4}h_N^
        {-1/2}}.
        \]
    \end{myproof}

    \subsubsection{Proof of \cref{aprop:steps}} We prove \cref{aprop:steps} in the
    remainder of this section. The first part is a direct application of
    Lemma A.2 in \citet{imbens2012optimal}, which is a routine approximation
    of the sum $\tilde A_{1N}$ with its integral counterpart.
    \begin{myproof}[Proof of \cref{aprop:steps}(1)]
        The claim follows directly from \cref{lemma:ik}, which is a
        restatement of Lemma A.2 in \citet{imbens2012optimal}. The
        inversion part follows from $1/(a+b) = 1/a + O(b/a^2)$.
    \end{myproof}
    
    Next, the proof of part (2) follows from bounds of the discrepancy
    between $\tilde A_{1N}$ and $\check A_{1N}$, detailed in 
    \cref{lemma:Sn_bounds}. 
    
    \begin{myproof}[Proof of \cref{aprop:steps}(2)]
        \Cref{lemma:Sn_bounds} directly shows that \[
        \tilde A_{1N} = \check A_{1N} + O_p(N^{-1/2})\begin{bmatrix}
            1/h_N & 1 \\ 
            1 & h_N
        \end{bmatrix}
        \]
        when we expand \[A_{1N} = \begin{bmatrix}
            S_{0N} & S_{1N} \\ 
            S_{1N} & S_{2N}
        \end{bmatrix}\]
        in the notation of \cref{lemma:Sn_bounds}. 
        The part about $\tilde A_{2N}$ follows similarly from 
        \cref{cor:Sn_bounds}. 
    \end{myproof}

    Next, the proof of part (3) follows from bounds of the discrepancy
    between $\hat A_{kN}$ and $\tilde A_{kN}$, detailed in 
    \cref{lemma:boundJ,lemma:boundY}. 
  \begin{myproof}[Proof of \cref{aprop:steps}(3)]
  Note that (1) and (2) implies that \[\tilde A_{1N} = \begin{bmatrix}
      O_p(1) & O_p(h_N) \\ 
      O_p(h_N) & O_p(h_N^2)
  \end{bmatrix}.\]
  
        \Cref{lemma:boundJ} shows that \[
        \hat A_{1N} = \tilde A_{1N} + O_p(N^{-1/2}) \begin{bmatrix}
            1 & h_N \\ 
            h_N & h_N^2
        \end{bmatrix}.
        \]
        and \cref{lemma:boundY} shows that \[
         \hat A_{2N} =  \tilde A_{2N} + O_p(N^{-1/2}) 
        \colvecb{2}{1}{h_N}.
        \]
        
        The inverse is then \[
         \hat A_{1N}^{-1} = \tilde A_{1N}^{-1} + O_p(N^{-1/2}) 
         \begin{bmatrix}
            1 & 1/h_N \\ 
            1/h_N & 1/h_N^2
        \end{bmatrix}
        \]
        
        Multiplying the terms out, we have that \[
        \hat A_{1N}^{-1} \hat A_{2N} = \tilde A_{1N}^{-1} \tilde A_{2N} +
        \colvecb{2}{
            N^{-1/2}  
        }{ 
         N^{-1/2}/h_N 
        }
        \]
        Scaling by $\sqrt{Nh_N}$ yields the bound $\sqrt{h_N}$ in (3).
    \end{myproof}
    
    Lastly, we consider the fourth claim. To that end, we recall that \[
    \E[Y_i(c) \mid J_i(c) = 1, R_{it_s} = r] = \E[Y_i(s_1) \mid J_i
    (c) = 1, R_{it_s} = r] \equiv \mu_+(r).
    \]
    Let $\epsilon_i = Y_i(c) - \mu_+(R_{it_s})$. Now, observe that $\E
    [\epsilon_i \mid R_{it_s}, J_{i}(c) = 1] = 0$. 
     We first do a Taylor expansion
 of $\mu_+$.
 \cref{as:cts_diff} implies that
\[
    \mu_+(r) = \mu_+(\rho(c)) + \mu'_+(\rho(c)) (r- \rho) + \frac{1}
    {2}\mu''_+(\rho(c)) 
    (r- \rho )^2 =
    \nu(r; c, \rho),
    \]
    where $\abs{\nu(r; c,\rho)} < B_D(r-\rho(c))^3 + B_\mu(c) (|r-\rho(c)|
    +
    |\rho(c) - \rho|)|\rho(c)- \rho|$ for
    some
    constant $B_\mu(c)$.  
    \begin{myproof}[Proof of \cref{aprop:steps}(4)]
          In the notation of \cref{lemma:Sn_bounds,lemma:Tn_bounds}, we can write
    $\tilde A_{2N} = \frac{1}{Nh_N} \sum_{i=1}^N J_i(c) I_{1i}^+
        (C_N, h_N) x_i(C_N) Y_i(c)$ as \[
        \colvecb{2}{
            \mu_+(c) S_{0N} + \mu_+'(c) S_{1N} + \frac{\mu_+''(c)}{2} S_
            {2N}
        }{\mu_+(c) S_{1N} + \mu_+'(c) S_{2N} + \frac{\mu_+''(c)}{2} S_
            {3N}} + \bar\nu_N + \colvecb{2}{T_{0N}}{T_{1N}},
        \]
        where the argument $\rho = \rho(C_N)$ for $S_{kN}$. Let \[
    \tilde B_{2N} = 
    \sqrt{Nh_N} \pr{ \colvecb{2}{
        \frac{\mu_+''(c)}{2} S_
            {2N}
        }{
     \frac{\mu_+''(c)}{2} S_
            {3N}} + \bar\nu_N + \colvecb{2}{T_{0N}}{T_{1N}}}.
    \]
    Let $\check B_{2N}$ be similarly defined. 
    Note that \[
    \tilde A_{1N}^{-1} \tilde A_{2N} = \colvecb{2}{\mu_+(c)}{\mu_+'(c)} +
    \frac{1}{\sqrt{Nh_N}} \tilde A_{1N}^{-1} \tilde B_{2N}
    \]
    and similarly \[
    \check A_{1N}^{-1} \check A_{2N} = \colvecb{2}{\mu_+(c)}{\mu_+'(c)} +
    \frac{1}{\sqrt{Nh_N}} \check A_{1N}^{-1} \check B_{2N}
    \]
    Thus it remains to show that \[\tilde B_{2N} = \check B_{2N} + 
    \colvecb{2}{O_p(h_N^{3/2} +
                N^
                {-1/4 }h_N^{-1/2})}{O_p(h_N^{5/2} + N^
                {-1/4} h_N^{1/2})}.\]
    The above claim follows immediately from the bounds in 
    \cref{lemma:Sn_bounds,lemma:Tn_bounds,lemma:bound_nu}.
    \end{myproof}

    \subsubsection{Central limit theorem and variance estimation}
    
    Under a Taylor expansion (\cref{as:cts_diff}) of $\mu_+(r)$, we have that, so long as
    $h_N = o(N^{-1/5})$, \[
    \sqrt{Nh_N} (\tilde \beta_0 - \mu_+(c)) = \underbrace{\frac{1}{
    \sqrt{Nh_N}} \sum_
        {i=1}^N \frac{\nu_2 - \nu_1 \frac{R_{it_1} - \rho(c)}{h_N}}
        {(\nu_0\nu_2 - \nu_1^2) f(\rho(c)) \P(J_i(c) = 1)}\cdot
        W_i(c, h_N) \cdot (Y_i(c) - \mu_+(R_{it_1}))}_{Z_N} + o_P
        (1)
        \numberthis \label{eq:influence_fn_rd}
    \]
    via a standard argument. See, for instance, \citet{imbens2008regression,hahn2001identification,imbens2012optimal}.
    
    \begin{theorem}
    \label{thm:variance_est}
        When $h_N = o(N^{-1/5})$, under \cref{as:ctsdensity,as:second_moment,as:cts_diff}, we have the following
        central
        limit theorem: \[
        \hat\sigma_N^{-1} (\check \beta_0 - \mu_+(c)) \dto \Norm
        (0, 1)
        \]
        where the variance estimate is \[
        \hat \sigma_N^{2} = \frac{4Nh_N}{N_+} \pr{\frac{1}{N_+} \sum_
        {i=1}^N W_i(C_N, h_N) Y_i
        (C_N)^2 - \hat \beta_0^2} \quad N_+ = \sum_{i=1}^N W_i(C_N, h_N).
        \]
    \end{theorem}
    \begin{myproof}
        The central limit theorem follows from \cref{lemma:clt}, which
        shows normality of $Z_N$ under Lyapunov conditions, and 
        \cref{lemma:variance_est}, which
        shows consistency of $\hat \sigma_N^2$. 
    \end{myproof}
    
    \subsubsection{Guide to the lemmas}
    
    We conclude the main text of this appendix section with a guide to the
    lemmas that are appended in the rest of the section 
    (\cref{asub:cltandvarest,asub:xbounds,asub:event,asub:discrepancy_j,asub:disc_y}). 
    The key to the bounds is placing ourselves in an event that is
    well-behaved, in the sense that the ordering of the sample cutoffs
    $C_N$ agrees with its population counterpart. This is dealt with in
    \cref{asub:event}. Under such an event, all
    but $\sqrt{N}$ of students' qualification statuses in sample disagree
    with those in population, yielding bounds related to $J_i$ 
    (\cref{asub:discrepancy_j}) and $J_i Y_i(C_N)$ (\cref{asub:disc_y}).
    Having dealt with $J_i(C_N, h_N) \neq J_i(c)$, we can bound the
    discrepancy due to $\rho(C_N) \neq \rho(c)$, and those are in 
    \cref{cor:Sn_bounds,lemma:Tn_bounds} in \cref{asub:xbounds}. Lastly,
    \cref{asub:cltandvarest} contains lemmas that are useful for the CLT
    and variance estimation parts of the argument. 
    
    \subsection{Placing ourselves on well-behaved events} 
    \label{asub:event}
    
    \begin{lemma}
    \label{lemma:nice_event_as}
    Let $0\le M_N \to \infty$ diverge. Let $A_N =
    A_N(M_N, h_N)$ be the
    following event: Let \[\lbr{t_s}(s, Q_s, c_s) \equiv \inf \br{r \in [0,1]: \frac{Q_s + r}{\bar q_s + 1} \ge
c_s} \text{ where $\inf_{[0,1]} \emptyset = 1$}.
\]
be the test-score space cutoff corresponding to students of discrete 
level $Q_s$.
\begin{enumerate}
   
        \item ($c$ and $C_N$ agree on when $\lbr{t_s}(s, q_s, c_s) \in (0,1)$) For any school
        $s$ and
        any $q \in \{0,\ldots, \bar q_s\}$,
        $\lbr{t_s}(s, q, C_{N}) \in (0, 1) $ if and only if $\lbr{t_s}(s, q, c)
        \in (0,1)$. If $\lbr{t_s}(s, q, C_{N}) \not\in (0, 1)$, then $\lbr{t_s}(s,         q, C_{N}) = \lbr{t_s}(s, q, c)$. 
        
        \item The cutoffs converge for every $s,q$: 
        \[
        \max_{s} \max_{q} \, \abs{\lbr{t_s}(s, q, c_s) - \lbr{t_s}(s, q, C_{s,N})}
        \le M_N N^{-1/2}.
        \]

        \item ($c$ and $C_N$ agree on all the ordering of $r_{t,s}$) For
        any schools $s_1, s_2$ which use the same test $t$, $r_
        {t,s_1}(C_N)$ and $r_{t,s_2}(C_N)$ are exactly ordered as $r_
        {t,s_1}(c)$ and $r_{t,s_2}(c)$. 
        
        \item For all $\succ, Q$,  $\rho(C_N) + h_N <
        \lbR{t_1}(s_1;
        \succ_i, Q_i, C_N)$ if and only if $\rho(c) < \lbR{t_1}(s_1;
        \succ_i, Q_i, c)$.
        
        \item Suppose $s_0$ is a test-score school that uses $t_1$. For all $\succ, Q$,
        $\rho(C_N) - h_N >  \lbr{t_1}(s_0; Q, \succ, C_N)$ if and only if $\rho(c) >
        \lbr{t_1}(s_0; Q, \succ, c) $
        
    \end{enumerate}    
        Under \cref{as:limitcutoff,as:interior} and $h_N \to 0$, $A_N$
        occurs almost surely eventually:
        \[
        \lim_{N\to\infty} \P(A_N) = 1.
        \]
    \end{lemma}
    \begin{myproof}
        Since finite intersections of eventually almost sure events are eventually almost
        sure,
        it suffices to show that the following types events individually occur with
        probability tending to one: \begin{enumerate}

            \item For any fixed $s$ and any $q \in \{0,\ldots, \bar
            q_s\}$, $\lbr{t_s}(s, q, C_{s,N}) \in (0, 1) $ if and only if $\lbr{t_s}(s, q,  c_s)
        \in (0,1)$.   If $\lbr{t_s}(s, q, C_{s,N}) \not\in (0, 1)$, then $\lbr{t_s}(s, q, C_{s,N}) = \lbr{t_s}(s, q, c_s)$.

        \begin{itemize}

            \item If $s$ is undersubscribed in population $c_s = 0$, then (3) in 
            \cref{as:interior} implies that eventually $C_{s,N} = c_s = 0$. On that event,
            $\lbr{t_s}(s, q,c) = \lbr{t_s}(s, q, C_N)$. 
            \item Otherwise, suppose $c_s \in (0,1)$. By (2) in \cref{as:interior}, $c_s$
            does not equal any exact $q/(\bar q_s + 1)$. Note that by 
            \cref{as:limitcutoff}, for any fixed $\epsilon > 0$, 
            $\P(C_{s,N} \in [c_s -
            \epsilon, c_s+ \epsilon]) \to 1$. We can choose $\epsilon$ such that for some
            unique
            $q$,  \[
            \frac{q}{\bar q_s + 1} < \frac{q + c_s - \epsilon}{\bar q_s + 1} \le \frac{q
            + c_s + \epsilon}{\bar q_s + 1} < \frac{q
            + 1}{\bar q_s + 1}. 
            \]
            Thus, for that $q$, $C_{s,N} \in [c_s -
            \epsilon, c_s+ \epsilon]$ implies that both $g^{-1}(q, C_N)$ and $g^{-1}(q,
            c)$ are interior. For all $q' < q$, both are equal to 1 and for all $q' > q$,
            both are equal to zero.

        \end{itemize}
        \item For fixed $s, q$, $\abs{\lbr{t_s}(s, q, c_s) - \lbr{t_s}(s, q, C_{s,N})}
        \le M_N N^{-1/2}$. Let $q_s^\dagger$ be the $q_s$ such that $\lbr{t_s}(s, q_s^\dagger, c_s)
        \in (0,1)$. 
        \begin{itemize}
            \item If $q \neq q_s^\dagger$, with probability tending to 1 $
            \abs{\lbr{t_s}(s, q, c_s) - \lbr{t_s}(s, q, C_{s,N})} = 0$. 
            \item If $q = q_s^\dagger$, then since $\max_s \abs{c_s - C_{s,N}} =
            O_p(N^
            {-1/2})$ and $V_{is}(q, \cdot)$ is affine, the preimage  is also
            $O_p(N^{-1/2})$ (uniformly over $s$).
        \end{itemize}
        
        \item For fixed schools $s_1, s_2$ which use the same test $t$,
        $r_
        {t,s_1}(C_N)$ and $r_{t,s_2}(C_N)$ are exactly ordered as $r_
        {t,s_1}(c)$ and $r_{t,s_2}(c)$. 
        
        \begin{itemize}
            \item Suppose $r_
        {t,s_1}(c) > r_{t,s_2}(c)$. Note that for any $\epsilon > 0$,
        $\P\bk{r_{t,s_1}(C_N) > r_
        {t,s_1}(c) - \epsilon} \to 1$ by \cref{as:limitcutoff}. Similarly, 
        $\P\bk{r_{t,s_2}(C_N) < r_
        {t,s_2}(c) + \epsilon} \to 1$. Therefore, we may take $\epsilon =
        \frac{r_
        {t,s_1}(c)  - r_{t,s_2}(c)}{2}.$
        \item Suppose $r_
        {t,s_1}(c) = r_{t,s_2}(c)$. Then by (4) in \cref{as:interior}, both
        schools are undersubscribed. In that case, (3) in 
        \cref{as:interior} implies that $r_
        {t,s_1}(C_N) = r_{t,s_2}(C_N)$ eventually almost surely.
        \end{itemize}
        \item For a fixed $\succ, Q$ and $h_N\to0$,
         $\rho(C_N) + h_N <
        \lbR{t_1}(s_1;
        {\succ}, Q, C_N)$ if and only if $\rho(c) < \lbR{t_1}(s_1;
        {\succ}, Q, c)$.
        \begin{itemize}
            \item By \cref{as:interior}, $\rho(c) \in (0,1)$, and hence $\rho(c) \neq \lbR{t_1}(s_1;
        {\succ}, Q, c)$ for any $\succ, Q$. The event in (2) implies \[
        |\rho(C_N) - \rho(c)| < M_N N^{-1/2}
        \]
        and $\abs{\lbR{t_1}(s_1;
        {\succ}, Q, C_N) - \lbR{t_1}(s_1;
        {\succ}, Q, c)} < M_N N^{-1/2}$. Under this event, since $h_N \to
        0$, we have $\rho(C_N) + h_N < \lbR{t_1}(s_1;
        {\succ}, Q, C_N)$ for all sufficiently large $N$. Hence if (2)
        occurs almost surely eventually, then (4) must also.  \qedhere
        \end{itemize}
        \item Suppose $s_0$ is a test-score school that uses $t_1$. For all $\succ, Q$,
        $\rho(C_N) - h_N >  \lbr{t_1}(s_0; Q, \succ, C_N)$ if and only if $\rho(c) >
        \lbr{t_1}(s_0; Q, \succ, c) $.

        \begin{itemize}
            \item The proof of this claim is similar to the proof of the last claim (4). 
        \end{itemize}

        \end{enumerate}
    \end{myproof}
    
        \begin{rmk}
            We work with nonstochastic sequences of the bandwidth parameter $h_N$. If
            the bandwidth parameter is a stochastic $H_N$, then we can modify by
            appending to $A_N$ the event $H_N < M_N h_N$ for some nonstochastic sequence
            $h_N$. If $H_N = O_p (h_N)$, then $\P(H_N < M_N h_N) \to 1$; as a result,
            our subsequent conclusions are not affected. $\blacksquare$
    \end{rmk}
    
\subsection{Bounding discrepancy in $J_i$}
\label{asub:discrepancy_j}
    By studying the implications
of the event $A_N$---all score cutoffs are induced by
 $C_N$ agrees with that induced by $c$ and all almost-sure-qualification
 statuses also
 agree---we immediately have the following result, which, roughly speaking, implies that
 the event $J_i(C_N) \neq J_i(c)$ is a subset of an event where $R_i$
 belongs to a set of Lebesgue
 measure at most $M_N N^{-1/2}$. 
    \begin{lemma}
    \label{lemma:nice_event}
        On the event $A_N$, \begin{enumerate}
            \item For all $i$, $I_i(c) = I_i(C_N).$
            \item For all $i$, $I_{1i}(c, 0) = I_{1i}(C_N, h_N)$.
            \item For all $i$, $I_{10i}(c) = I_{10i}(C_N)$.
            \item For $t \neq t_0,t_1$, $I_{ti}(c) \neq I_{ti}(C_N)$
            implies that (a) $\lbR{t}(s_0; Q_i, \succ_i, C_N) \in (0,1)$, 
            (b) \[\abs{\lbR{t}(s_0; Q_i, \succ_i, C_N) - \lbR{t}(s_0;
            Q_i,
            \succ_i, c)} \le M_N N^{-1/2},\]
            and (c) $R_{it}$ is between  $\lbR{t}(s_0; Q_i, \succ_i, C_N) $
            and $\lbR{t}(s_0;
            Q_i,
            \succ_i, c)$. 
            
            \item If $s_0$ is a test-score school using $t_0$ and  $t_1 = t_0$, then $
            I_{0i}(C_N) = I_{0i}(c).
            $
            \item 
            If $s_0$ is a test-score school using $t_0$ and $t_1 \neq t_0$, $
            I_{0i}(c) \neq I_{0i}(C_N)
            $ implies that (a) \[\lbR{t_0}(s_0; Q_i, {\succ_i}, C_N) \in 
            (0,1),\] (b) \[
            \abs{\lbR{t_0}(s_0; Q_i, \succ_i, C_N) - \lbR{t_0}(s_0;
            Q_i,
            \succ_i, c)} \le M_N N^{-1/2},
            \]
            (c) \[
            \abs{\lbr{t_0}(s_0; Q_i, \succ_i, C_N) - \lbr{t_0}(s_0;
            Q_i,
            \succ_i, c)} \le M_N N^{-1/2},
            \]
            and (d) either  $R_{it_0}$ is between $\lbR{t_0}(s_0; Q_i,
            \succ_i, C_N)$ and
            $\lbR{t_0}(s_0; Q_i, \succ_i, c)$, or $R_{it_0}$ is between $
            \lbr{t_0} (s_0; Q_i,
            {\succ_i}, C_N)$ and $\lbr{t_0}(s_0; Q_i, \succ_i, c)$.
            \item Under \cref{as:bounded_density}, for all $i$, $\pi_i(c) =
            0$ if and only if
            $\pi_i(C_N) = 0$. Moreover, if $\pi_i(c) > 0$, then \[
            |\pi_i(C_N) - \pi_i(c)| \le L B M_N N^{-1/2}.
            \]
        \end{enumerate}
    \end{lemma}
    \begin{myproof}
        Every claim is immediate given the definition of $A_N$ in 
        \cref{lemma:nice_event_as}.
    \end{myproof}
    
    \begin{cor}
    \label{cor:J_fact}
        On the event $A_N$, the disagreement $J_i(C_N, h_N) \neq J_i(c, h_N)$ implies
        that $(R_{it}: t\neq t_1) \in
        K({\succ_i}, Q_i; c)$, where $\mu_{\R^{T-1}}(K(\succ_i, Q_i; c)) \le
        TM_N N^
        {-1/2}$. Moreover, under \cref{as:interior}, for all sufficiently
        small $h_N, J_i(c, h_N) =
        J_i(c)$ does not depend on $h_N$. 
    \end{cor}
    
    \begin{myproof}
        $J_i(C_N, h_N) \neq J_i(c, h_N)$ implies that at least one of $I_i, I_{ti}, I_{0i}$ have a
        disagreement over $c$ and $C_N$. On $A_N$, disagreements of $I_{ti}$ imply that
        $R_{ti}$ is contained in a region of measure at most $M_N N^{-1/2}$. If $t_1 =
        t_0$, then the union of disagreements over $I_{ti}$ is a region of size at most
        $ (T-1)M_N N^{-1/2}$, and $I_i, I_{0i}$ has no disagreement. If $t_1 \neq t_0$,
        disagreements of $I_ {0i}$ imply that $R_{ it_0}$ is contained in a region of
        measure at most $2M_N N^{-1/2}$, and hence the union of disagreements is a
        region of size at most $T M_N N^{-1/2}$. Neither case implies
        anything about $R_{it_1}$. 
        
           The only part of $J_i$ that depends on $h_N$ is $I_1(c, h_N)$,
        which does not depend on $h_N$ when $h_N$ is sufficiently small due
        to \cref{as:interior}.
    \end{myproof}
    
    \begin{cor}
    \label{cor:inv_prop_bounded}
    Suppose $M_N = o(N^{1/2})$. On the event $A_N$, there exists some $\eta > 0$,
    independently of $M_N$, such that for all sufficiently large $N$, if $\pi_i(C_N) > 0
    $ then $\pi_i(C_N) \ge \eta$. Equivalently, $1/\pi_i(C_N) < 1/\eta$
    whenever defined.
    \end{cor}
    
    \begin{myproof}
        Let $\eta = \min \br{v = \pi_i(c) : v > 0} / 2 > 0$. Under $A_N$,
        $\pi_i(c) = 0$ if and only if $\pi_i(C_N) = 0$ and $\pi_i(C_N)$ is
        uniformly
        $o(1)$ away from $\pi_i(c)$. Hence for sufficiently large $N$, the
        discrepancy between $\pi_i(c)$ and $\pi_i(C_N)$ is bounded above by
        $\eta$, thereby $\pi_i(C_N) > \eta$ as long as $\pi_i(C_N) > 0$. 
    \end{myproof}

    \begin{lemma}
    \label{lemma:boundJ}
    Let $\Gamma_i \ge 0$ be some random variable at the student level where    \[
    \E[\Gamma_i  \mid R_i, {\succ_i}, Q_i, Z_i] < B_M < \infty
    \] almost surely.
    Under \cref{as:limitcutoff,as:interior,as:bounded_density}, assuming
    $N^{-1/2} = o(h_N)$,
    the discrepancy of the sample selection is of the following stochastic
    order:
        \[F_N \equiv \frac{1}{\sqrt{N} h_N} \sum_{i=1}^N \abs{J_i(C_N,
        h_N) - J_i
        (c, h_N)}
        I_{1i}^{+}(C_N, h_N) \Gamma_i = O_p(1).\]
    \end{lemma}
    \begin{myproof}
        On the event $A_N$, 
        $\abs{\rho(c) - \rho(C_N)} \le M_N N^{-1/2}$. Then, on $A_N$, 
        \begin{align*}
                \sqrt{N} h_N F_N & = \sum_{i=1}^N \abs{J_i(C_N) - J_i
        (c)}
        I_{1i}^{+}(C_N, h_N) \Gamma_i \\ &\le \sum_{i=1}^N \one\bk{(R_{it}
        : t\neq
        t_1)
        \in
        K(\succ_i, Q_i; c)} \one(R_{it_1} \in [\rho(c) - M_N N^
        {-1/2}, \rho(c) + M_N N^
        {-1/2} + h_N])  \Gamma_i
        \\&\equiv \sqrt{N} h_N G_N(M_N). 
        \end{align*}
        Hence, under \cref{lemma:nice_event_as}, for any sequence $M_N \to
        \infty $,  since the corresponding $\P(A_N) \to 1$, \[
        F_N = F_N A_N + o_p(1) \le G_N(M_N) A_N + o_p(1) \le G_N(M_N) +
        o_p(1).
        \]
        Since $G_N \ge 0$ almost surely, by Markov's inequality, 
        \cref{as:bounded_density}, and $N^{-1/2} = o(h_N)$, 
        \[
        G_N = O_p(\E[G_N]) = \frac{1}{\sqrt{N} h_N} \cdot O_p\pr{N \cdot (T
        M_N N^{-1/2}) \cdot (2M_N N^
        {-1/2}
        + h_N) \cdot B} \le M_N^2 O_p(1).
        \]
        Note that \begin{align*}
        \E[G_N] &= \frac{1}{\sqrt{N}h_N} N \E\bk{R_i \in \tilde K(\succ_i,
        Q_i; c)} \E[\Gamma_i \mid R_i \in \tilde K(\succ_i,
        Q_i; c)] \\ 
        &\le  \frac{\sqrt{N}}{h_N} B \cdot (T M_N N^{-1/2}) \cdot (2 M_N
        N^{-1/2} + h_N) \cdot  B_M \\ 
        &= O(M_N^2)
        \end{align*}
        Therefore, for any $M_N \to \infty$, no matter how slowly,  $
        F_N = O_p(M_N^2)$. This implies that $F_N = O_p(1)$.\qedhere
    \end{myproof}
    
    \subsection{Bounding discrepancy in terms involving $Y_i(C_N)$}
    \label{asub:disc_y}
    \begin{lemma}
    \label{lemma:boundY}
    Fix $M_N \to \infty$. Suppose that $\E[\abs{Y_i(s_1)} \mid R_i,
    \succ_i, Q_i, Z_i] < B_M < \infty$ almost surely. Then the difference
    \begin{align*}
         \abs[\bigg]{\sum_{i=1}^N J_i(C_N, h_N) I_1^+ (C_N, h_N) Y_i(C_N) -
        \sum_
        {i=1}^N J_i(c, h_N) I_1^+ (C_N, h_N) Y_i(c)} \le \Delta_{1N} + \Delta_{2N} + \Delta_{3N}
        \end{align*}
        where \begin{align*}
        \Delta_{1N} &= \sum_i I_i^{+}(C_N, h_N) J_i(c, h_N) \abs[\bigg]{
        \frac{\di_i(C_N)}{\pi_i(C_N)} - \frac{\di_i(c)}{\pi_i(c)}}  |Y_i(s_1)| \\ 
        \Delta_{2N} &= \sum_i I_i^{+}(C_N, h_N) \frac{\di_i(c)}{\pi_i(c)} 
        \abs{J_i(C_N, h_N) - J_i(c, h_N)} |Y_i(s_1)|  \\
        \Delta_{3N} &= \sum_i I_i^{+}(C_N, h_N) \abs{J_i(C_N, h_N) -
        J_i(c, h_N)}\abs[\bigg]{
        \frac{\di_i(C_N)}{\pi_i(C_N)} - \frac{\di_i(c)}{\pi_i(c)}} |Y_i(s_1)|.
        \end{align*}

        Moreover, under 
        \cref{as:limitcutoff,as:interior,as:bounded_density}, for $j=1,2,3$, $\frac{\Delta_
        {jN}}
        {\sqrt{N}h_N} = O_p
        (1)$. As a result, \[
         \abs[\bigg]{\sum_{i=1}^N J_i(C_N, h_N) I_1^+ (C_N, h_N) |Y_i(C_N)| -
        \sum_
        {i=1}^N J_i(c, h_N) I_1^+ (C_N, h_N) |Y_i(c)|} = O_p\pr{ N^
        {1/2} h_N}.
        \]
    \end{lemma}
    \begin{myproof}
    The part before ``moreover'' follows from adding and subtracting and
    triangle inequality. 
    
    To prove the claim after ``moreover,'' first, note that by 
        \cref{cor:inv_prop_bounded}, for all sufficiently large $N$, the
        inverse propensity weight $1/\pi_i < 1/\eta$. Immediately, then,
        $\Delta_{2N}, \Delta_{3N}$ are bounded above by \[
        \sum_i I_i^{+}(C_N, h_N) \abs{J_i(C_N, h_N) - J_i(c, h_N)} \cdot
        |Y_i(s_1)| = O_p(\sqrt{N}h_N)
        \]
        via \cref{lemma:boundJ}. 
    
        By the same argument where we bound $1/\pi_i$, \[
        \Delta_{1N} = \sum_{i=1}^N I_i^+(C_N, h_N) J_i(c, h_N) \abs[\bigg]
        {\frac{1}{\pi_i(C_N)} - \frac{1}{\pi_i(c)}} \di_i(c) |Y_i(s_1)| + O_p
        (
        \sqrt{N} h_N).
        \]        
        By \cref{lemma:nice_event}, \[
        \abs[\bigg]
        {\frac{1}{\pi_i(C_N)} - \frac{1}{\pi_i(c)}} < M_N N^{-1/2}.
        \]
        
        On $A_N$, since \[
        \sum_i I_i^+(C_N, h_N) J_i(c, h_N) \di_i(c) |Y_i(s_1) |\le  \sum_i
        \one(R_i \in \tilde K(\succ_i, Q_i; c)) |Y_i(s_1)|
        \]
        where $\sup_{\succ_i, Q_i} \mu\pr{K(\succ_i, Q_i; c)} = O(h_N)$. We
        have again by Markov's inequality and the bound on the conditional
        first moment of $Y_i(s_1)$, \[
        \sum_i I_i^+(C_N, h_N) J_i(c, h_N) \di_i(c) |Y_i(s_1)| = O_p(Nh_N). 
        \]
        Hence $\Delta_{1N} = O_p(\sqrt{N}h_N)$.
    \end{myproof}
   
   \subsection{Bounding terms involving $x_i$ and $I_{1i}^+$}
   \label{asub:xbounds} 
   \begin{lemma}
   \label{lemma:Sn_bounds}
       Suppose $N^{-1/2} = o(h_N)$.
   Consider \[S_{k,N}(\rho) \equiv \frac{1}{Nh_N} \sum_{i=1}^N J_i(c)
   (R_{it_1} - \rho)^k \one\pr{R_{it_1} \in [\rho, \rho + h_N]}.\] Then,
   under \cref{as:interior,as:bounded_density,as:limitcutoff}, \[
   \abs{S_{k,N}(\rho(C_N)) - S_{k,N}(\rho(c))} = O_p\pr{
   h_N^{k-1} N^{-1/2}
   } = o_p(h_N^k).
   \]
   \end{lemma}
   
   \begin{myproof}
          Suppose $N$ is sufficiently large such that $M_N N^{-1/2} < h_N$. 
    On the event
   $A_N$, \[
   \abs{S_{k,N}(\rho(C_N)) - S_{k,N}(\rho(c))} \le \sup \br{\abs{S_{k,N}
   (\rho) -
      S_{k,N}
      (\rho(c))} : \rho \in [\rho(c)-M_N N^{-1/2},
   \rho(c)+M_N N^{-1/2}]}.
   \] 
   For a fixed $\rho \in [\rho(c)-M_N N^{-1/2},
   \rho(c)+M_N N^{-1/2}]$, the difference \begin{align*}
   \abs{S_{k,N}(\rho(C_N)) - S_{k,N}(\rho(c))} \le \frac{1}{Nh_N} 
   {\sum_{i=1}^NJ_i(c) \one(R_{it_1} \in [\rho(c), \rho
   (c) + h_N]) \Delta_{1ik} } \\ +  \frac{1}{Nh_N} 
   {\sum_{i=1}^NJ_i(c) \one(R_{it_1} \in \Delta_{2})
   (R_{it_1}- \rho)^k } \\ 
   + \frac{1}{Nh_N} 
   {\sum_{i=1}^NJ_i(c) \one(R_{it_1} \in \Delta_{2})
   \Delta_{1ik} }
   \end{align*}
   where $\Delta_{1ik} = \abs{(R_{it_1} - \rho)^k - (R_{it_1} - \rho(c))^k}$
   and
   $\Delta_{2} = [\rho, \rho(c)] \cup [\rho + h_N, \rho(c) + h_N]$ if
   $\rho < \rho(c)$ and $[\rho(c), \rho] \cup [\rho(c) + h_N, \rho +
   h_N]$ otherwise. 
   
   Note that $\Delta_{1ik} = 0$ if $k=0$. If $k > 0$ then \[
   \Delta_{ik} < |\rho - \rho(c)| k (2M_N N^{-1/2} + h_N)^{k-1} < B_k
   M_N N^{-1/2} h_N^{k-1}
   \]  
   for some constants $B_k$, by the difference of two $k$\th{} powers
   formula. Let $B_0 = 0$, then the first term is bounded by \[
   B_k M_N N^{-1/2} h_N^{k-1} \cdot \frac{1}{Nh_N} 
   \sum_{i=1}^NJ_i(c) \one(R_{it_1} \in [\rho(c), \rho
   (c) + h_N]).
   \]
   The second term is bounded by \[
   B_k' h_N^k \frac{1}{Nh_N} 
   {\sum_{i=1}^NJ_i(c) \one(R_{it_1} \in \Delta_{2})}
   \]
   for some constants $B_k'$ where $B_0' = 1$. 
   The third term is bounded by \[
    B_k M_N N^{-1/2} h_N^{k-1} \frac{1}{Nh_N} 
   {\sum_{i=1}^NJ_i(c) \one(R_{it_1} \in \Delta_{2})}
   \]
   These bounds hold regardless of $\rho$, and hence taking the supremum
   over $\rho$ yields that, for any  $M_N \to \infty$, \begin{align*}
      \abs{S_{k,N}(\rho(C_N)) - S_{k,N}(\rho(c))} &= O_p\pr{
    B_k M_N N^{-1/2} h_N^{k-1} + h_N^{k-1} M_N N^{-1/2} + B_k M_N N^{-1}
    h_N^{k-2}
   } \\ &= O_p(M_N h_N^{k-1} N^{-1/2}).
   \end{align*}
   Hence $\abs{S_{k,N}(\rho(C_N)) - S_{k,N}(\rho(c))} = O_p\pr{h_N^{k-1}
   N^{-1/2}}$.
   \end{myproof}
   
   \begin{cor}
       \label{cor:Sn_bounds}
       The conclusion of \cref{lemma:Sn_bounds} continues to hold if each
       term of $S_{k,N}(\rho)$ is multiplied with some
       independent $\Gamma_i$ where
       $\E[|\Gamma_i| \mid R_i, \succ_i, Q_i, Z_i] < B_M < \infty$ almost
       surely.
   \end{cor}
   
   \begin{myproof}
       The bounds continue to hold where the right-hand side involves terms
       like \[
       \frac{1}{Nh_N} 
   {\sum_{i=1}^NJ_i(c) \one(R_{it_1} \in \Delta_{2})} |\Gamma_i|.
       \]
       The last step of the proof to \cref{lemma:Sn_bounds} uses Markov's
       inequality, which incurs a constant of $B_M$ since terms like \[
       \E[|\Gamma_i| \mid J_i(c) \one(R_{it_1} \in \Delta_{2}) = 1] \le
       B_M.
       \]
   \end{myproof}

   \begin{lemma}
   \label{lemma:Tn_bounds}
       Suppose $N^{-1/2} = o(h_N)$. Suppose $\epsilon_i$ are independent
       over $i$ with $\E [\epsilon_i \mid J_i(c), R_{it_1}] = 0$ and
       $\var[\epsilon_i \mid J_i(c), R_{it_1}] < B_V < \infty$ almost
       surely.
   Consider \[T_{k,N}(\rho) \equiv \frac{1}{Nh_N} \sum_{i=1}^N J_i(c,
   h_N)
   (R_{it_1} - \rho)^k \one\pr{R_{it_1} \in [\rho, \rho + h_N]}
   \epsilon_i.\] Then,
   under \cref{as:interior,as:bounded_density,as:limitcutoff}, for $k =
   0,1$, \[
   \abs{T_{k,N}(\rho(C_N)) - T_{k,N}(\rho(c))} = O_p\pr{
   N^{-1/4} \cdot N^{-1/2} \cdot h_N^{k-1}
   }.
   \]
   \end{lemma}
   \begin{myproof}
        Suppose $N$ is sufficiently large such that $M_N N^{-1/2} < h_N$. 
    On the event
   $A_N$, \[
   \abs{T_{k,N}(\rho(C_N)) - T_{k,N}(\rho(c))} \le \sup \br{\abs{T_{k,N}
   (\rho) -
      T_{k,N}
      (\rho(c))} : \rho \in [\rho(c)-M_N N^{-1/2},
   \rho(c)+M_N N^{-1/2}]}.
   \] 
   For a fixed $\rho \in [\rho(c)-M_N N^{-1/2},
   \rho(c)+M_N N^{-1/2}]$, the difference \begin{align*}
   \abs{T_{k,N}(\rho(C_N)) - T_{k,N}(\rho(c))} \le \frac{1}{Nh_N} 
   \abs[\bigg]{\sum_{i=1}^NJ_i(c) \one(R_{it_1} \in [\rho(c), \rho
   (c) + h_N]) \Delta_{1ik}  \epsilon_i}\\ 
   +  \frac{1}{Nh_N} 
   \abs[\bigg]{\sum_{i=1}^NJ_i(c) \one(R_{it_1} \in \Delta_{2})
   (R_{it_1}- \rho)^k  \epsilon_i} \\ 
   + \frac{1}{Nh_N} 
   \abs[\bigg]{\sum_{i=1}^NJ_i(c) \one(R_{it_1} \in \Delta_{2})
   \Delta_{1ik}   \epsilon_i}
   \end{align*}
    where $\Delta_{1ik} = \abs{(R_{it_1} - \rho)^k - (R_{it_1} - \rho(c))^k}$
   and
   $\Delta_{2} = [\rho, \rho(c)] \cup [\rho + h_N, \rho(c) + h_N]$ if
   $\rho < \rho(c)$ and $[\rho(c), \rho] \cup [\rho(c) + h_N, \rho +
   h_N]$ otherwise. 
   Note that $\Delta_{1ik} = 0$ if $k = 0$ and $\Delta_{1ik} = |\rho -
   \rho(c)| < M_N N^{-1/2}$ if $k = 1$.

   We first show that \[
   \abs[\bigg]{\sum_{i=1}^N \one(R_{it_1} \in \Delta_{2}) \eta_i} = O_p
   (N^{1/4}) \quad \eta_i \equiv J_i(c, h_N) \epsilon_i
   \] Note that the event \begin{align*}
   \sum_{i=1}^N \one(R_{it_1} \text{ is between $\rho(c)$ and $\rho$, for
   some $\rho\in [\rho(c)-M_N N^{-1/2},
   \rho(c)+M_N N^{-1/2}]$}) \\ <
   2N \cdot M_N \cdot B M_N N^{-1/2}
   = 2BM_N^2 N^{1/2} \equiv K_N
   \end{align*}
   occurs with probability tending to $1$, and so does the event \[
   \sum_{i=1}^N \one(R_{it_1} \text{ is between $\rho(c) + h_N$ and $\rho + h_N$, for
   some $\rho\in [\rho(c)-M_N N^{-1/2},
   \rho(c)+M_N N^{-1/2}]$}) <
   K_N.
   \]
   On both events, the sum\[
   \abs[\bigg]{\sum_{i=1}^N \one(R_{it_1} \in \Delta_{2}) \eta_i} \le \sup_{U_1
   <
   K_N} \abs[\bigg]{
   \sum_{1 \le u_1 \le U_1
   } \eta_1(u_1)} + \sup_{U_2 < K_N} \abs[\bigg]{\sum_{1\le u_2 \le U_2 }
   \eta_2(u_2)}
   \]
   where we label the observation such that $\eta_1(u)$ is the $u$\th{}
   $\eta_i$ with $R_{it_1}$ closest to $\rho(c)$
    and $\eta_2(u)$ is the $u$\th{} $\eta_i$ with $R_
   {it_1}$ closest to $\rho(c) + h_N$.
   Observe that $Z_{1U} \equiv \sum_{1 \le u \le U} \eta_1(u)$ is a
   martingale adapted to the filtration $\mathcal F_U = \sigma \br{(R_
   {it_1})_{i=1}^N,
   \eta_1(u) : u \le U}.$
   By Kolmogorov's maximal inequality, \[
   \P\pr{
    \sup_{U \le K_N} |Z_{1U}| \ge t
   } \le \frac{\E[Z_{1K_N}^2]}{t^2} \le \frac{K_N B_V}{t^2}.\] Similarly,
   we obtain the same bound
   for the terms involving $\epsilon_2(u)$. Hence \[
   \sup_{U_1
   <
   K_N} \abs[\bigg]{
   \sum_{1 \le u_1 \le U_1
   } \eta_1(u_1)} + \sup_{U_2 < K_N} \abs[\bigg]{\sum_{1\le u_2 \le U_2 }
   \eta_2(u_2)} = O_p(\sqrt{K_N}) = M_N O_p(N^{1/4}).
   \]
   Therefore, since for any arbitrarily slowly diverging $M_N$, the three
   events that we place ourselves on occurs
   with probability tending to 1, $
   \abs[\big]{\sum_{i=1}^N \one(R_{it_1} \in \Delta_{2}) \eta_i} = O_p
   (N^{1/4}).
   $
   
   Now, we bound the three terms on the RHS. The second term is
   bounded above by \[
   \frac{h_N^{k-1}}{N} \abs[\bigg]{\sum_{i=1}^N J_i(c, h_N) \one(R_{it_1}
   \in \Delta_2)
   \eta_i }  = O_p(h_N^{k-1} N^{-3/4}).
   \]
   The third term is also $O_p(h_N^{k-1} N^{-3/4})$ since $\Delta_{1ik} <
   M_N N^{-1/2} = O(1)$ uniformly over $i$. The first term is zero if
   $k=0$. If $k=1$, the first term is bounded above by \[
   M_N N^{-1/2-1}h_N^{-1} \sum_{i} \one(R_{it_1} \in [\rho(c), \rho(c) +
   h_N])\eta_i.
   \]
   Chebyshev's inequality suggests that \[
   \sum_{i} \one(R_{it_1} \in [\rho(c), \rho(c) +
   h_N])\eta_i = O_p\pr{\sqrt{N} \sqrt{\var(\one(R_{it_1} \in [\rho(c),
   \rho(c) +
   h_N])\eta_i)}} = O_p(\sqrt{Nh_N}),
   \]
   thus bounding the first term with $O_p\pr{N^{-1}h_N^{-1/2}} = o_p
   (h_N^{k-1} N^{-3/4})$. Hence, since the above bounds are uniform over
   $\rho \in [\rho(c)-M_N N^{-1/2},
   \rho(c)+M_N N^{-1/2}]$, the bound is $O_p(N^{-3/4} h_N^{k-1})$ on
   the difference $\abs{T_{k,N}(\rho(C_N)) - T_{k,N}(\rho(c))}$. 
   \end{myproof}
   
   \begin{lemma}
       Suppose $\nu(r; c,\rho)$ is such that \[\abs{\nu(r; c,\rho)} < B_D
       (r-\rho(c))^3 + B_\mu(c) (|r-\rho(c)|
    +
    |\rho(c) - \rho|)|\rho(c)- \rho|.\] 
    Then the difference \begin{align*}
     &\bar \nu_N(C_N) - \bar \nu_N(c) \\ &\equiv  \frac{1}{Nh_N}\sum_
     {i=1}^N
     J_i(c) I_{1i}^+
    (C_N, h_N) x_i(C_N) \nu(R_{it_1}; c, \rho(C_N)) - \frac{1}{Nh_N}\sum_{i=1}^N J_i(c) I_{1i}^+
    (c, h_N) x_i(c) \nu(R_{it_1}; c, \rho(c)) \\ 
    &= o_p
       (h_N
       N^{-1/2}),
    \end{align*}
    assuming $N^{-1/2} = o(h_N)$. 
    \label{lemma:bound_nu}
   \end{lemma}
   \begin{myproof}
       On $A_N,$ when $I_{1i}^+ = 1$, the $\nu$ terms are uniformly bounded
       by \[
       B_D (h_N + 2M_N N^{-1/2})^3 + 10 B_\mu(c) (h_N + M_N N^{-1/2}) M_N
       N^{-1/2} = O(h_NN^{-1/2} + h_N^3).
       \]
       Thus, by \cref{lemma:Sn_bounds}, the difference is bounded by \[
       O_p(h_NN^{-1/2} + h_N^3)\colvecb{2}{N^{-1/2}/h_N}{N^{-1/2}} = o_p
       (h_N
       N^{-1/2}).
       \]
       
   \end{myproof}
   
       \begin{lemma}[A modified version of Lemma A.2 in \citet{imbens2012optimal}]
       \label{lemma:ik}
        Consider $S_{k,N} = S_{k, N}(c)$ in \cref{lemma:Sn_bounds}. Then,
        under \cref{as:ctsdensity}, \[
        S_{k,N} = \P(J_i(c) = 1) \cdot f(\rho(c)) h_N^{k} \int_0^{1/2} t^j
        \,dt + o_p(h_N^k),
        \]
        and, as a result, 
        \[
        \begin{bmatrix}
            S_{0, N} & S_{1,N} \\ 
            S_{1,N} & S_{2,N}
        \end{bmatrix}^{-1} = \begin{bmatrix}
            a_2  & -a_1/h_N \\ 
            -a_1/h_N & a_2/ h_N^2
        \end{bmatrix} + \begin{bmatrix}
            o_p(1) & o_p(1/h_N) \\ 
            o_p(1/h_N) & o_p(1/h_N^2)
        \end{bmatrix}
        \]
        where the constants are\footnote{The constants $\nu_k$ depends on the
kernel choice, which we fix to be the uniform kernel $K(x) = \one(x <
1/2)$.} \[
a_k = \frac{\nu_k}{\P(J_i(c) = 1)f(\rho(c))(\nu_0 \nu_2 - \nu_1^2)} = 
\frac{12/(k+1)}{\P(J_i(c) = 1)f(\rho(c))}
\quad \nu_k = \int_0^
{1}
t^k \,dt = \frac{1}{k+1}
\]
and $ f(\rho(c)) = p(R_{it_1} = \rho(c) \mid J_i(c) = 1)$ is the conditional
density of
the running variable at the cutoff point. 
    \end{lemma}
   \begin{myproof}
       The presence of $J_i(c)$ adds $\P(J_i(c) = 1)$ to the final result,
       via conditioning on $J_i(c) = 1$. The rest of the result
       follows directly from Lemma A.2 in \citet{imbens2012optimal} when
       working with the joint distribution conditioned on $J_i(c) = 1$.
   \end{myproof}

\subsection{Central limit theorem and variance estimation}
\label{asub:cltandvarest}

\begin{lemma}
\label{lemma:avar}
    Let \[Z_N = 
    \frac{1}{\sqrt{Nh_N}} \sum_{i=1}^N W_i(c, h_N) \frac{4-6\frac{R_{it_1} -
    \rho(c)}{h_N}}{\P(J_i(c) = 1)f(\rho(c))} \epsilon_i
    \]
    Then, under \cref{as:ctsdensity,as:second_moment}, \[\var(Z_N) \to \frac{4}{\P(J_i(c) = 1)f(\rho(c))}
    \sigma_+^2\]
    as $N\to \infty$.
\end{lemma}
\begin{myproof}
    It suffices to compute the limit \begin{align*}
    &\E\bk{\frac{\one(R_{it_1} \in [\rho(c), \rho(c) + h_N])}{h_N} \pr{2-3
    \frac{R_{it_1} -
    \rho(c)}{h_N}}^2 \epsilon_i^2 \mid J_i(c) = 1} \\ 
    &= \E\bk{\frac{\one(R_{it_1} \in [\rho(c), \rho(c) + h_N])}{h_N} \pr{2-3
    \frac{R_{it_1} -
    \rho(c)}{h_N}}^2 \E[\epsilon_i^2 \mid J_i(c) = 1, R_{it_1}] \mid
    J_i(c) = 1} \\ 
    &= \frac{1}{h_N}\int_{\rho(c)}^{\rho(c) + h_N} \pr{2-3\frac{r-\rho(c)}
    {h_N}}^2\sigma_+^2(r) f(r) \,dr \tag{Denote the conditional variance
    with $\sigma_+^2$} \\
    &= \int_0^1 (2-3v)^2 \sigma_+^2(\rho(c) + h_Nv) f(\rho(c) + h_Nv) \,dv
    \\ 
    &\to \sigma_+^2 f(\rho(c)) \cdot \int_0^1 (2-3v)^2\,dv \tag{Dominated
    convergence and continuity} \\ 
    &= \sigma_+^2 f(\rho(c)).
    \end{align*} 
    Thus, the limiting variance is \[
    \frac{1}{Nh_N} \cdot N \cdot \P(J_i(c) = 1) \cdot \frac{4h_N}{\P(J_i(c)
    = 1)^2
    f(\rho(c))^2}{} (\sigma_+^2 f(\rho(c)) + o(1)) \to \frac{4}{\P(J_i(c)
    = 1) f(\rho(c))}.
    \]
\end{myproof}

\begin{lemma}[Lyapunov]
    \label{lemma:clt}
    Let \[Z_N = 
    \frac{1}{\sqrt{Nh_N}} \sum_{i=1}^N W_i(c, h_N) \frac{4-6\frac{R_{it_1} -
    \rho(c)}{h_N}}{\P(J_i(c) = 1)f(\rho(c))} \epsilon_i \equiv \sum_{i=1}^N
    Z_{N,i}.
    \]
    Then, under \cref{as:second_moment,as:ctsdensity} $N\E|Z_{N,i}|^{2+\varepsilon} \to 0$ where
    $\varepsilon$ is given
    in \cref{as:second_moment}. Hence \[
    Z_N \dto \Norm\pr{0, \frac{4}{\P(J_i(c) = 1)f(\rho(c))}
    \sigma_+^2}.
    \]
\end{lemma}
\begin{myproof}
    The part after ``hence'' follows directly from the Lyapunov CLT for
    triangular arrays. 
    
    Now, \[
    \E|Z_{N,i}|^{2+\varepsilon} = \frac{\P
    (J_i(c) = 1)}{Nh_N (Nh_N)^{\varepsilon/2}}\cdot \E\bk{\one(R_{it_1}
    \in
    [\rho(c), \rho(c) + h_N]) \cdot \pr{2-3
    \frac{R_{it_1} -
    \rho(c)}{h_N}}^{2+\varepsilon} \epsilon_i^{2+\varepsilon} \mid J_i(c) =
    1}
    \]
    Since the $2+\varepsilon$ moment of $\epsilon_i$ is uniformly bounded,
    and $R_{it_1} -
    \rho(c) < h_N$ whenever $\one(R_{it_1}
    \in
    [\rho(c), \rho(c) + h_N]) = 1$, the above is bounded above by \[
    B_{CLT} \frac{1}{N(Nh_N)^{\varepsilon/2}} = o(1/N)
    \]
    for some constant $B_{CLT}$.
\end{myproof}

\begin{lemma}[WLLN for triangular arrays, \citet{durrett2019probability}
Theorem 2.2.11]
\label{lemma:wllndurrett}
    For each $n$ let $X_{n,k}$ be independent for $1 \le k \le n$. Let
    $b_n > 0$ with $b_n \to\infty$. Let $\bar X_{n,k} = X_{n,k}\one
    \pr{|X_{n,k} \le b_n|}$. Suppose that as $n\to \infty$, 
    \begin{enumerate}
        \item $\sum_k \P\br{|X_{n,k}| > b_n} \to 0$ 
        \item $b_n^{-2} \sum_{k=1}^n \E[\bar X_{n,k}^2] \to 0$. 
    \end{enumerate}
    Let $S_n = \sum_{k} X_{n,k}$ and let $\mu_n = \E[\bar X_{n,k}]$,
        then \[
        \frac{1}{b_n}(S_n - \mu_n) \pto 0.
        \]
\end{lemma}

\begin{lemma}[Variance estimation]
    \label{lemma:variance_est}
    Let $N_+$ be the number of observations with $J_i(C_N) = 1$ and $R_
    {it_1}
    \in
    [\rho(C_N), \rho(C_N) + h_N])$. Then, under 
    \cref{as:ctsdensity,as:second_moment}, and that $\hat \beta_0 = \mu_+
    (\rho(c)) + o_p(1),$ \[
       \frac{Nh_N}{N_+} \pr{\frac{1}{N_+} \sum_{i=1}^N W_i(C_N, h_N) Y_i
       (C_N)^2 - \hat
                 \beta_0^2} \pto \frac{\sigma_+^2}{\P(J_i(c) = 1) f(\rho
                 (c))}.
    \]
    
\end{lemma}
\begin{myproof}
    Note that 
    \[
    \frac{1}{Nh_N} N_+ = \frac{1}{Nh_N}\sum_i W_i(C_N, h_N) = \frac{1}
    {Nh_N} \sum_i W_i(c, h_N) + o_p(1) = \P(J_i(c)=1)f(\rho
    (c)) + o_p(1) \tag{\cref{lemma:boundJ,lemma:Sn_bounds}}
    \]    
    
    By \cref{lemma:boundY,cor:Sn_bounds}, we have that \begin{align*}
        \frac{1}{Nh_N}\sum_{i=1}^N W_i(C_N, h_N) Y_i^2(C_N) &= \frac{1}
    {Nh_N}\sum_{i=1}^N W_i(c, h_N) Y_i^2(c) + o_p(1) \\ &= \P(J_i(c) = 1)\E
    \bk{
    \frac{\one(R_{it_1}
    \in
    [\rho(c), \rho(c) + h_N])}{h_N} Y_i(c)^2 \mid J_i(c) = 1
    } + o_p(1) \\ 
    & \to \P(J_i(c) = 1) \E[Y_i(c)^2 \mid J_i(c) = 1, R_{it_1} = \rho(c)] f
    (\rho(c)).
    \end{align*}
    The second equality follows from \cref{lemma:wllndurrett}, which
    requires some justification.
    Barring that, the
    claim follows via Slutsky's theorem, noting that $\hat \beta_0 =
    \mu_+(\rho(c)) + o_p(1)$.
    
    To show the second equality above, let $X_{k,N} = W_k(c, h_N) Y_i^2(c)$
    and let $b_N = Nh_N$. Note that by
    Markov's inequality and \cref{as:second_moment}, \[
    \P\pr{
    X_{k,N} > b_N
    } = \P\pr{W_k(c, h_N) Y_i(c)^{2+\varepsilon} > b_N^{1+\varepsilon/2}}
    \lesssim
    \frac{\E[W_k(c, h_N)]}{b_N^{1+\varepsilon/2}} \lesssim \frac{h_N}{b_N^
    {1+\varepsilon/2}}.
    \]
    Thus the first condition of \cref{lemma:wllndurrett} is satisfied:\[
    \sum_{k} \P[X_{k,N} > b_N] \lesssim b_N/b_N^{1+\varepsilon/2} \to 0.
    \]
    Note that $\E[\bar X_{k,N}] \lesssim h_N$ since $\E[X_{k,N} \mid X_
    {k,N} \neq 0] < \infty$. 
    Note that \citep[Lemma 2.2.13][]{durrett2019probability} \[
    \E[\bar X_{k,N}^2] = \int_0^{b_N} 2y \P(X_{k,N} > y) \,dy \lesssim
    \int_0^{b_N} 2y \frac{h_N}{y^{1+\varepsilon/2}} \,dy
    \]
    via the same Markov's inequality argument. Calculating the integral
    shows that \[
    b_N^{-2}\sum_k \E[\bar X_{k,N}^2] \to 0
    \]
    and thus the second condition follows. The implication of 
    \cref{lemma:wllndurrett} is that \[
    \frac{1}
    {Nh_N}\sum_{i=1}^N W_i(c, h_N) Y_i^2(c) = \E\bk{
    \frac{1}
    {Nh_N}\sum_{i=1}^N W_i(c, h_N) Y_i^2(c) \one(Y_i^2(c) < Nh_N)
    } + o_p(1).
    \]
    Since $\E[Y_i^2(c) \mid J_i(c) = 1, R_{it_1} = r] < B_V < \infty$, \[
    \E\bk{
    \frac{1}
    {Nh_N}\sum_{i=1}^N W_i(c, h_N) Y_i^2(c) \one(Y_i^2(c) < Nh_N)
    } = \E\bk{
    \frac{1}
    {Nh_N}\sum_{i=1}^N W_i(c, h_N) Y_i^2(c)
    } + o(1),
    \]
    concluding the proof.
\end{myproof}

\end{document}